\documentclass{aa}

\usepackage{graphicx}
\usepackage{amsmath}	
\usepackage{amssymb}	
\usepackage{txfonts}
\usepackage{longtable}
\usepackage[normalem]{ulem}
\usepackage{tablefootnote}

\usepackage[title]{appendix}
\usepackage{blindtext}

\usepackage{natbib,twoopt}
\usepackage[breaklinks=true]{hyperref} 
\bibpunct{(}{)}{;}{a}{}{,}             
\makeatletter
  \newcommandtwoopt{\citeads}[3][][]{\href{http://adsabs.harvard.edu/abs/#3}%
    {\def\hyper@linkstart##1##2{}%
     \let\hyper@linkend\@empty\citealp[#1][#2]{#3}}}
  \newcommandtwoopt{\citepads}[3][][]{\href{http://adsabs.harvard.edu/abs/#3}%
    {\def\hyper@linkstart##1##2{}%
     \let\hyper@linkend\@empty\citep[#1][#2]{#3}}}
  \newcommandtwoopt{\citetads}[3][][]{\href{http://adsabs.harvard.edu/abs/#3}%
    {\def\hyper@linkstart##1##2{}%
     \let\hyper@linkend\@empty\citet[#1][#2]{#3}}}
  \newcommandtwoopt{\citeyearads}[3][][]%
    {\href{http://adsabs.harvard.edu/abs/#3}
    {\def\hyper@linkstart##1##2{}%
     \let\hyper@linkend\@empty\citeyear[#1][#2]{#3}}}
  \newcommandtwoopt{\citetaliasads}[3][][]%
    {\href{http://adsabs.harvard.edu/abs/#3}
    {\def\hyper@linkstart##1##2{}%
     \let\hyper@linkend\@empty\citetalias[#1][#2]{#3}}}
  \newcommandtwoopt{\citepaliasads}[3][][]%
    {\href{http://adsabs.harvard.edu/abs/#3}
    {\def\hyper@linkstart##1##2{}%
     \let\hyper@linkend\@empty\citepalias[#1][#2]{#3}}}
\makeatother

\hypersetup{
    colorlinks = true,
    citecolor = blue,
    linkcolor = blue,
    urlcolor = blue
}

\usepackage{placeins}

\newcommand{\msun}{M$_{\odot}$}
\newcommand{\rsun}{R$_{\odot}$}

\usepackage{adjustbox} 
\usepackage{float} 
\usepackage{xtab} 
\usepackage{booktabs} 
\usepackage[dvipsnames]{xcolor} 

\usepackage{tablefootnote} 
\usepackage{subfigure} 

\makeatletter
\renewcommand*{\@fnsymbol}[1]{\ensuremath{\ifcase#1\or *\or \dagger\or \ddagger\or
   \mathsection\or \mathparagraph\or \|\or **\or \dagger\dagger
   \or \ddagger\ddagger \else\@ctrerr\fi}}
\makeatother 

\defcitealias{2022A&A...668A..89U}{Paper I} 

\begin{document}

   \title{Volume-limited sample of low-mass red giant stars, the progenitors of  hot subdwarf  stars}

   \subtitle{ll. Sample validation\thanks{Based on observations collected with the CORALIE échelle spectrograph on the 1.2-m Euler Swiss telescope at La Silla observatory, European Southern Observatory (ESO) through the Chilean telescope time under program ID CN2019-58, CN2020B-77, CN2022A-82 and CN2023B-83.}
}

\author{Diego Benitez-Palacios\inst{1}, Murat Uzundag\inst{2}, Maja Vu\v{c}kovi\'{c}\inst{1}, Eduardo Arancibia-Rojas\inst{1}, Alex Dur\'{a}n-Reyes\inst{1},\\ Joris Vos\inst{3}, Alexey Bobrick\inst{4}, M\'{o}nica Zorotovic\inst{1} and Mat\'ias I. Jones\inst{5}}

\institute{Instituto de F\'isica y Astronom\'ia, Universidad de Valpara\'iso, Gran Breta\~na 1111, Playa Ancha, Valpara\'iso 2360102, Chile
\\\email{\url{diego.benitezp@postgrado.uv.cl}}
\and
Institute of Astronomy, KU Leuven, Celestijnenlaan 200D, B-3001 Leuven, Belgium  
\\\email{\url{muratuzundag.astro@gmail.com}}
\and
Astronomical Institute of the Czech Academy of Sciences, CZ-25165, Ond\v{r}ejov, Czech Republic 
\and
Technion - Israel Institute of Technology, Physics Department, Haifa, Israel 32000 
\and
European Southern Observatory, Alonso de C\'ordova 3107, Vitacura, Casilla,19001, Santiago, Chile }
           
\date{}

\titlerunning{Volume-limited sample of low-mass red giant stars}
\authorrunning{Benitez-Palacios et al.}

  \abstract
   {
Binary hot subdwarf B (sdB) stars are typically produced from low-mass red giant branch (RGB) stars that have lost almost all their envelopes through binary mass transfer while still fusing helium in their cores. Particularly, when a low-mass red giant enters stable Roche lobe overflow (RLOF) mass transfer near the tip of the RGB, a long-period sdB binary may be formed. 
   }
   { 
We aim to extend our previous volume-limited sample of 211 stars within 200 pc, a homogeneous sample of low-mass red giants, predicted progenitors of wide sdB binaries, 
to 500 pc and validate it. Additionally, our goal is to provide the distribution of stellar parameters for these stars.
   }
   {We refined our original 500 pc sample by incorporating Gaia DR3 parallax values and interstellar extinction measurements. Next, we collected multi-epoch high-resolution spectra for 230 stars in the volume-limited sample using the CORALIE échelle spectrograph from 2019 to 2023. To confirm or discard binarity, we combined astrometric parameters from Gaia with the resulting radial velocity variations. We derived the distribution of stellar parameters using evolutionary models and employed the equivalent evolutionary phase to verify the evolutionary stage of the stars in our sample. Finally, we compared our stellar parameters with the literature.}
   {The derived stellar parameters confirmed that 82\% of stars in our sample are indeed in the RGB phase, while 18\% are red clump (RC) contaminants. This was expected due to the overlapping of RGB and RC stars in the colour-magnitude diagram. Additionally, 75\% of the confirmed RGB stars have a high probability of being part of a binary system. Comparison with the literature shows good overall agreement with a scatter $\lesssim 15\%$ in stellar parameters, while the masses show somewhat higher dispersion ($\sim 20\%$).}
   {We have obtained the most complete volume-limited sample of binary RGB star candidates within 500 pc. These systems are likely progenitors of hot subdwarfs and other classes of stripped helium stars.}
   

   \maketitle

\section{Introduction}

Hot subdwarf B (sdB) stars are core helium-burning (CHeB) stars located at the hot end of the horizontal branch (HB), the so-called extreme horizontal branch (EHB). They are characterized by surface temperatures in the range of $20\,000 - 40\,000$ K and surface gravities between $5 \lesssim \log g \lesssim 6$ dex (cm/s$^2$) 
(see \citeads{2016PASP..128h2001H} for a full review).

Unlike typical HB stars, sdBs have a very thin hydrogen-rich envelope ($\sim0.01$\msun) which is insufficient to sustain H shell burning, suggesting that sdBs are remnant cores of red giant branch (RGB) stars that ignited He and lost most of their envelopes at the same time \citepads{2016PASP..128h2001H}.

Observational evidence shows that a significant fraction of sdBs are found in close binary systems (orbital period < 10 d) with WD or M dwarf companions (e.g. \citeads{2001MNRAS.326.1391M}; \citeads{2004Ap&SS.291..321N}; \citeads{2011MNRAS.415.1381C}; \citeads{2003AJ....126.1455S}), as well as in long-orbital-period (> 250 d) binary systems with F, G or K type companions (e.g. \citeads{2012A&A...548A...6V}; \citeads{2013A&A...559A..54V}; \citeads{2017A&A...605A.109V}). Hence, binary evolution is considered the main formation mechanism for sdBs. Indeed, \citetads{2020A&A...642A.180P} argued that binary interaction is always required to form these systems.

\citeads{2002MNRAS.336..449H}, \citeyearads{2003MNRAS.341..669H} explored several formation channels for sdBs through binary interactions and their relative importance through binary population synthesis (BPS) studies. They identified three main formation channels that can explain the entire distribution of sdBs, namely common envelope (CE) evolution \citepads{1976IAUS...73...75P}, stable Roche lobe overflow (RLOF) evolution \citepads{2002MNRAS.336..449H}, and double white dwarf merger \citepads{1984ApJ...277..355W}, accounting for close, wide, and single sdBs, respectively. However, they predicted a period distribution of systems forming through the stable RLOF channel ranging from 10--500 days, while observation programs have discovered periods exceeding 1000 days (\citeads{2012ApJ...758...58B}; \citeads{ 2012MNRAS.421.2798D}; \citeads{2012ASPC..452..163O}; \citeads{2012A&A...548A...6V}; \citeads{2013A&A...559A..54V}; \citeads{2014ASPC..481..265V}; \citeads{2017A&A...605A.109V}).

\citetads{2013MNRAS.434..186C} revisited the models of \citetads{2002MNRAS.336..449H} incorporating a more sophisticated treatment of angular momentum loss and atmospheric RLOF, showing that the final mass--orbital period relation increases with composition, resulting in models with periods up to 1100 days for solar composition and up to 1600 days when atmospheric RLOF is considered.

However, observations of long-period sdB systems also show high eccentricity, indicating a need for improvement in the models. \citetads{2015A&A...579A..49V} proposed three eccentricity pumping mechanisms that could be responsible for the observed distribution. While two of these mechanisms (phase-dependent RLOF and interaction between a circumbinary disk and the binary) could potentially result in eccentric binaries, their models fail to reproduce the observed trend between period and eccentricity. Additionally, the orbit must be eccentric prior to mass transfer, with the proposed mechanisms enhancing this eccentricity.

\citetads{2019MNRAS.482.4592V} discovered a correlation between the orbital period and the mass ratio (defined as $M_{\text{sdB}}/M_{\text{MS}}$) in observed wide sdB binaries, indicating lower mass ratios at longer orbital periods. They also identified a correlation between the initial mass ratio at the onset of RLOF and the core mass of the sdB progenitor using theoretical models. This correlation was used to test the stability of the RLOF, assuming that the companion has not accreted any material during the mass transfer phase. This assumption, supported by observational evidence \citepads{2018MNRAS.473..693V}, yielded the maximum initial companion mass and the lowest mass ratio of the binary at the onset of RLOF. They concluded that the initial mass ratio decreases with increasing core mass.

A significant step forward in understanding wide sdB systems was taken by \citetads{2020A&A...641A.163V}. Through a statistically significant BPS study using the \textit{Modules for Experiments in Stellar Astrophysics} (MESA; \citeads{2011ApJS..192....3P}) code, they achieved an excellent match with the observed period-mass ratio correlation without explicit parameter tuning. Furthermore, their study revealed a strong agreement with the observed period-metallicity correlation, highlighting the influence of the Milky Way's metallicity history on the properties of post-mass transfer binaries. Additionally, they predicted several new correlations that connect the observed sdB binaries with their progenitors.

Based on the predictions made by \citetads{2020A&A...641A.163V}, an observational campaign to search for the progenitors of wide sdB + MS binaries was initiated by \citetads[][, hereafter \citetalias{2022A&A...668A..89U}]{2022A&A...668A..89U}. Low-mass RGB stars were selected from Gaia DR2 using colour-magnitude cuts to exclude contaminants (e.g. stars with UV and IR excess) and quality criteria ensuring parallax and flux uncertainties below 10\%. The low parallax uncertainty enabled reliable distance estimates (calculated as the inverse of parallax), allowing the identification of stars within a 500 pc volume. High-resolution spectroscopy was combined with Gaia eDR3 astrometric indicators to classify binary systems with orbital periods of 100–900 days.

The present study is the continuation of the work initiated in \citetaliasads{2022A&A...668A..89U}, where stars in the selected region were originally observed up to a distance of 200 pc. Our primary aim is to extend the observed sample up to a volume of 500 pc and validate the sample by providing a comprehensive analysis of the physical properties of the stars in the sample. The long-term goal of the project is to solve the orbital parameters of the selected low-mass RGB stars in order to confirm their binarity. Furthermore, having a volume-limited sample of binary low-mass RGB stars, which are possible progenitors of sdB systems, together with the recently published 500 pc sample of sdBs  \citepads{2024A&A...686A..25D} will allow to perform a direct comparison of both populations in the same volume. Combining observational results with simulations from theoretical BPS studies will further help to understand the physics of mass transfer in the stable RLOF case.

The paper is organized as follows. In sec. \ref{sec:Methods}, we
describe the observations and data reduction process; we also explain the method used to compute radial velocities (RVs) and stellar parameters. In sec. \ref{sec:Results}, we present our results and in sec. \ref{sec:Discussion}, we discuss their implications. Finally, in sec. \ref{conclusion}, we summarize our results and give an outlook for the future.

\section{Methods}
\label{sec:Methods}

In this section, we describe the current status of our observations and the data reduction process, followed by the methodology employed to compute RVs and stellar parameters.

\subsection{Observations and data reduction}
\label{subsec:observations}

We have updated the selected targets outlined in \citetaliasads{2022A&A...668A..89U} by incorporating Gaia data release 3 (DR3; \citeads{2023A&A...674A...1G}) parallaxes and interstellar reddening. For the latter, we used the \texttt{Combined19} 3D dust maps, which combines the maps from \citetads{2003A&A...409..205D}, \citetads{2006A&A...453..635M}, and \citetads{2019ApJ...887...93G}, as implemented in the Python package \texttt{mwdust}\footnote{\url{https://github.com/jobovy/mwdust}} \citepads{2016ApJ...818..130B}. Details of these computations are provided in Appendix \ref{Appendix:update_sample}. Updating the new parallaxes from DR3 and applying interstellar reddening excluded about 10\% of the stars in the original sample \citepaliasads{2022A&A...668A..89U}.

Spectroscopic observations of the low-mass RGB candidates are ongoing using the CORALIE échelle spectrograph \citepads{2001Msngr.105....1Q} mounted at the Swiss 1.2-meter Leonhard Euler telescope at La Silla observatory in Chile. CORALIE has a resolving power of R$\sim$60\,000, allowing for a long-term radial velocity precision up to 5 m\,s$^{-1}$ \citepads{2010A&A...511A..45S}. The spectrograph is fed by two fibres, one centred on the target and the other on a reference lamp for wavelength calibration, which can be either a Thorium-Argon (ThAr) lamp or a Fabry-Perot (FP) interferometer.

We extended the observations outlined in \citetaliasads{2022A&A...668A..89U} from 200 to 500 pc. Five observing runs have been conducted so far, gathering a total of 419 high-resolution spectra for 235 different stars (including 5 standard stars). A summary of the observing runs is provided in Table \ref{tablespec1}.

We reduced the data using the customized $\tt CERES$ pipeline \citepads{2017PASP..129c4002B}, which performs all the extraction processes from basic bias, to order tracing, wavelength calibration, and computation of precise RVs using the cross-correlation function (CCF) technique \citepads{1967ApJ...148..465G}. 

\begin{table}
\setlength{\tabcolsep}{2.5pt}
\renewcommand{\arraystretch}{1.1}
\centering
\caption{Observing log of the spectroscopic data obtained for the low-mass red giant stars studied in this work.}
\begin{tabular}{cccc}
\hline \hline
  Instrument & Date   & Range & S/N \\
               &        &  (\AA) &   \\
\hline

CORALIE &  15-16-17 June 2019        &   3800-6800   &  30-80    \\
CORALIE &  18-19-20 November 2021    &   3800-6800   &  50-100   \\
CORALIE &  4 April 2022 &   3800-6800   &   40-60   \\
CORALIE &  14-16-17 July 2023 &   3800-6800   &   10-60   \\
CORALIE &  6-7-8-9 October 2023 &   3800-6800   &   25-65   \\
\hline 
\label{tablespec1}
\end{tabular}
\end{table}

\subsection{Radial velocity measurements}
\label{subsec:RV}

$\tt CERES$ computes the CCF for each extracted order using one of three possible stellar masks (G2, K5, and M5) chosen by the user. We used the G2 mask for all our targets. The CCFs are then combined using a weighted sum based on the median S/N of each order.

The CCF reaches its minimum value close to the radial velocity of the observed star. The actual RV is computed by fitting a Gaussian to the CCF, and the resulting mean is taken to be the RV of the star. 

The uncertainties are calculated using empirical scaling relations determined via Monte Carlo simulations that employ the mean S/N per resolution element close to the Mg triplet zone as priors. The scaling relations are combined with the dispersion of the Gaussian fit to the CCF, $\sigma_{ccf}$, and the continuum S/N at $5130$ \AA, denoted as $SN_{5130}$. The exact equation is \citepads{2017PASP..129c4002B}: 

\begin{equation}\label{ec:CCF_error}
\text{RV error} = b + \frac{a(1.6 + 0.2\sigma_{ccf})}{SN_{5130}},
\end{equation}

\noindent where $a$ and $b$ are the scaling parameters found with the Monte Carlo simulations. The RVs of all our targets are presented in Table \ref{tab:RV_Diego}, where we included the date and S/N of each observation, as well as the category for each star. The category is based on the robust binary classification method proposed in \citetaliasads{2022A&A...668A..89U} where the renormalized unit weight error (RUWE) and astrometric excess noise (AEN), taken from Gaia eDR3 \citepads{2021A&A...649A...1G}, are combined with radial velocity variations ($\Delta$RV) between different epochs to classify the targets as:

\begin{enumerate}
\item If at least two of the following conditions are fulfilled: RUWE $\geq 1.4$, AEN $\geq 0.4$, and $\Delta$RV $\geq 0.1$.

\item Only one of the three parameters exceeds the threshold.

\item None of the thresholds are exceeded.

\end{enumerate}

Objects categorized as 1 or 2 have a high probability of belonging to binary systems. 

\subsection{Derivation of stellar parameters}
\label{subsec:spec}

We used $\tt SPECIES$, a publicly available\footnote{\url{https://github.com/msotov/SPECIES/wiki}} code mostly written in Python, to derive stellar parameters. The code retrieves photometric and astrometric data from several catalogues and uses parallaxes and proper motions to estimate the evolutionary state of stars \citepads{2007MNRAS.380.1230C}. Giants are assigned an initial metallicity ([Fe/H]$_{\text{initial}}$) of zero, while initial effective temperature ($T_{\text{eff, initial}}$) is calculated using colour-temperature relations from \citetads{1999A&AS..140..261A}. Initial surface gravity ($\log$g$_{\text{initial}}$) is set by equation (1) from \citetads{2018A&A...615A..76S}. These parameters, together with Fe I and Fe II line equivalent widths, form an initial atmospheric model that is then iteratively refined using the 2017 version of $\tt MOOG$ \citepads{2012ascl.soft02009S} and ATLAS9 atmospheric models.  \citepads{2003IAUS..210P.A20C}. Macroturbulence (v$_{\text{mac}}$) is derived from \citetads{2016A&A...592A.156D}, and rotational velocity (v$\sin i$) is determined by fitting synthetic profiles to absorption lines.

\begin{figure}[h!]
    \centering
    \includegraphics[height=0.8\linewidth, width=\linewidth]{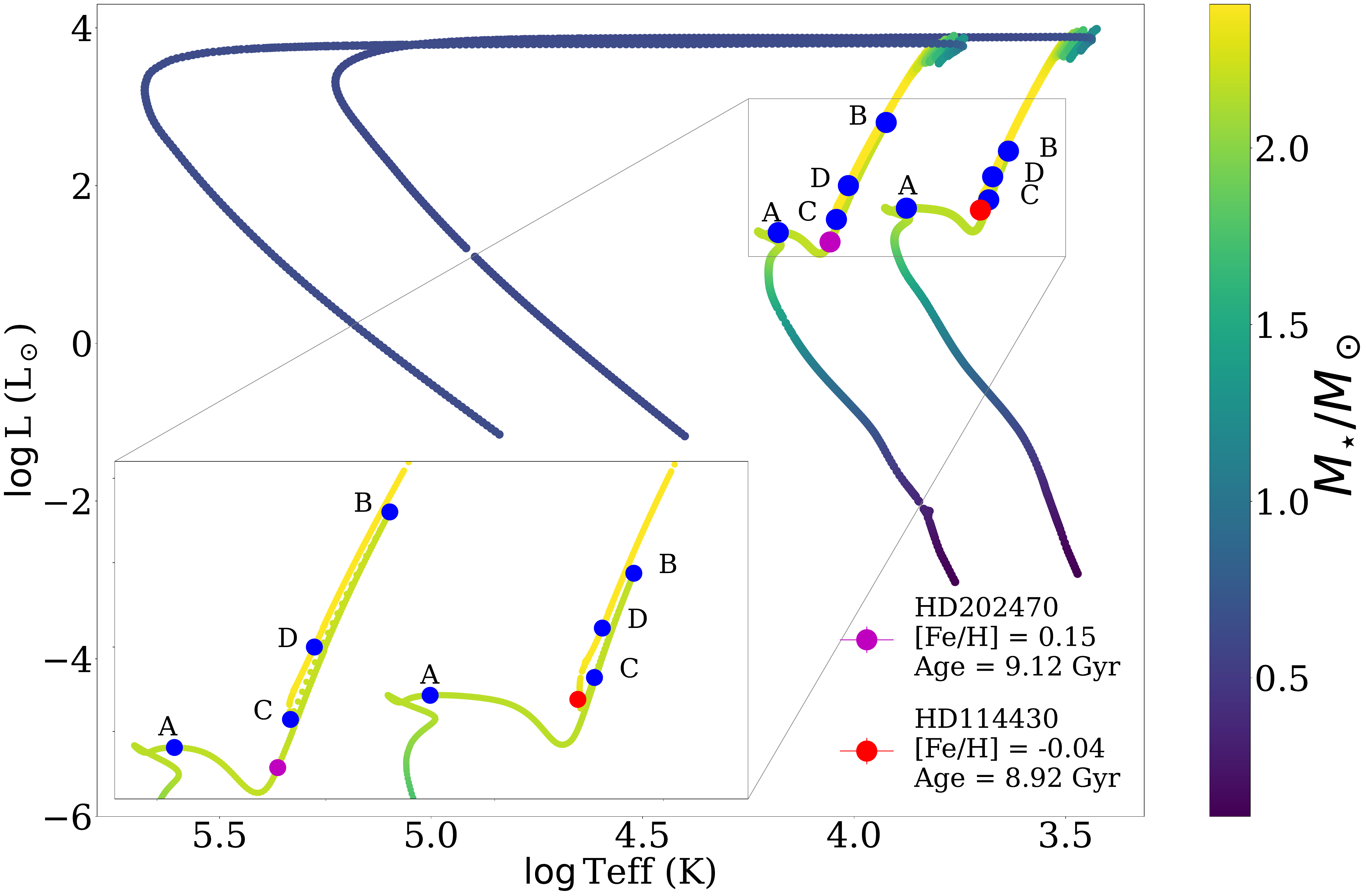}
    \caption{Isochrones calculated using $\tt SPECIES$ for the stars HD114430 (red point) and HD202470 (magenta point). The isochrone of HD202470 has been displaced by 10\% to the left for clarity. The blue dots indicate the location of the TAMS (marked as A), RGBtip (B), ZACHeB (C), and TACHeB (D). The inset shows a zoom to the location of the stars, where it can be noted that HD202470 is still ascending the RGB phase while HD114430 has already passed the ZACHeB; this is reflected in their EEPs (494 and 661, respectively). The metallicity and age used to create each isochrone are shown in the figure. The size of the error bars is smaller than the size of the points.}
    \label{fig: isochrones}
\end{figure}

For determining physical parameters such as mass, radius, age, luminosity, and evolutionary stage, $\tt SPECIES$ uses the $\tt isochrones$\footnote{\url{https://github.com/timothydmorton/isochrones}} package \citepads{2015ascl.soft03010M}. Based on the derived atmospheric parameters, $\tt isochrones$ generates stellar tracks using MESA Isochrones and Stellar Tracks (MIST; \citeads{2016ApJ...823..102C}) models and interpolates them using $\tt MultiNest$\footnote{\url{https://github.com/JohannesBuchner/PyMultiNest}} \citepads{2009MNRAS.398.1601F}. A Bayesian approach is then employed to select the best model. Detailed descriptions of this computation process are provided in \citetads{2018A&A...615A..76S} and \citetads{2021A&A...647A.157S}.

One useful parameter output by $\tt SPECIES$ is the equivalent evolutionary phase (EEP). As mentioned in \citetads{2016ApJS..222....8D}, $\tt MIST$ interpolates among a set of stellar tracks to construct isochrones by identifying specific evolutionary stages through physical conditions. These stages are known as the `primary' EEPs. The most relevant primary EEPs for this work are:

\begin{itemize}
\item  Terminal age main sequence (TAMS): When the central Hydrogen mass fraction ($X_c$) decreases to $X_c = 10^{-12}$.

\item The tip of the RGB (RGBTip): point at which
the stellar luminosity reaches a maximum (or $T_{\text{eff}}$ a minimum) after core H burning is complete but before core He burning has progressed significantly. Specifically, the central He mass fraction ($Y_c$) must satisfy the relation $Y_c > Y_{c,\text{TAMS}}- 0.01$.

\item The zero-age core He burning (ZACHeB): Marks the onset of sustained core He burning, indicating the HB phase. It is identified as a minimum in the core temperature while $Y_c > Y_{c,\text{RGBTip}} - 0.03$.

\item Terminal age core He burning (TACHeB): When the central He mass fraction decreases to $Y_c = 10^{-4}$, marking the end of core He burning.
\end{itemize}

After the identification of these points, the segment between two adjacent primary EEPs is further divided into equally spaced `secondary' EEPs according to a distance metric function.

Following \citetads{2021A&A...647A.157S}, we consider 454 < EEP < 631 to be the RGB phase and 631 $\le$ EEP $\le$ 707 to be the HB phase. $\tt SPECIES$ also infers the probability of a star to be either in the RGB or the HB
phase using the EEPs. As an example, in Fig. \ref{fig: isochrones}, we show isochrones computed for the stars HD114430 and HD202470, including the metallicity and age used for each isochrone. We can see that HD202470 (magenta point in the figure) is in the early stages of the RGB phase, way before reaching the tip (point B). This is reflected in its EEP value, 494, which is just slightly above the value used to identify the TAMS (454). On the other hand, HD114430 (red point) has an EEP value of 661; therefore, this star has just recently passed the ZACHeB stage (point C), as can be noted in Fig. \ref{fig: isochrones}.

\section{Results}
\label{sec:Results}

Five of the observed stars (29PSC, HD103433, HD177668, HD219470, and HD220096) suffered from errors in the fit when using $\tt SPECIES$. A visual inspection of the spectra did not reveal any evident reason for the failure. 
While we do list their RVs computed using $\tt CERES$, we decided to exclude them from further analysis. 
Additionally, 33 RGB binary systems with orbital periods between 100\,d and 900\,d were identified in \citetaliasads{2022A&A...668A..89U} by crossmatching the entire RGB sample with different surveys within a volume of 200 pc (\citeads{2016A&A...593A.133B}; \citeads{2004A&A...421..241S}; \citeads{2011A&A...536A..71J}; \citeads{2008AJ....135..209M}; \citeads{2016AJ....152...19W}; \citeads{2023A&A...674A..39G}), seven of which are part of our observed sample. We extended this search to 500\,pc and found 230 new potential binaries within our sample by cross-matching it with the non-single star (NSS) catalogue from Gaia DR3\footnote{\url{https://cdsarc.cds.unistra.fr/viz-bin/cat/I/357}} \citepads{2023A&A...674A..34G}. The NSS catalogue provides orbital solutions and classifications for spectroscopic binaries based on RV time-series. It includes single-lined spectroscopic binaries (SB1), circular solutions (SB1C), and systems exhibiting trends or stochastic behaviour. This catalogue, built using robust pipelines and extensive validation, represents the first Gaia release with orbital solutions derived from spectroscopic data. For a comprehensive description of the data processing and methodologies, refer to \citetads{2025A&A...693A.124G}. We have already observed 57 matching stars. The full list is presented in Tables \ref{tab:cross_NSS_twobody} and \ref{tab:cross_NSS_acc}. 

\subsection{Results from \texttt{CERES}}
\label{subsec:Results from ceres}

Table \ref{tab:RV_Diego} listed the full 419 RV measurements for the 235 different stars, 131 of them having multi-epoch spectra. The first five listed stars are the RV standards observed in the different runs, followed by the seven known binaries found in \citetaliasads{2022A&A...668A..89U}. Subsequent stars are ordered first by category, which represents the probability of a star being part of a binary system, then by number of epochs, and finally by alphabetical order. Multiple epochs for the same star are ordered by date.

From stars in Table \ref{tab:RV_Diego}, 112  are assigned category 1, 66 of which already have multiple epoch observations. From the remaining targets, 60 are categorized as category 2, 16 of which already have multi-epoch observations. The remaining 58 targets are assigned category 3.

\subsection{Results from \texttt{SPECIES}}
\label{subsec:Results from species}

\begin{figure*}[h!]
\centering
\includegraphics[scale=.42]{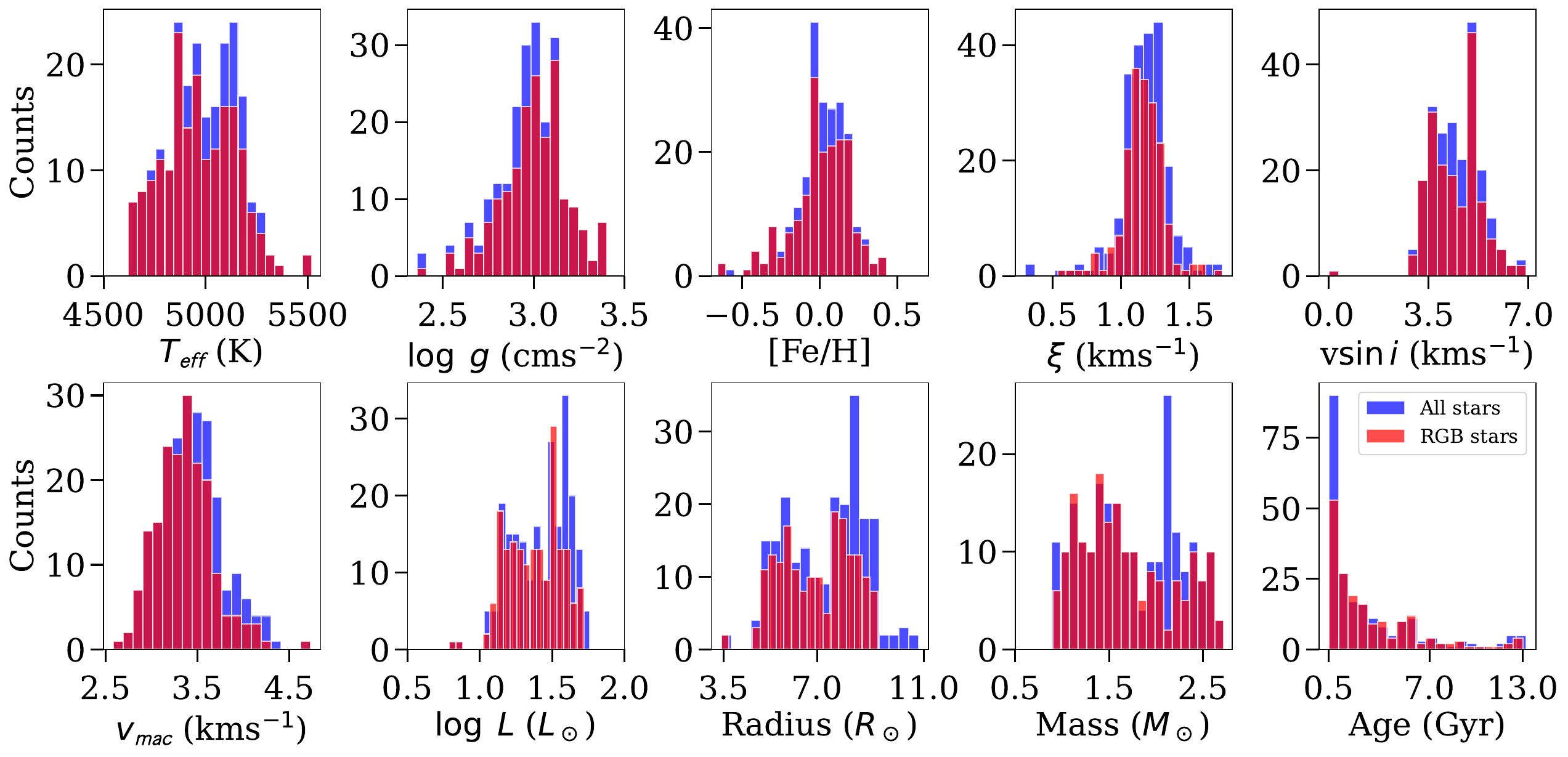}
\caption{Distribution of stellar parameters calculated using $\tt SPECIES$ for all stars in our sample. The RGB stage represented in red are stars with 454 < EEP < 631.}
\label{fig:dist_species}
\end{figure*}

\begin{figure*}[h!]
\centering
\subfigure{\includegraphics[height=0.4\textwidth, width=0.45\textwidth]{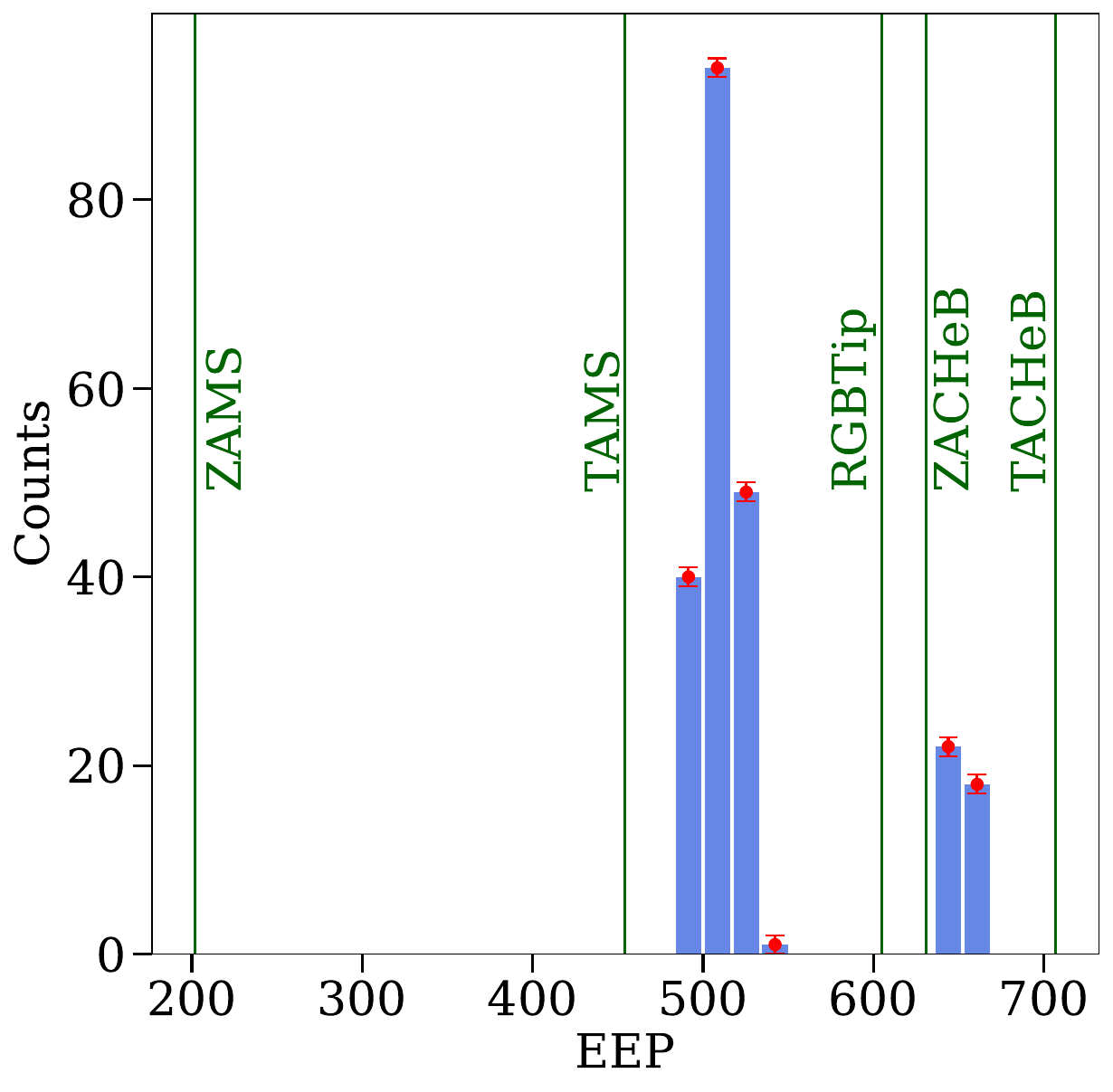}}
\hspace{0.05\textwidth}
\subfigure{\includegraphics[height=0.4\textwidth, width=0.45\textwidth]{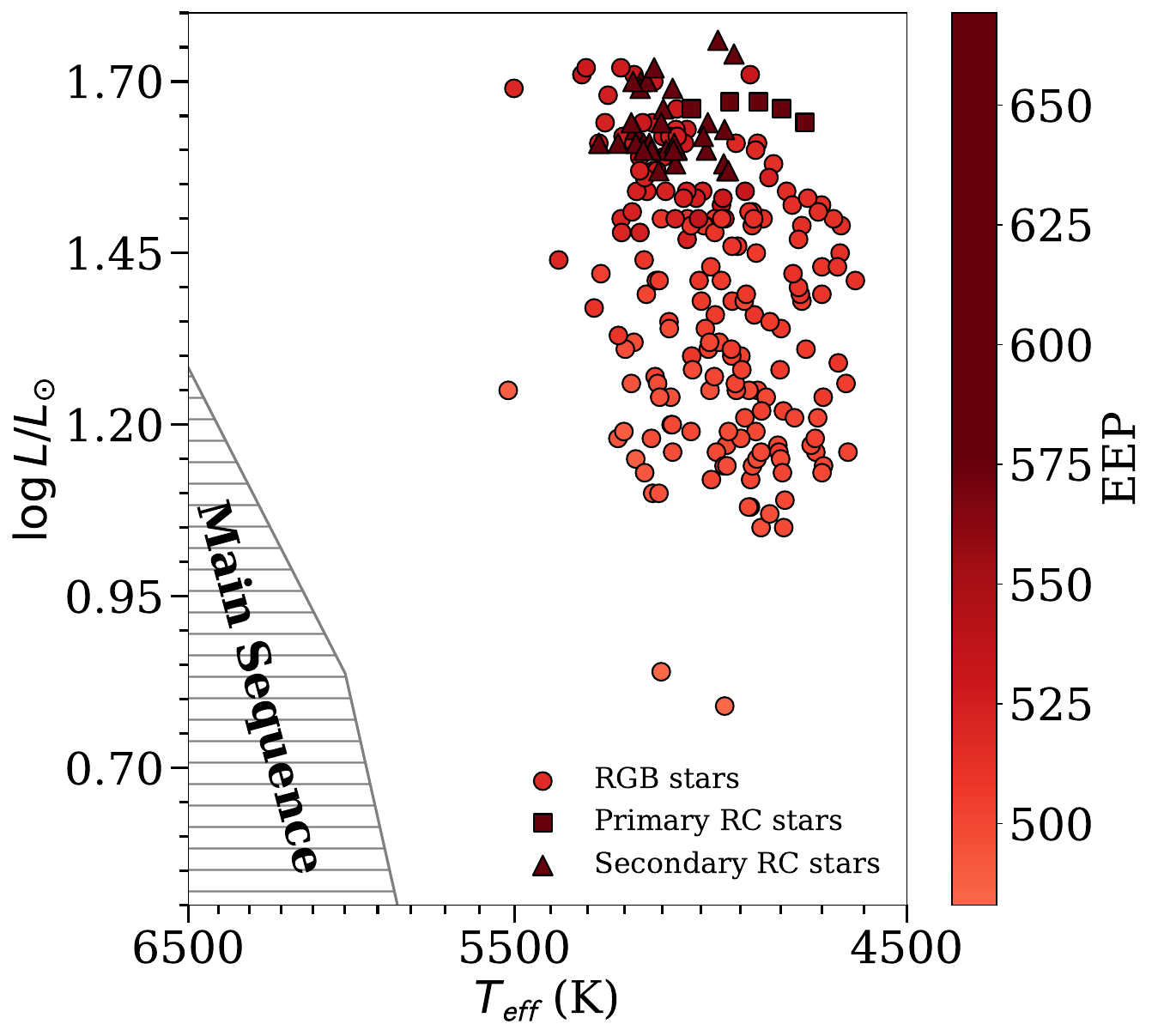}}
\caption{\textbf{Left:} Distribution of EEPs for the targets in our sample, 82\% of the stars are on the RGB phase while the remaining 18\% are core He burning stars. We did not find any MS stars in our sample. \textbf{Right:} HRD for stars in our sample constructed based on the $T_{\text{eff}}$ and $L$ given by $\tt SPECIES$ and colour-coded by the EEPs. Circles represent RGB stars, while squares and triangles are used for primary and secondary RC stars, respectively (see the text for definitions and references). The temperature axis is logarithmic. A theoretical MS constructed with MESA from TAMS to ZAMS is shown as the grey area for comparison. Only a portion of the MS band is shown.}
\label{fig:EEP}
\end{figure*}

The distribution of stellar parameters calculated with $\tt SPECIES$ for the stars in our sample is shown in Fig. \ref{fig:dist_species}, where a total of 20 bins were used for each histogram, and listed in Table \ref{tab: species_all}. We measured the distribution of EEPs for stars in our sample (left panel of Fig. \ref{fig:EEP}, where we used a bin width of 15) and distinguished between RGB and core He burning stars accordingly. The red histograms in Fig. \ref{fig:dist_species} show the distributions of stellar parameters for the identified RGB stars in our sample, while the distributions for the complete sample are shown as the blue histograms. To confirm the validity of the EEPs, we constructed a Hertzsprung–Russell diagram (HRD) with the stellar parameters computed by $\tt SPECIES$ for stars in our sample and mapped the colour of each target to the value of its EEP (right panel of Fig. \ref{fig:EEP}). We also use different symbols for RGB and CHeB stars in Fig. \ref{fig:EEP} (right panel). RGB stars are represented by circles, while CHeB stars are distinguished by squares and triangles. Squares denote stars with masses $\sim 1.0$ \msun, corresponding to low-mass Red Clump (RC) stars that ignited helium under degenerate conditions, known as primary RC stars \citepads{2016ARA&A..54...95G}. Triangles represent more massive RC stars (M$\sim 2.0$ \msun), for which helium ignition occurred under non-degenerate conditions; these are referred to as secondary RC stars \citepads{1999MNRAS.308..818G}. As a comparison, a theoretical MS band from the zero-age main sequence (ZAMS) to the TAMS constructed using MESA is shown as the grey area in the right panel of Fig. \ref{fig:EEP}. The band was constructed from a grid of models with solar composition and masses ranging from 1 \msun\, to 15 \msun\, in intervals of 1 \msun.

\section{Discussion}
\label{sec:Discussion}

\subsection{Potential binaries}

In total, 169 ($\sim 75\%$) of the stars in the observed sample have a high probability of being part of a binary system (110 classified as 1 and 59 as 2), and 109 ($\sim 48\%$) of these already have multi-epoch observations. Follow-up observations after having the final sample will focus primarily on these stars to construct RV curves, derive orbital parameters, and finally confirm binarity.

As mentioned in sec. \ref{sec:Results}, we have found 230 new potential binaries within our sample by cross-matching it with the NSS catalogue from Gaia DR3 (see Tables \ref{tab:cross_NSS_twobody} and \ref{tab:cross_NSS_acc}). 119 of these objects are from the \texttt{nss\_acceleration\_astro} catalogue; hence, they do not have orbital parameters measured and are categorized as potential binaries due to having a non-linear proper motion which is compatible with an acceleration solution. We have already observed 39 of them. From the remaining 111 stars, 43 are from the \texttt{nss\_non\_linear\_spectro} catalogue; therefore, they also lack orbital parameters. However, they are consistent with non-single-star orbital models for spectroscopic binaries, and we have already observed three of them. The remaining 68 objects are from the \texttt{nss\_two\_body\_orbit} catalogue; these objects do have orbital parameter measurements, and we have observed 15 of them. 

\subsection{Identified red giants}
\label{subsec:Identified Red Giants}

Fig. \ref{fig:EEP} reveals that all stars in our sample are indeed beyond the MS. The HDR diagram in the right panel of the figure shows good agreement with the hypothesis that the observed targets are RGB stars. Since stars with EEP $>$ 631 overlap with the identified RGB stars, they are most probably RC contaminants, which are expected given their characteristic colours and luminosity \citepads{2016ARA&A..54...95G}. We expect to find secondary RC stars within our observed sample, as they are less luminous than primary RC stars \citepads{1999MNRAS.308..818G}. In total, 184 ($\sim$ 82\%) are on the RGB, while the remaining 40 ($\sim$ 18\%) stars are RC contaminants, among which only 2\% are primary RC stars and 16\% are secondary RC stars, as expected.

Of the 184 identified RGB stars, 88 are categorized as 1, 50 as 2, and 46 as 3. In total, $\sim$75\% of the identified RGB stars have a high probability of being part of a binary system. Once orbital parameters are determined, this sample will be useful as a prior for BPS studies aiming to test the stability of the mass transfer in the RLOF. Due to these results, we consider the selection criteria proposed in \citetaliasads{2022A&A...668A..89U} as valid to identify low-mass RGB stars in binary systems, which are possible progenitors of wide sdB stars.

\subsection{Stellar parameters}
\label{subsec:Stellar parameters}

Fig. \ref{fig:dist_species} shows the distributions of stellar parameters for the stars in our sample. We will discuss in detail these results for the RGB sample (red histograms) and mention the differences with the complete sample (blue histograms) only when a deviation is significant.

The effective temperature distribution has a mean of 5000 K and a standard deviation of 177 K. However, it is not entirely symmetrical, showing a drop at about 5000 K, followed by a quick rise. This pattern suggests a bimodal distribution rather than a Gaussian, likely due to the presence of stars in the late sub-giant or early RGB stage, which still have a relatively high temperature.

In contrast, the $\log g$ distribution is more symmetrical, with a mean of 3.0 dex (cgs) and a standard deviation of 0.2 dex. However, there is a small asymmetry towards higher $\log g$ values, which is again consistent with the presence of sub-giant stars.

The metallicity distribution shows a pronounced peak at $-0.07$ dex. However, the overall distribution is quite symmetrical around $0.0$, with the presence of a few metal-poor stars. This can be explained by the fact that RGB stars are systematically older than typical unevolved stars in the solar neighbourhood. Therefore, RGB stars in our sample reflect the lower metallicity the Galaxy had at the time of their formation.

The microturbulent velocity distribution is quite symmetric, concentrated around a mean of 1.16 km\,s$^{-1}$, with a standard deviation of 0.16 km\,s$^{-1}$, while the distribution of rotational velocity is clearly non-symmetric, with a mean of 4.48 km\,s$^{-1}$ and a standard deviation of 0.88 km\,s$^{-1}$. One star rotates at $v\sin i = 0$ km\,s$^{-1}$, which is only possible if the inclination is $\sim 0$. However, this is an artefact from the code since the measurement error is also $0$. Inspecting the star (BD+004462), we realized that the code could not fit the lines used to derive the rotational velocity. There is no obvious reason for this failure from the visual inspection, we will investigate it further in the future. 

Macroturbulent velocity shows a symmetric distribution around a mean of $3.37$ km\,s$^{-1}$ with a standard deviation of $0.32$ km\,s$^{-1}$ and a small asymmetry towards higher values. Regarding luminosity, the distribution shows a mean of $\log L\,(L_\odot) = $1.38 with a standard deviation of 0.19.

The radius distribution also seems to present a bimodality. For stars with a radius lower than $\sim 7.5$ \rsun, the mean is 5.96 \rsun, and the standard deviation is 0.80 \rsun, while for stars with radii larger than $\sim 7.5$ \rsun, the mean and standard deviation are 8.28 \rsun\, and 0.48 \rsun. Notably, CHeB stars exhibit larger radii than RGB stars in the sample, which is expected given their higher luminosity and similar surface temperature. Observations of RC stars through interferometry and spectroscopy indicate that they can easily reach sizes up to 10 \rsun \citepads{2018A&A...616A..68G}.

The mass distribution also appears to be bimodal, with masses $\lesssim 2$\msun\, having a mean of 1.42 \msun\, and a standard deviation of 0.28 \msun\,, while for masses $\gtrsim 2$\msun\, the distribution has a mean of 2.38 \msun\, with a standard deviation of 0.19\msun. However, there is no physical reason to think that RGB stars should have two different populations at low and intermediate-mass. Furthermore, one has to be very careful when interpreting the mass distribution of stars in an evolving stage. It might be tempting to say that one expects to have many more low-mass stars due to the behaviour of the initial mass function. However, stellar main-sequence lifetimes, as well as the duration of the evolved phases, RGB and CHeB in our case, depend non-trivially on the initial mass. Therefore, the mass distribution of the RGB stars should not be proportional or depend in any obvious way on the initial mass function. Given that our sample is not yet complete, a deeper analysis of the mass distribution will have to wait until we compile the full 500 pc sample.

On the other hand, CHeB stars also exhibit a bimodality in the mass distribution. This bimodality arises from our selection criteria (see eq. (1) to (4) in \citeads{2022A&A...668A..89U}), which excluded most low-mass primary RC stars that ignited helium under degenerate conditions \citepads{2016ARA&A..54...95G}, except for a few with masses $\sim 1$ \msun, which are the faintest and reddest among this group. Conversely, the second and more pronounced peak originates from secondary RC stars, where helium ignition occurred under non-degenerate conditions \citepads{1999MNRAS.308..818G}. As expected, the peak of the mass distribution for CHeB stars lies between $\sim 2$ and $\sim 2.3$ \msun, which is consistent with the expected distribution for secondary RC stars \citepads{2016ARA&A..54...95G}. For reference, the locations of the primary and secondary RC on the Gaia DR2 colour-magnitude diagram are shown in Fig. 10 of \citetads{2018A&A...616A..10G}.

\begin{figure*}[h!]
\centering
\subfigure{\includegraphics[height=0.35\linewidth, width=0.35\linewidth]{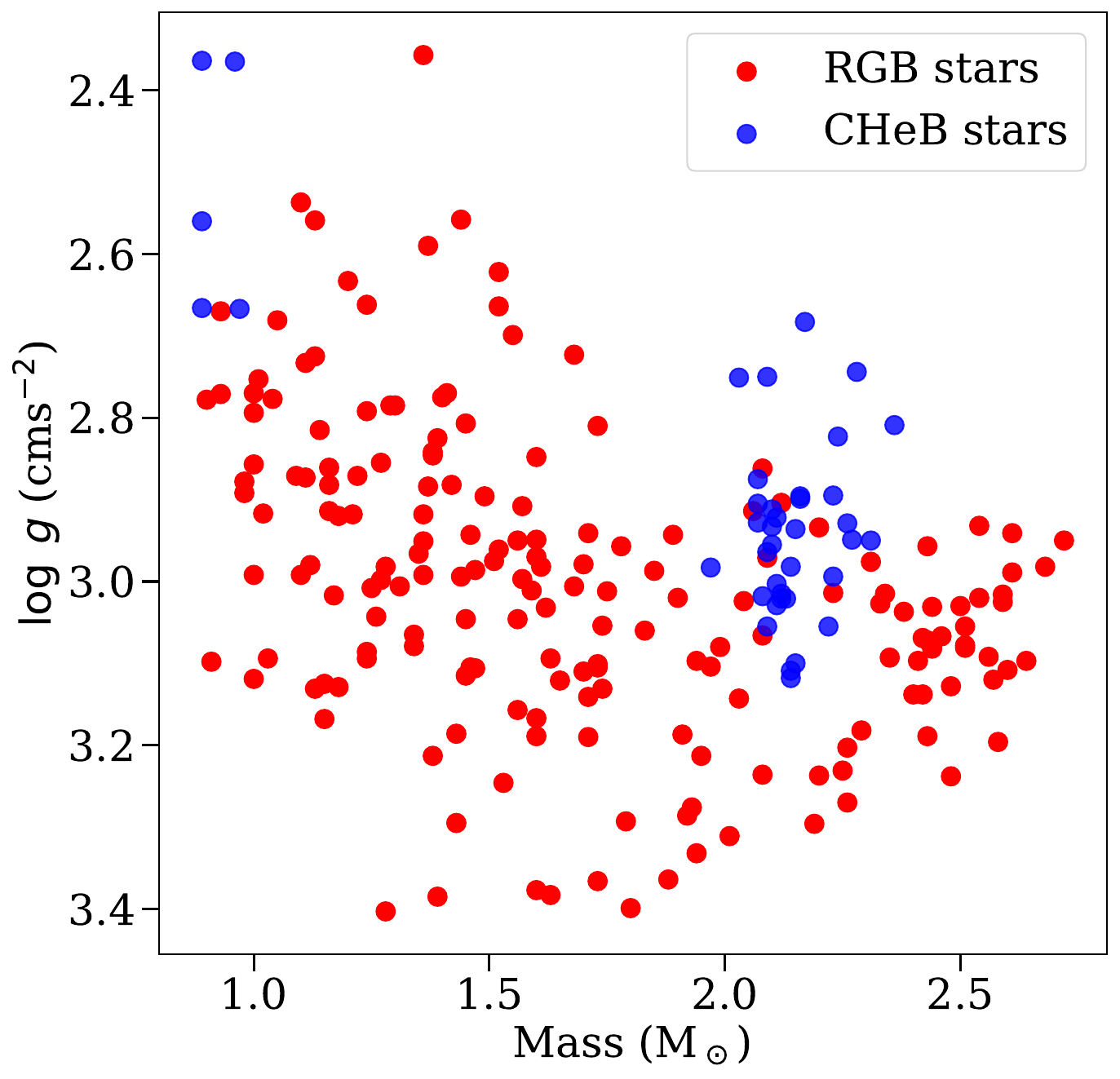}}
\hspace{0.05\textwidth}
\subfigure{\includegraphics[height=0.35\linewidth, width=0.35\linewidth]{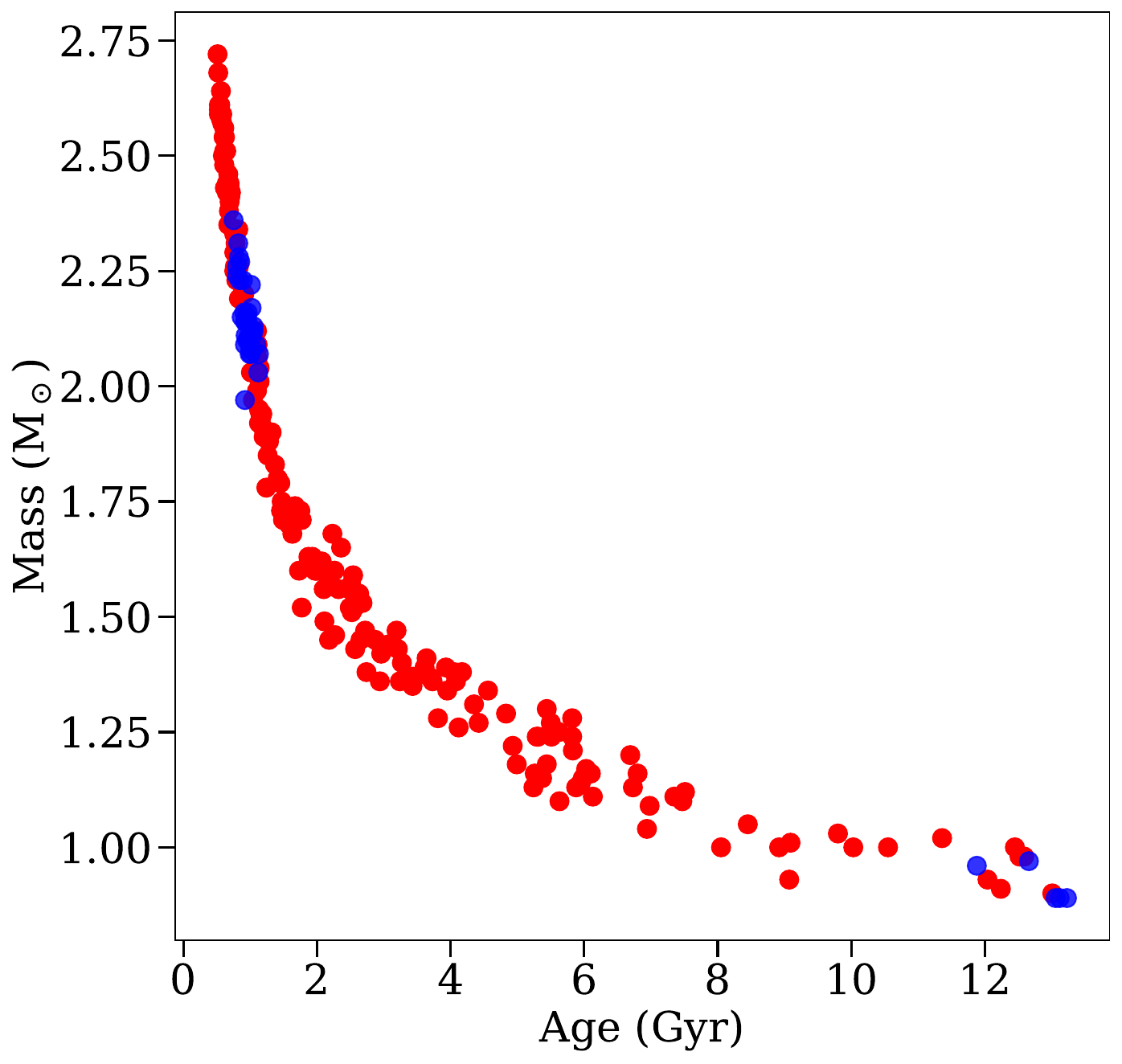}}
\caption{\textbf{Left:} Correlation between mass and surface gravity for stars in our sample. Two populations can be seen for the CHeB stars at around $\sim1$ and $\sim$ 2 \msun, which is in accordance with the results of \citetads{2016ARA&A..54...95G}.  
\textbf{Right:} Correlation between mass and age showing that high-mass stars are younger than low-mass stars, as expected.}
\label{fig:mass_correlations}
\end{figure*}

Concerning the age distribution, there is a significant presence of relatively young stars in the sample. However, very old stars are also present. This observation aligns with the range of masses. Moreover, the right panel of Fig. \ref{fig:mass_correlations} confirms that the more massive stars are the younger and low-mass stars the older, as they should be. However, estimating the age of stars, particularly during the giant phases, is challenging and subject to significant uncertainties (\citeads{2015MNRAS.451.2230M}; \citeads{2021AJ....161..100W}; \citeads{2024A&A...690A.323V}). Therefore, while these results serve as a rough check for consistency, they cannot be interpreted as precise indicators of stellar age.

\begin{figure}[ht!]
\centering
\includegraphics[height=0.7\linewidth, width=0.8\linewidth]{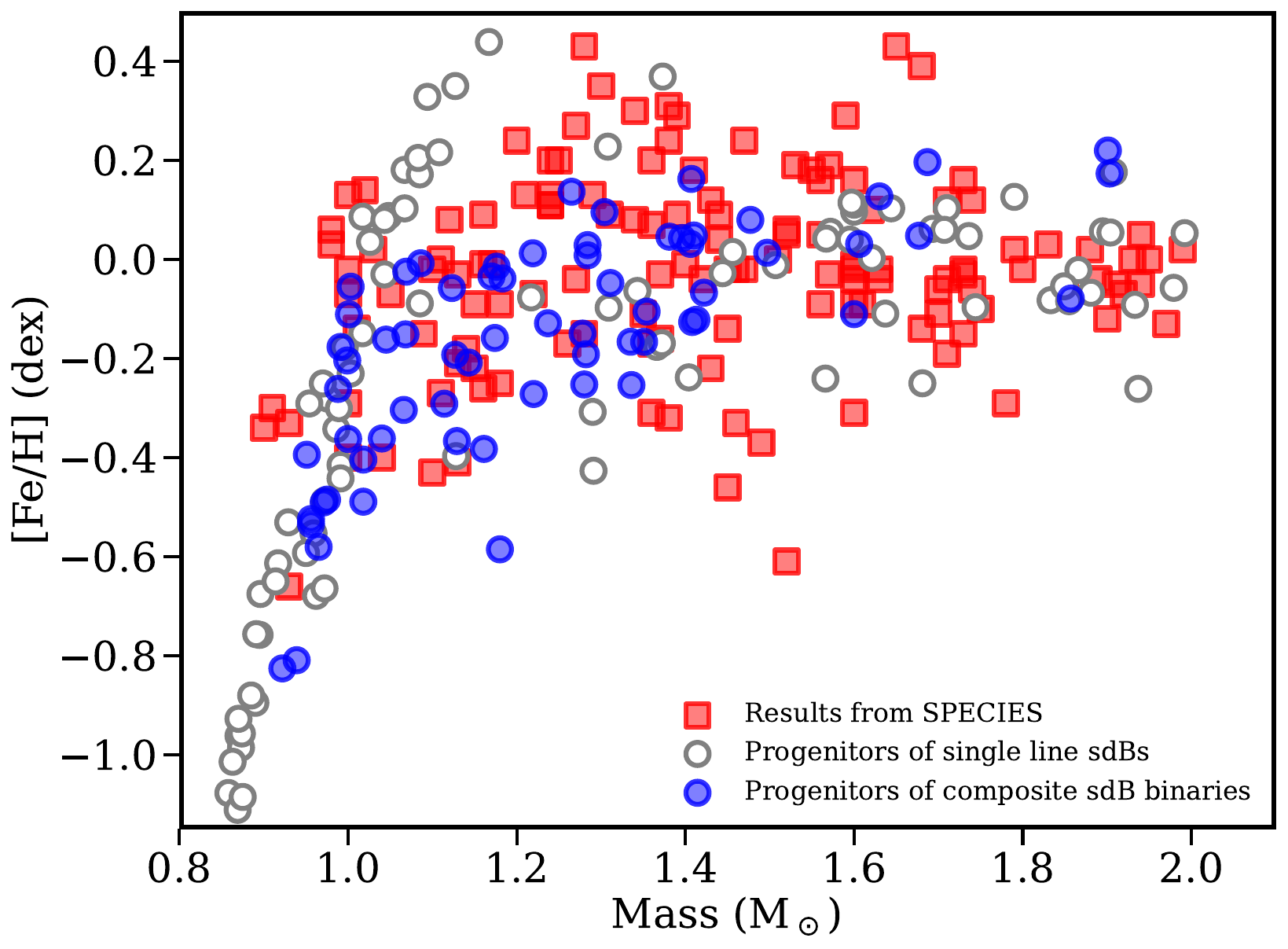}
\caption{Correlation between metallicity and mass for sdB progenitors compared with the theoretical predictions from \citetads{2020A&A...641A.163V}. Filled blue circles are systems recognizable as composite sdB binary systems. Open grey circles are those where only one component is visible. Filled red squares are the $\tt SPECIES$ results for RGB stars in our sample with mass $< 2$ \msun.}
\label{fig: mass-metallicity}
\end{figure}

An essential prediction made by \citetads{2020A&A...641A.163V} is the correlation between mass and metallicity for progenitors of wide sdB binaries. They performed BPS analysis with MESA to model wide sdB binaries focusing on progenitors with initial masses of 0.7–2.0\,\,\msun\, to ensure degenerative He ignition. Galactic chemical evolution is integrated, accounting for metallicity variations that influence RGB radii and orbital periods. Our results, as depicted in Fig. \ref{fig: mass-metallicity}, align well with their predictions, showing that wide sdB progenitor masses increase with metallicity. For the comparison, we excluded targets with masses exceeding $2.0$ \msun, since \citetads{2020A&A...641A.163V} limited their study to this mass value.

\subsection{Comparison with the literature}
\label{subsec:comparisons}

\begin{figure*}[h!]
\centering
\subfigure{\includegraphics[height=0.32\linewidth, width=0.34\linewidth]{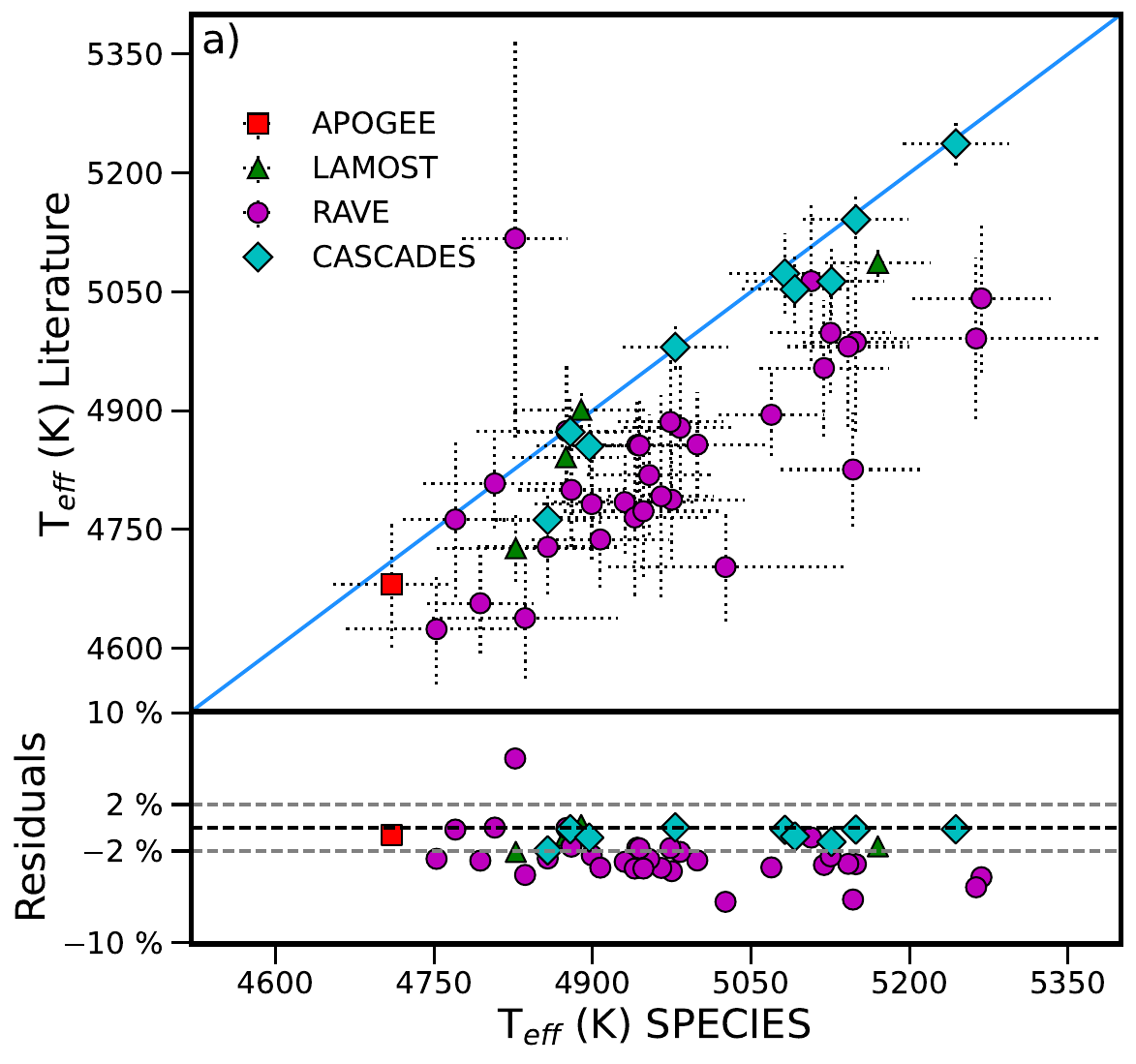}}
\subfigure{\includegraphics[height=0.32\linewidth, width=0.34\linewidth]{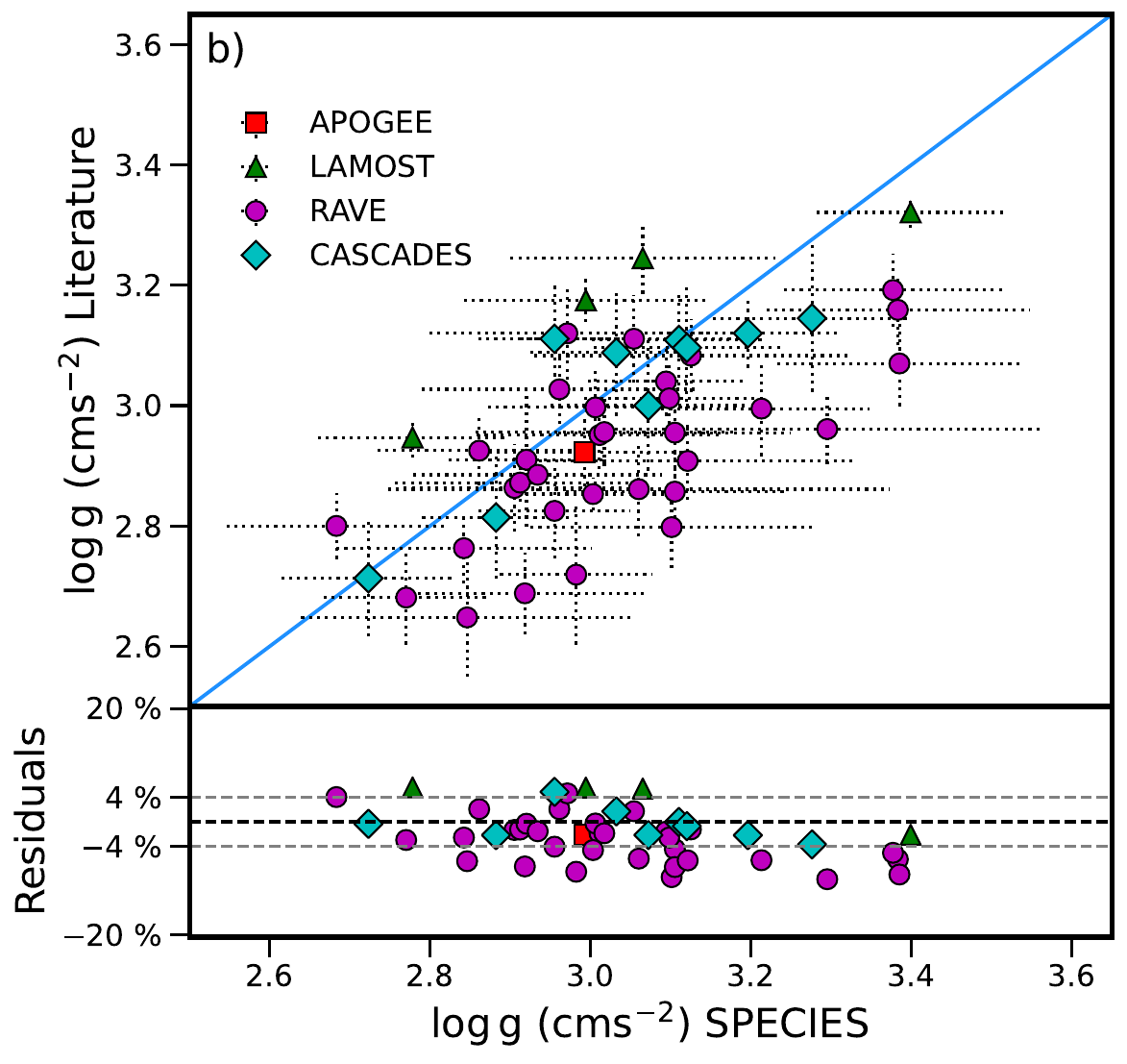}}
\subfigure{\includegraphics[height=0.32\linewidth, width=0.34\linewidth]{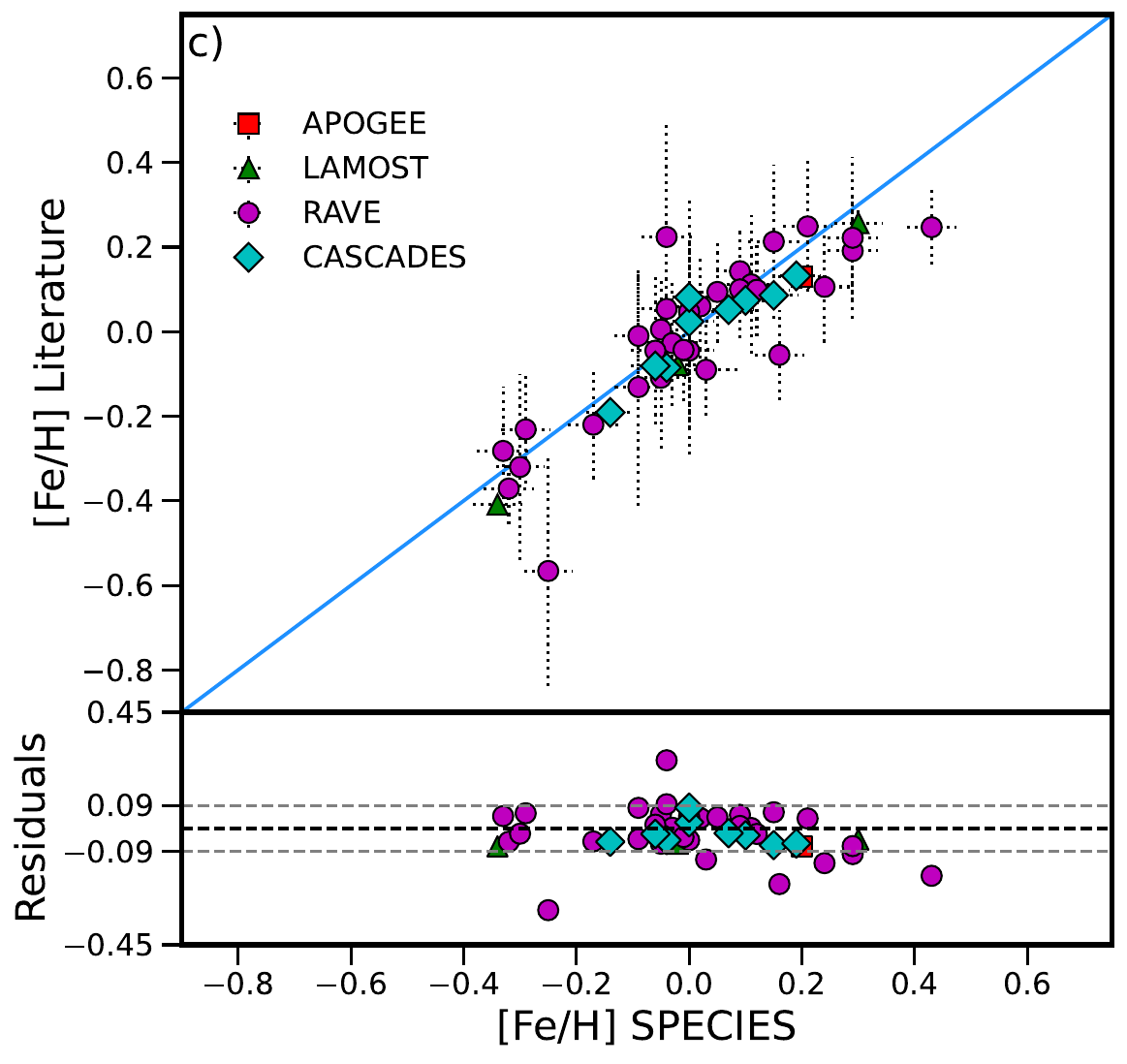}}
\subfigure{\includegraphics[height=0.32\linewidth, width=0.34\linewidth]{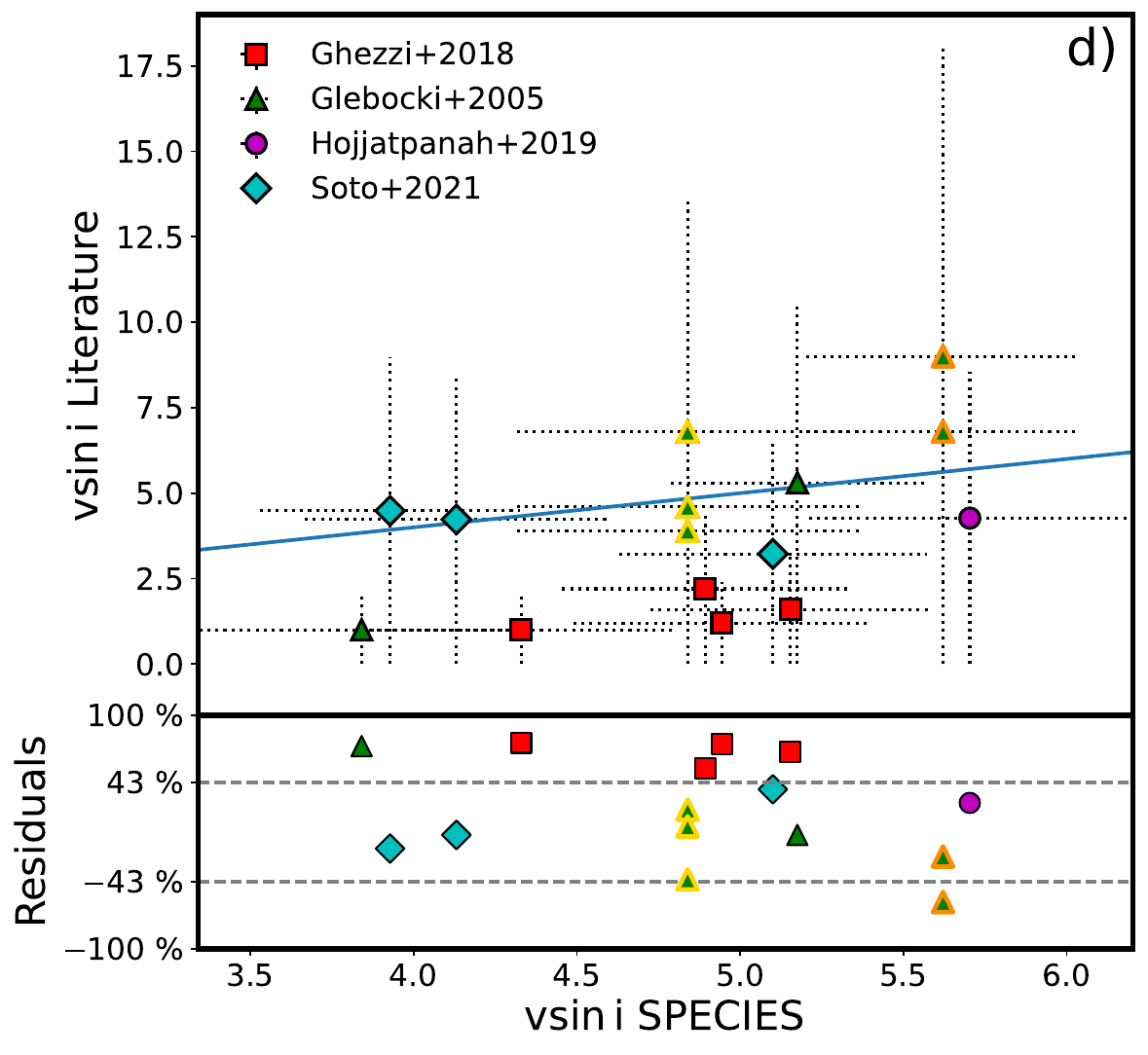}}
\caption{Comparison of stellar parameters obtained with \texttt{SPECIES} and different catalogues labelled as Literature. The solid blue line in each plot represents the one-to-one relation. The dashed grey lines in the lower panel show the 1$-\sigma$ dispersion of the residuals.}
\label{fig:comparison_large_surveys}
\end{figure*}

We crossmatched our observed sample with existing spectroscopic surveys using TOPCAT \citepads{2005ASPC..347...29T}. In total, we found 38 coincidences with the survey of surveys \citepads{2022A&A...659A..95T}. One of them comes from the Apache Point Observatory Galactic Evolution Experiment survey \citepads[APOGEE;][]{2017AJ....154...94M}, four are from the Large sky Area Multi-Object
fiber Spectroscopic Telescope survey \citepads[LAMOST;][]{2012arXiv1206.3569Z}, and 33 come from the RAdial Velocity Experiment survey \citepads[RAVE;][]{2006AJ....132.1645S}. We have also found nine coincidences with the CORALIE radial-velocity search for companions around evolved stars \citepads[CASCADES;][]{2022A&A...657A..87O}, four with the catalogue of stellar rotational velocities \citepads{2005yCat.3244....0G}, one with the catalogue for the ESPRESSO blind radial velocity exoplanet survey \citepads{2019A&A...629A..80H}, three with \citetads{2021A&A...647A.157S}, and four with \citetads{2018ApJ...860..109G}. These catalogues provided us with the possibility of comparing atmospheric parameters as well as rotational velocities. The comparison results are shown in Fig. \ref{fig:comparison_large_surveys}.

Panels a)–-c) in Fig. \ref{fig:comparison_large_surveys} show matches with APOGEE (red squares), LAMOST (green triangles), RAVE (purple circles), and CASCADES (cyan diamonds). Panel d) includes matches with \citetads{2018ApJ...860..109G} (red squares), \citetads{2005yCat.3244....0G} (green triangles), \citetads{2019A&A...629A..80H} (purple circle), and \citetads{2021A&A...647A.157S} (cyan diamonds). The blue line in each panel represents the one-to-one relation. Residuals are computed as $(X_\text{Literature} - X_{\tt SPECIES})/X_{\tt SPECIES}$, where $X$ represents a stellar parameter and `Literature' refers to the matched surveys, except for [Fe/H]. For [Fe/H], only the difference 
([Fe/H]$_\text{Literature}$ - [Fe/H]$_{\tt SPECIES}$) is calculated as many values are near zero, making relative residuals unrealistically large. Grey dashed lines indicate the 1-$\sigma$ dispersions in the residuals. Green triangles in panel d) with coloured borders indicate the same stars. HD167768 (gold borders) and 
HD209154 (orange borders) have multiple measurements in \citetads{2005yCat.3244....0G}, derived using different 
techniques\footnote{Descriptions of these techniques can be found in the CDS catalogue 
\href{https://vizier.cds.unistra.fr/viz-bin/VizieR?-source=III/244}{III/244}}, illustrating systematic differences in rotational velocity measurements even within the same survey.

From the comparison, parameters derived with SPECIES are generally consistent with measurements from the literature. While effective temperatures derived by SPECIES are slightly higher, as shown in Fig. \ref{fig:comparison_large_surveys}, panel a), the residuals remain within a 10\% dispersion. Measurements of surface gravity and metallicity closely follow the one-to-one relation, as seen in panels b) and c). In contrast, rotational velocities show significant inconsistencies. This is expected, as the limited sample size prevents a robust statistical comparison. Also, the fact that there is a considerable dispersion even within the same survey together with the large errors on each measurement does not allow to draw significant conclusions from panel d). Nevertheless, the figure highlights the challenges associated with accurately measuring rotational velocities.

Our measurements show remarkably good agreement with those from CASCADES, although the small sample size (nine matches) prevents robust conclusions from being drawn. This agreement remains an important validation, given that both studies used the same instrument (CORALIE) and followed a similar methodology to derive stellar parameters. The main difference is that SPECIES uses its own algorithm for measuring equivalent widths, while CASCADES used ARES (\citeads{2007A&A...469..783S}, \citeyearads{2015A&A...577A..67S})\footnote{SPECIES also used ARES in its first version; the custom EW algorithm was introduced in the second version, which is used in this work.}. Additionally, CASCADES derived luminosities, radii, and masses using evolutionary tracks from \citetads{2004ApJ...612..168P}, while SPECIES relies on MIST tracks. Since these additional parameters can only be compared with CASCADES and due to the limited number of matches, we exclude the comparisons from the main text; they are available in Appendix \ref{Appendix_CASCADES}.

Due to the limited statistics for almost all of the matches, statistical analyses were only possible with the RAVE sample. The median fractional residuals, along with the scatter of the measurements shown in Fig. \ref{fig:comparison_large_surveys}, are -2.84$\pm$2.23\% for T$_{\text{eff}}$, -2.78$\pm$4\% for $\log$g, and 0.02$\pm$0.09 for [Fe/H].

Apart from spectroscopy, asteroseismology is one of the most powerful tools to derive stellar parameters and probe stellar interiors. The technique is based on measuring stellar oscillations via Fourier analysis. Specifically, the frequency at maximum oscillation power, $\nu_{\text{max}}$, and large frequency separation, $\Delta \nu$, are measured from precise light curves obtained with space-based missions such as Kepler \citepads{2010Sci...327..977B}, K2 \citepads{2014PASP..126..398H} and the Transiting Exoplanet Survey Satellite \citepads[TESS,][]{2015JATIS...1a4003R}. Using these parameters, scaling relations can be used to measure absolute mass and radius (\citeads{1986ApJ...306L..37U}; \citeads{1991ApJ...368..599B}; \citeads{1995A&A...293...87K}; \citeads{2011A&A...530A.142B}) for stars that exhibit oscillations similar to those observed in the Sun (solar-like oscillators), including RGB stars (see \citeads{2021RvMP...93a5001A} for a review).

\begin{table}[ht!]
\setlength{\tabcolsep}{5.0pt}
\renewcommand{\arraystretch}{1.2}
\centering
\caption{Comparison of $\tt SPECIES$ results (A) with results from \citetads{2024A&A...682A...7B} (B).}
\label{tab:speciesxBeck2024}
\begin{tabular}{ccccc}
\hline
\hline
Star & HD270913 & HD40525 & HD45616 & CD-66436 \\
\hline
T$_\text{eff,A}$ (K) & 4836$\pm 88$ & 5103$\pm 58$ & 4994$\pm 75$ & 4875$\pm 85$ \\
T$_\text{eff,B}$ (K) & 5245$\pm 80$ & 5746$\pm 80$ & 5619$\pm 80$ & 5586$\pm 80$ \\
M$_\text{A}$ (\msun) & 1.7$^{+0.2}_{-0.1}$ & 2.23$\pm 0.07$ & 2.22$^{+0.06}_{-0.1}$ & 1.2$\pm 0.1$ \\
M$_\text{B}$ (\msun) & 1.9$\pm 0.4$ & 3.0$\pm 0.7$ & 3.3$\pm 0.6$ & 2$\pm 6$ \\
R$_\text{A}$ (\rsun) & 6.3$\pm 0.2$ & 8.9$\pm 0.1$ & 9.1$\pm 0.1$ & 4.89$^{+0.09}_{-0.08}$ \\
R$_\text{B}$ (\rsun) & 6.7$\pm 0.5$ & 10.6$\pm 0.8$ & 10.7$\pm 0.7$ & 6$ \pm11$ \\
\hline
\end{tabular}
\end{table}

\citetads{2024A&A...682A...7B} cross-matched the NSS catalogue from Gaia DR3 with catalogues of confirmed solar-like oscillators in the MS and RGB phase from the Kepler mission and stars in the southern continuous viewing zone of TESS, finding a total of 954 new binary-system candidates hosting a solar-like oscillating RGB (909) or either a MS or a subgiant (45) star in the \texttt{nss\_two\_body\_orbit} catalogue, and 937 binary candidates within the \texttt{nss\_non\_linear\_spectro} and \texttt{nss\_acceleration\_astro} catalogues. They calculated masses and radii for the oscillating stars using asteroseismic scaling relations and solar values as a base reference. We found six matches with their work and our observed sample. However, for one star (CD-79305), they do not provide seismic measurements; hence, we cannot compare it with our results. The comparison with their work and results from $\tt SPECIES$ is listed in Table \ref{tab:speciesxBeck2024}. \citetads{2024A&A...682A...7B} obtained relatively higher temperatures in their work. This is reflected in the obtained masses and radii, which are also slightly higher than the ones obtained using $\tt SPECIES$. However, all the masses and radii agree within their respective errors. Notice the big errors in the seismic measurements for the star CD-66436, this is a reflection of the high error in its $\Delta\nu$ value, $\Delta\nu = 12 \pm 11\,\mu$Hz.

\citetads{2021ApJ...919..131H} detected 158\,000 red giants using long-cadence (30 min) TESS data and measured their $\nu_{\text{max}}$ using machine-learning techniques. They then derived stellar parameters from seismic scaling relations. Additionally, Gaia eDR3 parallaxes and \texttt{mwdust} were used to estimate distances and extinctions. Afterward, effective temperature, surface gravity, metallicity, and extinction values were interpolated from the MIST bolometric correction tables to obtain bolometric corrections (see sec. 5.4 of \citeads{2016ApJ...823..102C}), which were applied to apparent magnitudes in order to calculate absolute magnitudes and luminosities. Colour-effective temperature relations were subsequently used to estimate effective temperatures. Finally, radii were determined using the Stefan–Boltzmann relation.

\citetads{2024ApJS..271...17Z} computed $\nu_{\text{max}}$ and $\Delta \nu$ for 8651 solar-like oscillators using 2-minute cadence TESS light curves. They then derived stellar parameters from seismic scaling relations. In addition, they incorporated spectroscopic effective temperatures, surface gravities, and metallicities from Gaia DR3 RVS spectra, combining these with apparent magnitudes for spectral energy distribution (SED) fitting to derive extinction and bolometric fluxes. Using Gaia parallaxes, they determined luminosities, and with spectroscopic temperatures, they calculated radii.

\begin{figure*}[h!]
\centering
\subfigure{\includegraphics[height=0.27\linewidth, width=0.27\linewidth]{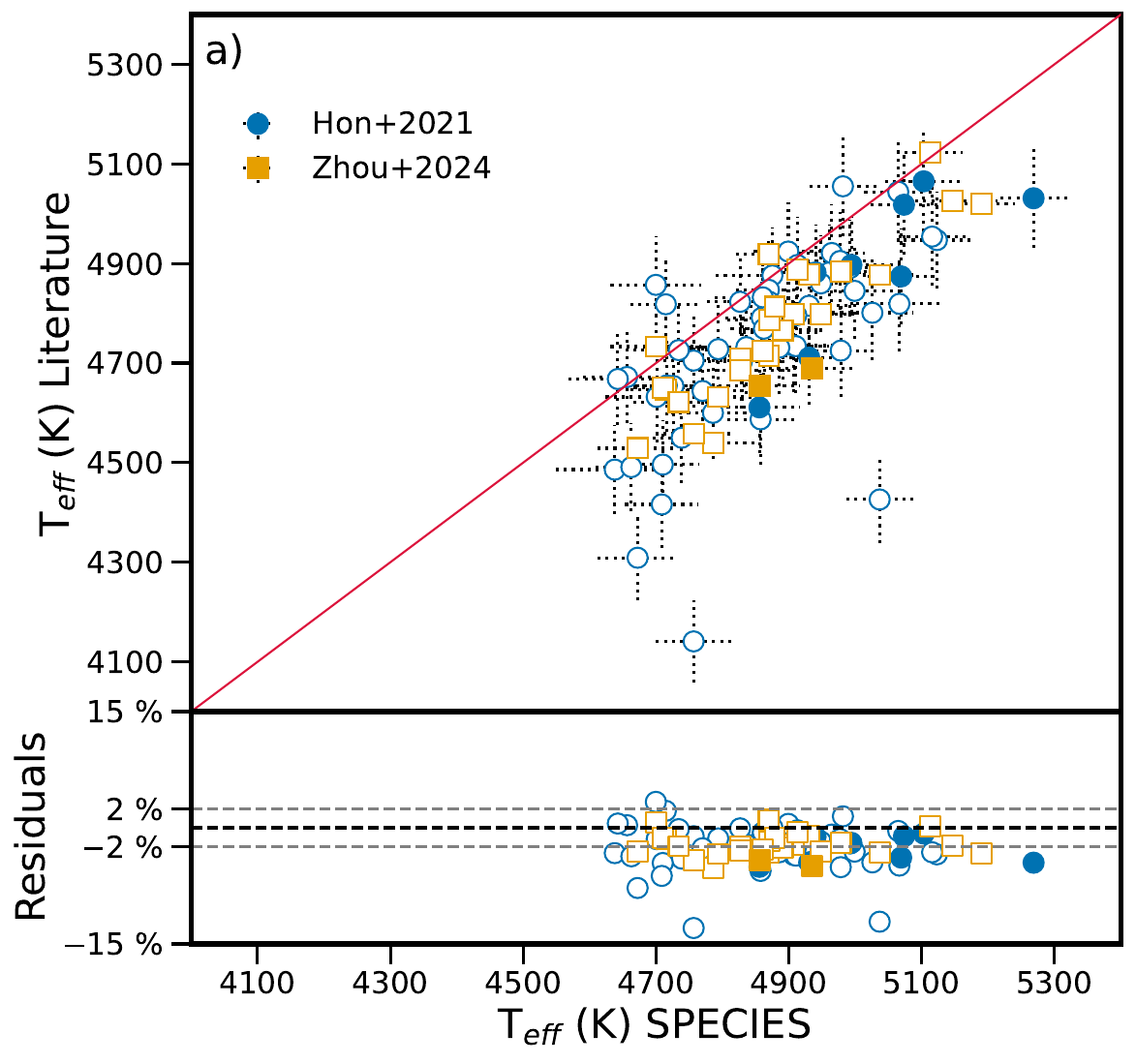}}
\subfigure{\includegraphics[height=0.27\linewidth, width=0.27\linewidth]{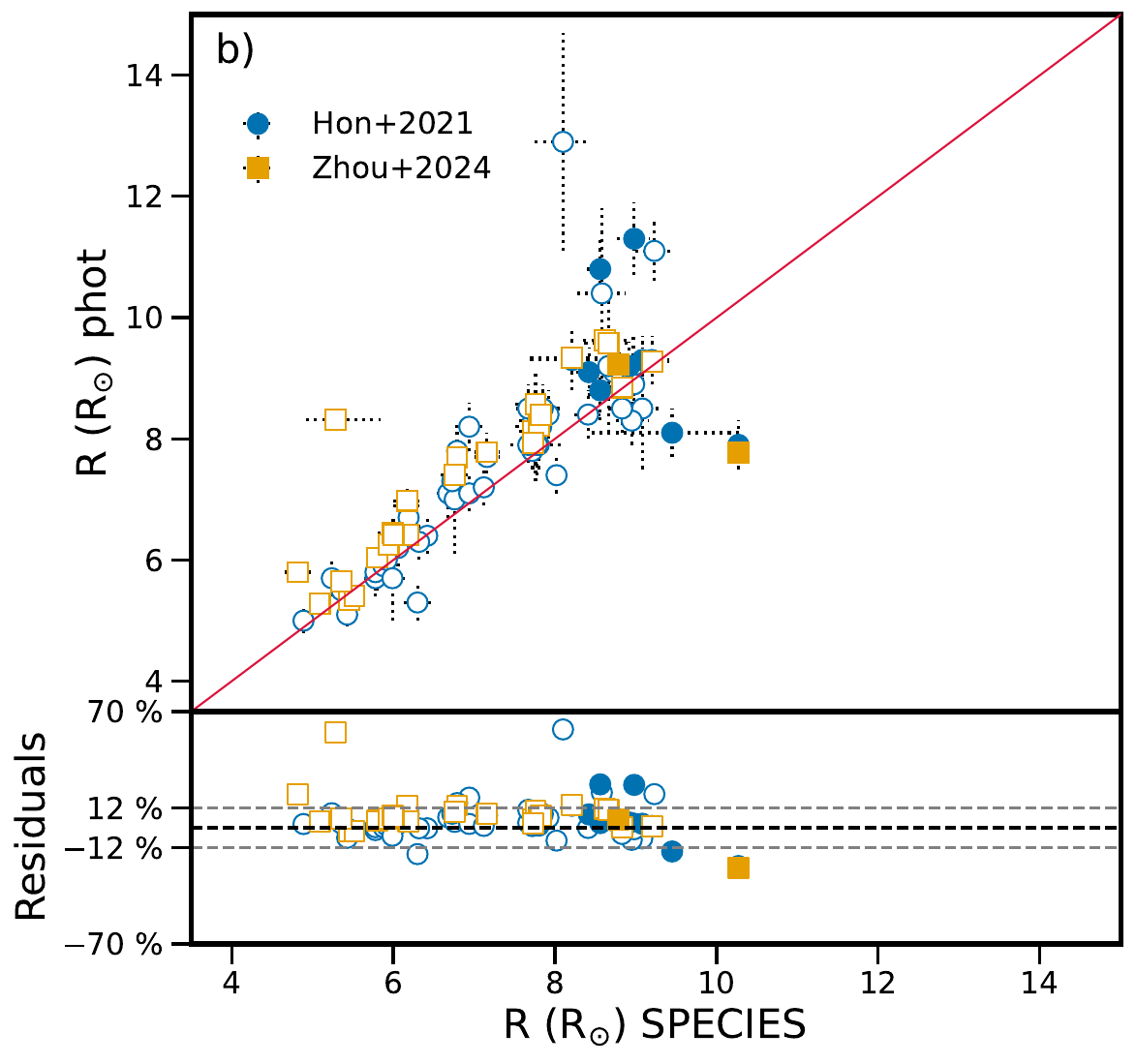}}
\subfigure{\includegraphics[height=0.27\linewidth, width=0.27\linewidth]{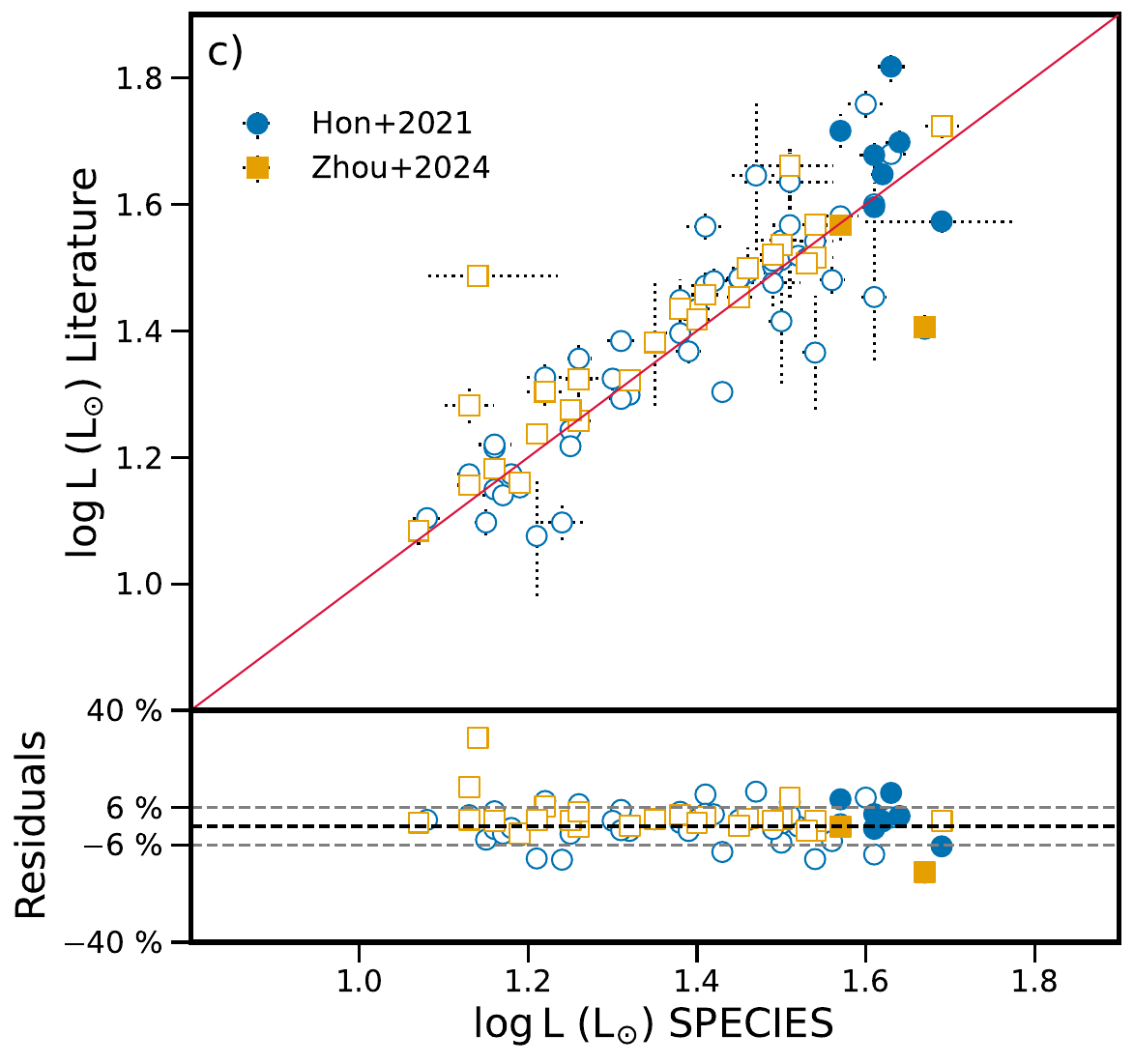}}
\subfigure{\includegraphics[height=0.27\linewidth, width=0.27\linewidth]{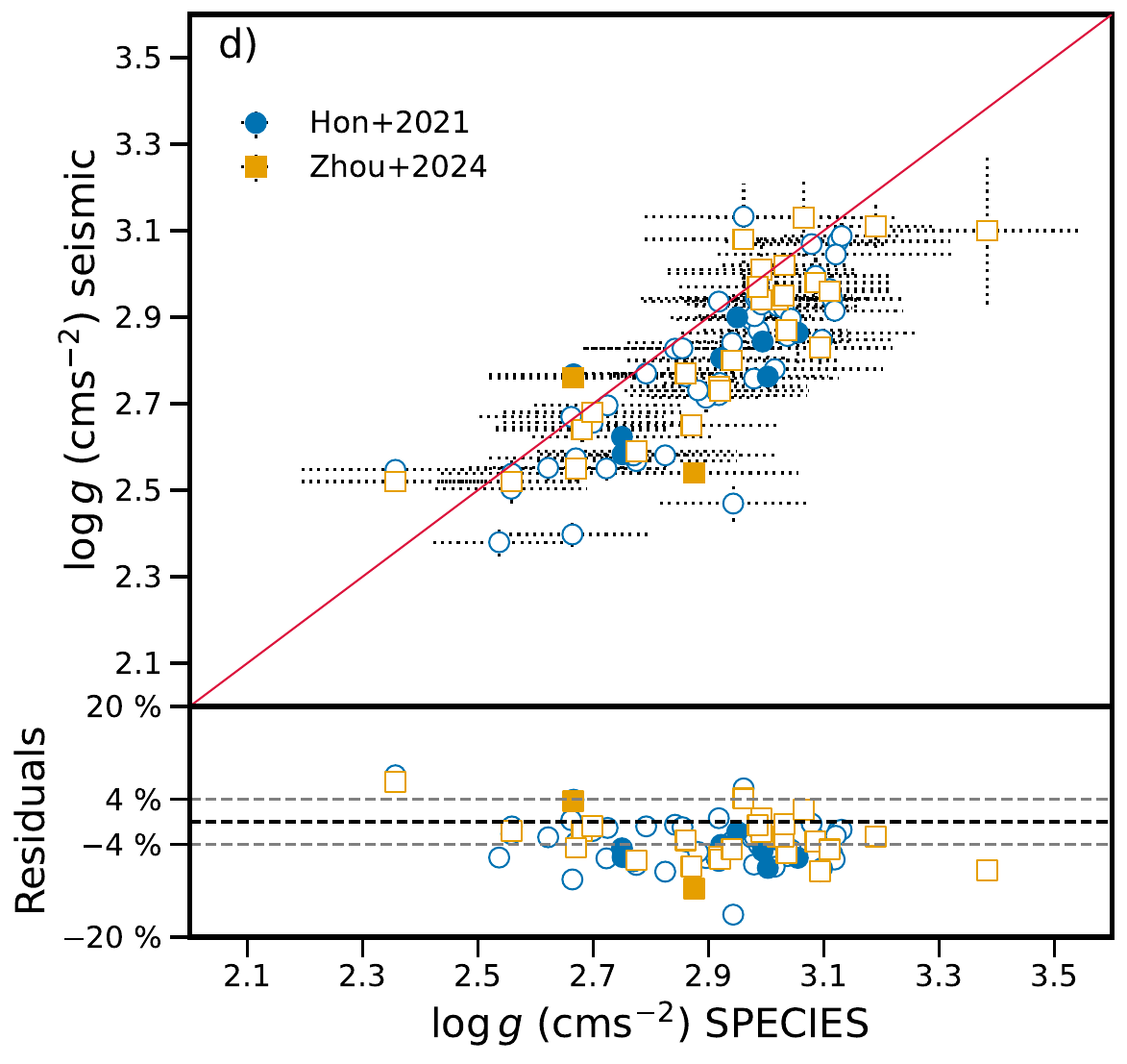}}
\subfigure{\includegraphics[height=0.27\linewidth, width=0.27\linewidth]{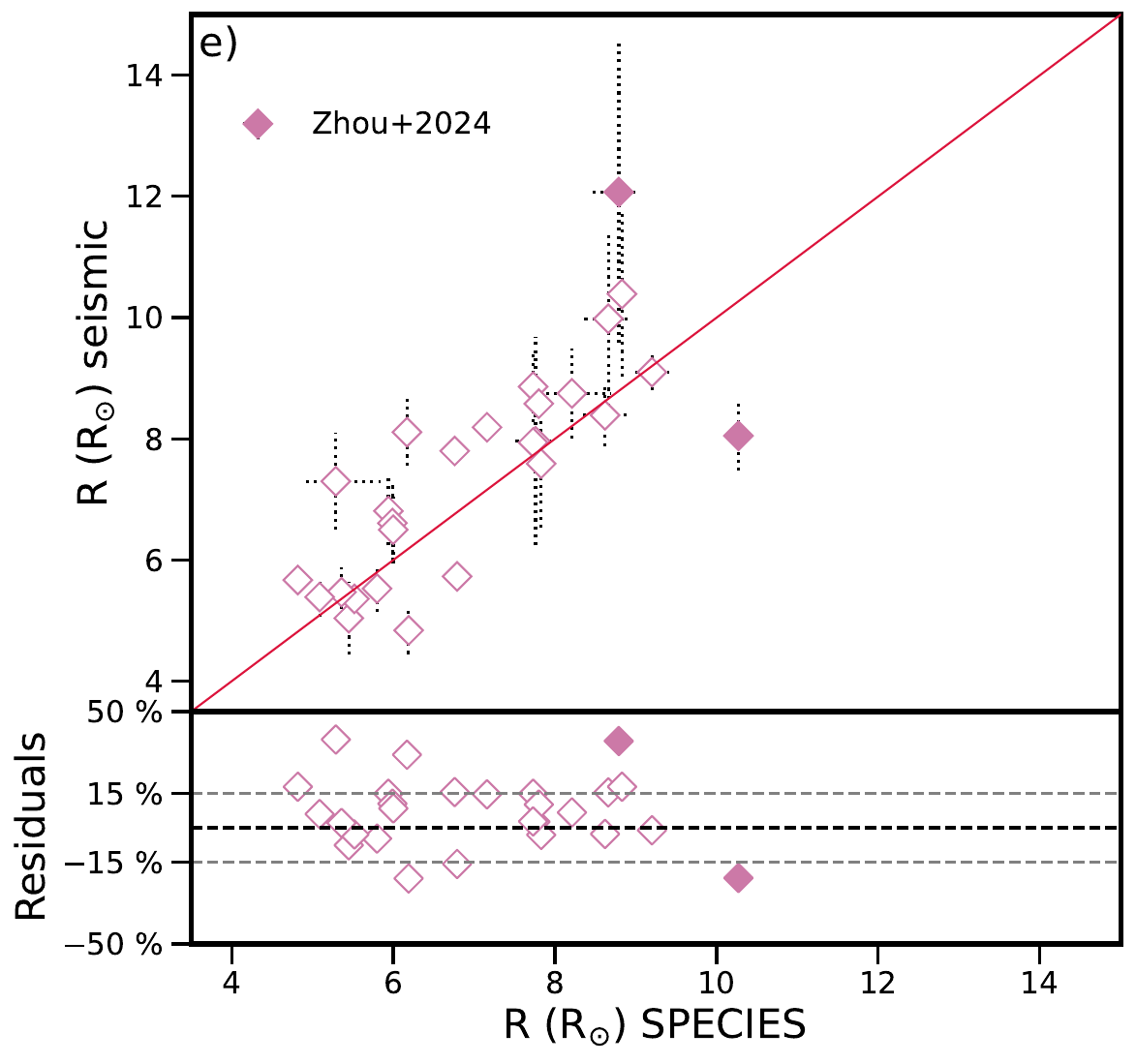}}
\subfigure{\includegraphics[height=0.27\linewidth, width=0.27\linewidth]{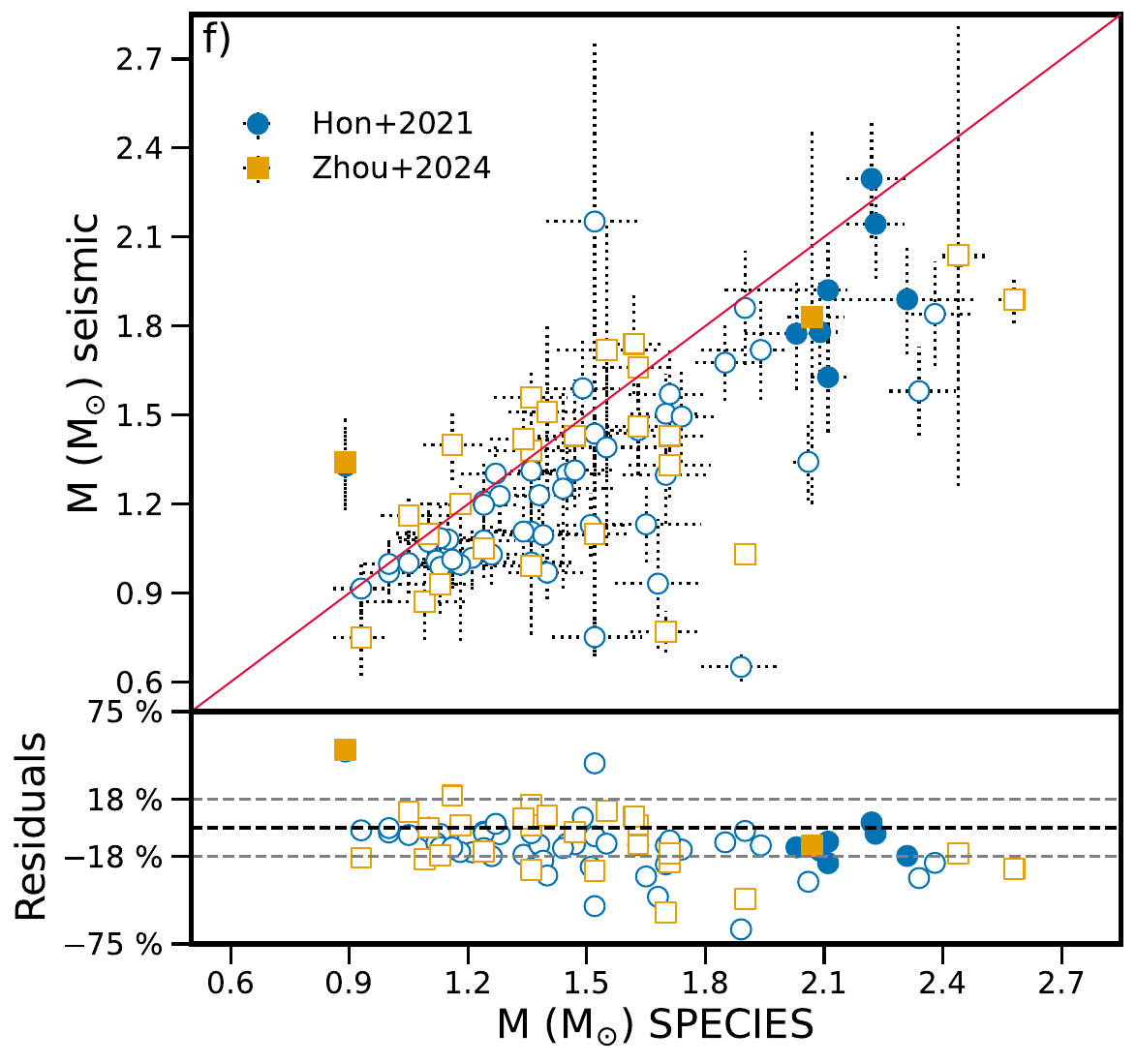}}
\caption{Comparison of stellar parameters from \citetads{2021ApJ...919..131H}, \citetads{2024ApJS..271...17Z}, and $\tt SPECIES$. Open symbols represent objects with an EEP consistent with an RGB star, while filled symbols correspond to those with an EEP consistent with CHeB stars. The solid red line in the upper panels shows the one-to-one relation. `Literature' refers to the studies by \citetads{2021ApJ...919..131H} and \citetads{2024ApJS..271...17Z}, `seismic' denotes parameters derived through asteroseismology in both studies. Radii are categorised into photo-geometric (from both studies) and seismic (from \citetads{2024ApJS..271...17Z} only). The dashed grey lines in the lower panels show the $1-\sigma$ dispersion of the residuals.}
\label{fig:SPECIES_Hon_Zhou}
\end{figure*}

We found 60 matched targets with the sample of \citetads{2021ApJ...919..131H} and 27 with the one of \citetads{2024ApJS..271...17Z}, 20 targets are present in both works. Fig. \ref{fig:SPECIES_Hon_Zhou} shows the comparison between our results and those from \citetads{2021ApJ...919..131H} and \citetads{2024ApJS..271...17Z}. From the 60 matched targets with \citetads{2021ApJ...919..131H}, 52 have EEPs consistent with the RGB phase (empty symbols in Fig. \ref{fig:SPECIES_Hon_Zhou}) and the remaining 8 are CHeB stars (filled symbols in Fig. \ref{fig:SPECIES_Hon_Zhou}). In the case of \citetads{2024ApJS..271...17Z}, one of the matched targets was identified as a CHeB star. Fractional residuals for each parameter were computed again as $(X_\text{Literature} - X_{\tt SPECIES})/X_{\tt SPECIES}$, where $X$ represents a stellar parameter and `Literature' refers to the studies by \citetads{2021ApJ...919..131H} and \citetads{2024ApJS..271...17Z}. The distribution of the residuals was analysed for each parameter, and the results are summarized in Table \ref{tab: resid_SPECIES_Hon_Zhou}. Parameters derived through asteroseismology are labelled as `seismic'. We distinguish radii derived from seismic scaling relations (R$_{\text{seismic}}$) from those obtained through photo-geometric methods (R$_{\text{phot}}$) which combine photometry and parallaxes — bolometric corrections and colour-effective temperature relations for \citetads{2021ApJ...919..131H} and SED fitting for \citetads{2024ApJS..271...17Z}.

\begin{table}[h!]
\setlength{\tabcolsep}{8.5pt}
\renewcommand{\arraystretch}{1.1}
	\centering
	\caption{Comparison of $\tt SPECIES$ results with other works.}
	\begin{tabular}{ccccc}
		\hline
		& \multicolumn{2}{c}{Hon+2021} & \multicolumn{2}{c}{Zhou+2024}\\
		\hline
		Parameter & MFR & Scatter & MFR & Scatter\\
		T$_\text{eff}$ &	-2.09\% &	2.86\% &	-2.43\% &	1.51\% \\
		$\log g_{\text{seismic}}$ &	-3.93\% &	3.79\% &	-2.67\% &	4.13\% \\
		$\log$L &	1.53\% &	6.08\% &	1.99\% &	7.22\% \\
		R$_{\text{seismic}}$ &	--- &	--- &	6.57\% &	14.86\% \\
		R$_{\text{phot}}$ &	2.60\% &	11.87\% &	5.41\% & 12.42\% \\
		M &	-11.95\% &	16.97\% &	-10.43\% &	21.10\% \\
		\hline
		\hline
	\end{tabular}
    \tablefoot{MFR stands for median fractional residual}
	\label{tab: resid_SPECIES_Hon_Zhou}
\end{table}

Panel (a)
in Fig. \ref{fig:SPECIES_Hon_Zhou} show that $\tt SPECIES$ gives larger values for 
$T_{\text{eff}}$.  
This is also evident in Table \ref{tab: resid_SPECIES_Hon_Zhou} by the relatively large negative median fractional residual for this parameter.  
Systematic biases in $T_{\text{eff}}$ determination for RGB stars using different approaches is a known issue in the literature (\citeads{2018A&A...612A..68S}; \citeads{2023A&A...670A.107H}; \citeads{2023ApJS..264...41Y}; \citeads{2024A&A...690A.323V}). This effect appears in panel (a) of Fig. \ref{fig:SPECIES_Hon_Zhou}, where the comparison between the spectroscopic $T_{\text{eff}}$ from SPECIES and the photometric values from \citetads{2021ApJ...919..131H} shows larger dispersion. In contrast, the dispersion is smaller when comparing with the $T_{\text{eff}}$ from \citetads{2024ApJS..271...17Z}, which is also derived from spectroscopy. These systematics can bias the estimated mass and thus the age (\citeads{2015MNRAS.451.2230M}; \citeads{2021AJ....161..100W}). However, investigating this discrepancy further is out of the scope of this work, and according to Table \ref{tab: resid_SPECIES_Hon_Zhou}, the scatter is $< 5\%$. Hence, $T_{\text{eff}}$ calculated by $\tt SPECIES$ is in good agreement with the literature.

Panel (b) in Fig. \ref{fig:SPECIES_Hon_Zhou} shows a comparison between the radii derived by SPECIES (using synthetic isochrones) and those obtained through photo-geometric methods. The figure shows good agreement, with a median fractional residual of 3.93\% and a scatter of 11.76\%. Dispersion increases for R $\boldsymbol{\gtrsim}$ 8\rsun, although the sample size is too small to draw robust conclusions for this regime.

Additionally, panel (c) in Fig. \ref{fig:SPECIES_Hon_Zhou} shows good agreement in luminosity, which is expected due to the strong correlation between radius and luminosity. Results from $\tt SPECIES$ have relatively lower errors. The median fractional residual is 1.88\%, with a scatter of 6.37\%. It is also notable that the apparent overestimation seen in values of $T_{\text{eff}}$ calculated by $\tt SPECIES$ is not as large for luminosity and photo-geometric radius. This may come from the method used by $\tt SPECIES$ to estimate parameters. While $T_{\text{eff}}$ and $\log g$ are derived from colours and spectral equivalent widths, luminosity, and other physical parameters, are determined by fitting a stellar model using $\tt isochrones$. This could also explain why errors appear smaller for physical parameters compared to atmospheric ones.

Surface gravity, shown in Fig. \ref{fig:SPECIES_Hon_Zhou}, panel (d), is derived from seismic scaling relations for both \citetads{2021ApJ...919..131H} and \citetads{2024ApJS..271...17Z}. There is again an apparent overestimation in the values from $\tt SPECIES$. However, note that seismic surface gravities are significantly more precise than spectroscopic ones. With a median fractional residual of -3.41\%, a scatter of 3.91\%, and most residuals within one standard deviation, we can say that $\tt SPECIES$ surface gravities show good agreement with the literature.

Considering the seismic radius from \citetads{2024ApJS..271...17Z}, compared to those derived with SPECIES in Fig. \ref{fig:SPECIES_Hon_Zhou} panel (e), the median fractional residual is 6.57\%, with a scatter of 14.86\%. The comparison between seismic and photo-geometric radii with SPECIES appears similar, though photo-geometric values show better agreement. However, uncertainties in the seismic radii are larger than those in the photo-geometric radii, as shown in panels (b) and (e) of Fig. \ref{fig:SPECIES_Hon_Zhou}.

For mass, panel (f) in Fig. \ref{fig:SPECIES_Hon_Zhou} shows good agreement at the low mass end. However, for $M \gtrsim 1.25$\msun, there is a notable discrepancy between the seismic masses and our results, with $\tt SPECIES$ systematically overestimating the mass. Mass discrepancy is a known issue in RGB stars (\citeads{2018A&A...612A..68S}; \citeads{2024A&A...690A.323V}), as well as in intermediate and high-mass stars (\citeads{1992LNP...401...21H}; \citeads{2010A&A...524A..98W}; \citeads{2020A&A...637A..60T}). As we mentioned, a bias in effective temperature can lead to a bias in the mass, and since we obtained overall higher effective temperatures using $\tt SPECIES$ in comparison to \citetads{2021ApJ...919..131H} and \citetads{2024ApJS..271...17Z}, this could lead to the apparent overestimation in mass. These discrepancies can be used to test and calibrate the evolutionary models currently used to model RGB stars, since, for example, mixing, enhancement of $\alpha$ elements, and boundary conditions have a huge impact on constraining stellar parameters during the RGB phase (\citeads{2018A&A...612A..68S}; \citeads{2015MNRAS.451.2230M}; \citeads{2024A&A...690A.323V}). Moreover, mixing also plays a crucial role in the formation of sdBs \citepads{2024MNRAS.52711184A}. Therefore, we plan to investigate these discrepancies further with more data and refined models to have more robust constraints on the physics of sdB formation.

The median fractional residual for mass is -11.27\%, with a scatter of 18.44\%, larger than for other stellar parameters, and can reach up to 21.10\% when we compare our results only with those from \citetads{2024ApJS..271...17Z}. However, it is not clear if the discrepancy comes from $\tt SPECIES$ overestimating the mass or the asteroseismic relations underestimating it since both approaches depend at a certain level on assumptions in the underlying physics of the model. The better approach to calibrate mass measurements would be using eclipsing binaries, for which the dynamic mass is measured using Kepler's third law and is, therefore, model-independent. There have been recent efforts to find and study eclipsing binaries with RGB components (e.g. \citeads{2025OJAp....8E..18R}). Unfortunately, comparisons of dynamic masses with seismic or evolutionary ones are not yet available for the RGB phase. Therefore, we cannot calibrate our measurements.  However, most residuals in Fig. \ref{fig:SPECIES_Hon_Zhou}(e) are within one standard deviation, indicating overall good agreement.

In general, our estimates align well with \citetads{2021ApJ...919..131H} and \citetads{2024ApJS..271...17Z}. Even though the dispersion in mass is higher, the overall results are still within one standard deviation. Therefore, we conclude that the results from $\tt SPECIES$ are valid.

\section{Conclusions}
\label{conclusion}

We continued the spectroscopic observations of potential progenitors of wide sdB binaries initiated in \citetaliasads{2022A&A...668A..89U}, including parallax measurements from Gaia DR3 and interstellar reddening. We extended the observed volume-limited sample from 200 to 500 pc. We collected, reduced, and analysed 415 high-resolution CORALIE spectra of 230 low-mass RGB binary candidates using the $\tt CERES$ \citepads{2017PASP..129c4002B} and $\tt SPECIES$ (\citeads{2018A&A...615A..76S}; \citeads{2021A&A...647A.157S}) pipelines. The main results from this work are:

\begin{itemize}

\item Updating the parallaxes from Gaia DR2 to DR3 and including reddening measurements further cleaned the original sample outlined in \citetaliasads{2022A&A...668A..89U} by about $\sim$10\%.

\item Stellar parameters for five of the 230 stars analysed here could not be calculated due to errors with $\tt SPECIES$. Therefore, we provided their RVs calculated by $\tt CERES$ and excluded them from the subsequent analysis. We will investigate these stars and the potential reason for the errors in a future work.

\item From the remaining 225 stars, 168 are categorized as high-priority potential binary system members and 57 as low-priority binary system members, based on the classification method described in \citetaliasads{2022A&A...668A..89U}.

\item Approximately 82\% of stars in our sample are RGB stars, while the remaining 18\% are on the CHeB phase. These CHeB stars are comprised of old low-mass primary RC stars that ignited helium degenerately (2\% ) and younger, more massive secondary RC stars that ignited helium under non-degenerate conditions (16\%).

\item 75\% of the identified RGB stars have a high probability of being part of a binary system, validating our sample.

\item We confirmed the theoretical prediction of \citetads{2020A&A...641A.163V} regarding the correlation between sdB progenitor mass and metallicity.

\item Comparison with the literature shows good overall agreement. The matches with available spectroscopic surveys show a quite low scatter in the residuals ($\lesssim 10\%$ in general) except for rotational velocities. However, it is not possible to draw a robust conclusion for this last parameter due to the poor number statistics and large individual errors. From asteroseismic results, scatter in the residuals is $\lesssim 15\%$, except for mass, which showed a scatter of $\sim 20\%$. Similar discrepancies are also present in other works (\citeads{1992LNP...401...21H}; \citeads{2015MNRAS.451.2230M}; \citeads{ 2018A&A...612A..68S}; \citeads{2020A&A...637A..60T}; \citeads{2023A&A...670A.107H}; \citeads{2024A&A...690A.323V}).

\end{itemize}

Future work will focus on completing multi-epoch observations for all stars in the 500 pc volume-limited sample and confirming binary systems through RV curve analysis. This, in combination with BPS studies, will help constrain the physics of mass transfer in the stable RLOF case. Additionally, our results can be used in investigating the observed discrepancies between different estimation methods, shedding light on poorly understood processes in the red giant phase for low-mass stars.

\section*{Data availability}
\label{sec:data_availability}

Tables \ref{tab:RV_Diego}, \ref{tab:cross_NSS_twobody}, \ref{tab:cross_NSS_acc} and \ref{tab: species_all} are only available in electronic form at the CDS via anonymous ftp to \url{cdsarc.u-strasbg.fr} (130.79.128.5) or via \url{http://cdsweb.u-strasbg.fr/cgi-bin/qcat?J/A+A/}.

\begin{acknowledgements}
We thank the referee for her or his helpful comments, which led to an improved presentation of our results.
D.B. gratefully acknowledges support from the Centro de Astrofísica de Valparaíso Proyect Cidi N$^\circ$21.
A.D. acknowledges financial support from ANID-Subdirección de Capital Humano/Magíster Nacional/2024-22241719. 
D.B., M.V., E.A, A.D., and M.Z. acknowledge support from the Fondecyt Regular through grant No. 1211941.
M.U. gratefully acknowledges funding from the Research Foundation Flanders (FWO) by means of a junior postdoctoral fellowship (grant agreement No. 1247624N). 
This work has made use of data from the European Space Agency (ESA) mission Gaia (https://www.cosmos.esa.int/Gaia), processed by the Gaia Data Processing and Analysis Consortium (DPAC, https://www.cosmos.esa.int/web/Gaia/dpac/consortium). Funding for the DPAC
has been provided by national institutions, in particular, the institutions participating in the Gaia Multilateral Agreement.

\end{acknowledgements}

\bibliographystyle{aa}
\bibliography{myrefs}

\begin{thebibliography}{95}
\expandafter\ifx\csname natexlab\endcsname\relax\def\natexlab#1{#1}\fi

\bibitem[{{Aerts}(2021)}]{2021RvMP...93a5001A}
{Aerts}, C. 2021, Reviews of Modern Physics, 93, 015001

\bibitem[{{Alonso} {et~al.}(1999){Alonso}, {Arribas}, \&
  {Mart{\'\i}nez-Roger}}]{1999A&AS..140..261A}
{Alonso}, A., {Arribas}, S., \& {Mart{\'\i}nez-Roger}, C. 1999, \aaps, 140, 261

\bibitem[{{Arancibia-Rojas} {et~al.}(2024){Arancibia-Rojas}, {Zorotovic},
  {Vu{\v{c}}kovi{\'c}}, {Bobrick}, {Vos}, \&
  {Piraino-Cerda}}]{2024MNRAS.52711184A}
{Arancibia-Rojas}, E., {Zorotovic}, M., {Vu{\v{c}}kovi{\'c}}, M., {et~al.}
  2024, \mnras, 527, 11184

\bibitem[{{Barlow} {et~al.}(2012){Barlow}, {Wade}, {Liss}, {{\O}stensen}, \&
  {Van Winckel}}]{2012ApJ...758...58B}
{Barlow}, B.~N., {Wade}, R.~A., {Liss}, S.~E., {{\O}stensen}, R.~H., \& {Van
  Winckel}, H. 2012, \apj, 758, 58

\bibitem[{{Beck} {et~al.}(2024){Beck}, {Grossmann}, {Steinwender}, {Schimak},
  {Muntean}, {Vrard}, {Patton}, {Merc}, {Mathur}, {Garcia}, {Pinsonneault},
  {Rowan}, {Gaulme}, {Allende Prieto}, {Arellano-C{\'o}rdova}, {Cao},
  {Corsaro}, {Creevey}, {Hambleton}, {Hanslmeier}, {Holl}, {Johnson}, {Mathis},
  {Godoy-Rivera}, {S{\'\i}mon-D{\'\i}az}, \& {Zinn}}]{2024A&A...682A...7B}
{Beck}, P.~G., {Grossmann}, D.~H., {Steinwender}, L., {et~al.} 2024, \aap, 682,
  A7

\bibitem[{{Belkacem} {et~al.}(2011){Belkacem}, {Goupil}, {Dupret}, {Samadi},
  {Baudin}, {Noels}, \& {Mosser}}]{2011A&A...530A.142B}
{Belkacem}, K., {Goupil}, M.~J., {Dupret}, M.~A., {et~al.} 2011, \aap, 530,
  A142

\bibitem[{{Bluhm} {et~al.}(2016){Bluhm}, {Jones}, {Vanzi}, {Soto}, {Vos},
  {Wittenmyer}, {Drass}, {Jenkins}, {Olivares}, {Mennickent},
  {Vu{\v{c}}kovi{\'c}}, {Rojo}, \& {Melo}}]{2016A&A...593A.133B}
{Bluhm}, P., {Jones}, M.~I., {Vanzi}, L., {et~al.} 2016, \aap, 593, A133

\bibitem[{{Borucki} {et~al.}(2010){Borucki}, {Koch}, {Basri}, {Batalha},
  {Brown}, {Caldwell}, {Caldwell}, {Christensen-Dalsgaard}, {Cochran},
  {DeVore}, {Dunham}, {Dupree}, {Gautier}, {Geary}, {Gilliland}, {Gould},
  {Howell}, {Jenkins}, {Kondo}, {Latham}, {Marcy}, {Meibom}, {Kjeldsen},
  {Lissauer}, {Monet}, {Morrison}, {Sasselov}, {Tarter}, {Boss}, {Brownlee},
  {Owen}, {Buzasi}, {Charbonneau}, {Doyle}, {Fortney}, {Ford}, {Holman},
  {Seager}, {Steffen}, {Welsh}, {Rowe}, {Anderson}, {Buchhave}, {Ciardi},
  {Walkowicz}, {Sherry}, {Horch}, {Isaacson}, {Everett}, {Fischer}, {Torres},
  {Johnson}, {Endl}, {MacQueen}, {Bryson}, {Dotson}, {Haas}, {Kolodziejczak},
  {Van Cleve}, {Chandrasekaran}, {Twicken}, {Quintana}, {Clarke}, {Allen},
  {Li}, {Wu}, {Tenenbaum}, {Verner}, {Bruhweiler}, {Barnes}, \&
  {Prsa}}]{2010Sci...327..977B}
{Borucki}, W.~J., {Koch}, D., {Basri}, G., {et~al.} 2010, Science, 327, 977

\bibitem[{{Bovy} {et~al.}(2016){Bovy}, {Rix}, {Green}, {Schlafly}, \&
  {Finkbeiner}}]{2016ApJ...818..130B}
{Bovy}, J., {Rix}, H.-W., {Green}, G.~M., {Schlafly}, E.~F., \& {Finkbeiner},
  D.~P. 2016, \apj, 818, 130

\bibitem[{{Brahm} {et~al.}(2017){Brahm}, {Jord{\'a}n}, \&
  {Espinoza}}]{2017PASP..129c4002B}
{Brahm}, R., {Jord{\'a}n}, A., \& {Espinoza}, N. 2017, \pasp, 129, 034002

\bibitem[{{Brown} {et~al.}(1991){Brown}, {Gilliland}, {Noyes}, \&
  {Ramsey}}]{1991ApJ...368..599B}
{Brown}, T.~M., {Gilliland}, R.~L., {Noyes}, R.~W., \& {Ramsey}, L.~W. 1991,
  \apj, 368, 599

\bibitem[{{Castelli} \& {Kurucz}(2003)}]{2003IAUS..210P.A20C}
{Castelli}, F. \& {Kurucz}, R.~L. 2003, in IAU Symposium, Vol. 210, Modelling
  of Stellar Atmospheres, ed. N.~{Piskunov}, W.~W. {Weiss}, \& D.~F. {Gray},
  A20

\bibitem[{{Chen} {et~al.}(2013){Chen}, {Han}, {Deca}, \&
  {Podsiadlowski}}]{2013MNRAS.434..186C}
{Chen}, X., {Han}, Z., {Deca}, J., \& {Podsiadlowski}, P. 2013, \mnras, 434,
  186

\bibitem[{{Choi} {et~al.}(2016){Choi}, {Dotter}, {Conroy}, {Cantiello},
  {Paxton}, \& {Johnson}}]{2016ApJ...823..102C}
{Choi}, J., {Dotter}, A., {Conroy}, C., {et~al.} 2016, \apj, 823, 102

\bibitem[{{Collier Cameron} {et~al.}(2007){Collier Cameron}, {Wilson}, {West},
  {Hebb}, {Wang}, {Aigrain}, {Bouchy}, {Christian}, {Clarkson}, {Enoch},
  {Esposito}, {Guenther}, {Haswell}, {H{\'e}brard}, {Hellier}, {Horne},
  {Irwin}, {Kane}, {Loeillet}, {Lister}, {Maxted}, {Mayor}, {Moutou}, {Parley},
  {Pollacco}, {Pont}, {Queloz}, {Ryans}, {Skillen}, {Street}, {Udry}, \&
  {Wheatley}}]{2007MNRAS.380.1230C}
{Collier Cameron}, A., {Wilson}, D.~M., {West}, R.~G., {et~al.} 2007, \mnras,
  380, 1230

\bibitem[{{Copperwheat} {et~al.}(2011){Copperwheat}, {Morales-Rueda}, {Marsh},
  {Maxted}, \& {Heber}}]{2011MNRAS.415.1381C}
{Copperwheat}, C.~M., {Morales-Rueda}, L., {Marsh}, T.~R., {Maxted}, P.~F.~L.,
  \& {Heber}, U. 2011, \mnras, 415, 1381

\bibitem[{{Dawson} {et~al.}(2024){Dawson}, {Geier}, {Heber}, {Pelisoli},
  {Dorsch}, {Schaffenroth}, {Reindl}, {Culpan}, {Pritzkuleit}, {Vos},
  {Soemitro}, {Roth}, {Schneider}, {Uzundag}, {Vu{\v{c}}kovi{\'c}}, {Antunes
  Amaral}, {Istrate}, {Justham}, {{\O}stensen}, {Telting}, {Djupvik}, {Raddi},
  {Green}, {Jeffery}, {Kepler}, {Munday}, {Steinmetz}, \&
  {Kupfer}}]{2024A&A...686A..25D}
{Dawson}, H., {Geier}, S., {Heber}, U., {et~al.} 2024, \aap, 686, A25

\bibitem[{{Deca} {et~al.}(2012){Deca}, {Marsh}, {{\O}stensen}, {Morales-Rueda},
  {Copperwheat}, {Wade}, {Stark}, {Maxted}, {Nelemans}, \&
  {Heber}}]{2012MNRAS.421.2798D}
{Deca}, J., {Marsh}, T.~R., {{\O}stensen}, R.~H., {et~al.} 2012, \mnras, 421,
  2798

\bibitem[{{dos Santos} {et~al.}(2016){dos Santos}, {Mel{\'e}ndez}, {do
  Nascimento}, {Bedell}, {Ram{\'\i}rez}, {Bean}, {Asplund}, {Spina},
  {Dreizler}, {Alves-Brito}, \& {Casagrande}}]{2016A&A...592A.156D}
{dos Santos}, L.~A., {Mel{\'e}ndez}, J., {do Nascimento}, J.-D., {et~al.} 2016,
  \aap, 592, A156

\bibitem[{{Dotter}(2016)}]{2016ApJS..222....8D}
{Dotter}, A. 2016, \apjs, 222, 8

\bibitem[{{Drimmel} {et~al.}(2003){Drimmel}, {Cabrera-Lavers}, \&
  {L{\'o}pez-Corredoira}}]{2003A&A...409..205D}
{Drimmel}, R., {Cabrera-Lavers}, A., \& {L{\'o}pez-Corredoira}, M. 2003, \aap,
  409, 205

\bibitem[{{Feroz} {et~al.}(2009){Feroz}, {Hobson}, \&
  {Bridges}}]{2009MNRAS.398.1601F}
{Feroz}, F., {Hobson}, M.~P., \& {Bridges}, M. 2009, \mnras, 398, 1601

\bibitem[{{Gaia Collaboration} {et~al.}(2023{\natexlab{a}}){Gaia
  Collaboration}, {Arenou}, {Babusiaux}, {Barstow}, {Faigler}, {Jorissen},
  {Kervella}, {Mazeh}, {Mowlavi}, {Panuzzo}, {Sahlmann}, {Shahaf}, {Sozzetti},
  {Bauchet}, {Damerdji}, {Gavras}, {Giacobbe}, {Gosset}, {Halbwachs}, {Holl},
  {Lattanzi}, {Leclerc}, {Morel}, {Pourbaix}, {Re Fiorentin}, {Sadowski},
  {S{\'e}gransan}, {Siopis}, {Teyssier}, {Zwitter}, {Planquart}, {Brown},
  {Vallenari}, {Prusti}, {de Bruijne}, {Biermann}, {Creevey}, {Ducourant},
  {Evans}, {Eyer}, {Guerra}, {Hutton}, {Jordi}, {Klioner}, {Lammers},
  {Lindegren}, {Luri}, {Mignard}, {Panem}, {Randich}, {Sartoretti}, {Soubiran},
  {Tanga}, {Walton}, {Bailer-Jones}, {Bastian}, {Drimmel}, {Jansen}, {Katz},
  {van Leeuwen}, {Bakker}, {Cacciari}, {Casta{\~n}eda}, {De Angeli},
  {Fabricius}, {Fouesneau}, {Fr{\'e}mat}, {Galluccio}, {Guerrier}, {Heiter},
  {Masana}, {Messineo}, {Nicolas}, {Nienartowicz}, {Pailler}, {Riclet}, {Roux},
  {Seabroke}, {Sordo}, {Th{\'e}venin}, {Gracia-Abril}, {Portell}, {Altmann},
  {Andrae}, {Audard}, {Bellas-Velidis}, {Benson}, {Berthier}, {Blomme},
  {Burgess}, {Busonero}, {Busso}, {C{\'a}novas}, {Carry}, {Cellino}, {Cheek},
  {Clementini}, {Davidson}, {de Teodoro}, {Nu{\~n}ez Campos}, {Delchambre},
  {Dell'Oro}, {Esquej}, {Fern{\'a}ndez-Hern{\'a}ndez}, {Fraile}, {Garabato},
  {Garc{\'\i}a-Lario}, {Haigron}, {Hambly}, {Harrison}, {Hern{\'a}ndez},
  {Hestroffer}, {Hodgkin}, {Jan{\ss}en}, {Jevardat de Fombelle}, {Jordan},
  {Krone-Martins}, {Lanzafame}, {L{\"o}ffler}, {Marchal}, {Marrese},
  {Moitinho}, {Muinonen}, {Osborne}, {Pancino}, {Pauwels}, {Recio-Blanco},
  {Reyl{\'e}}, {Riello}, {Rimoldini}, {Roegiers}, {Rybizki}, {Sarro}, {Smith},
  {Utrilla}, {van Leeuwen}, {Abbas}, {{\'A}brah{\'a}m}, {Abreu Aramburu},
  {Aerts}, {Aguado}, {Ajaj}, {Aldea-Montero}, {Altavilla}, {{\'A}lvarez},
  {Alves}, {Anders}, {Anderson}, {Anglada Varela}, {Antoja}, {Baines}, {Baker},
  {Balaguer-N{\'u}{\~n}ez}, {Balbinot}, {Balog}, {Barache}, {Barbato},
  {Barros}, {Bartolom{\'e}}, {Bassilana}, {Becciani}, {Bellazzini},
  {Berihuete}, {Bernet}, {Bertone}, {Bianchi}, {Binnenfeld}, {Blanco-Cuaresma},
  {Blazere}, {Boch}, {Bombrun}, {Bossini}, {Bouquillon}, {Bragaglia},
  {Bramante}, {Breedt}, {Bressan}, {Brouillet}, {Brugaletta}, {Bucciarelli},
  {Burlacu}, {Butkevich}, {Buzzi}, {Caffau}, {Cancelliere}, {Cantat-Gaudin},
  {Carballo}, {Carlucci}, {Carnerero}, {Carrasco}, {Casamiquela}, {Castellani},
  {Castro-Ginard}, {Chaoul}, {Charlot}, {Chemin}, {Chiaramida}, {Chiavassa},
  {Chornay}, \& {Comoretto}}]{2023A&A...674A..34G}
{Gaia Collaboration}, {Arenou}, F., {Babusiaux}, C., {et~al.}
  2023{\natexlab{a}}, \aap, 674, A34

\bibitem[{{Gaia Collaboration} {et~al.}(2018){Gaia Collaboration}, {Babusiaux},
  {van Leeuwen}, {Barstow}, {Jordi}, {Vallenari}, {Bossini}, {Bressan},
  {Cantat-Gaudin}, {van Leeuwen}, {Brown}, {Prusti}, {de Bruijne},
  {Bailer-Jones}, {Biermann}, {Evans}, {Eyer}, {Jansen}, {Klioner}, {Lammers},
  {Lindegren}, {Luri}, {Mignard}, {Panem}, {Pourbaix}, {Randich}, {Sartoretti},
  {Siddiqui}, {Soubiran}, {Walton}, {Arenou}, {Bastian}, {Cropper}, {Drimmel},
  {Katz}, {Lattanzi}, {Bakker}, {Cacciari}, {Casta{\~n}eda}, {Chaoul}, {Cheek},
  {De Angeli}, {Fabricius}, {Guerra}, {Holl}, {Masana}, {Messineo}, {Mowlavi},
  {Nienartowicz}, {Panuzzo}, {Portell}, {Riello}, {Seabroke}, {Tanga},
  {Th{\'e}venin}, {Gracia-Abril}, {Comoretto}, {Garcia-Reinaldos}, {Teyssier},
  {Altmann}, {Andrae}, {Audard}, {Bellas-Velidis}, {Benson}, {Berthier},
  {Blomme}, {Burgess}, {Busso}, {Carry}, {Cellino}, {Clementini}, {Clotet},
  {Creevey}, {Davidson}, {De Ridder}, {Delchambre}, {Dell'Oro}, {Ducourant},
  {Fern{\'a}ndez-Hern{\'a}ndez}, {Fouesneau}, {Fr{\'e}mat}, {Galluccio},
  {Garc{\'\i}a-Torres}, {Gonz{\'a}lez-N{\'u}{\~n}ez}, {Gonz{\'a}lez-Vidal},
  {Gosset}, {Guy}, {Halbwachs}, {Hambly}, {Harrison}, {Hern{\'a}ndez},
  {Hestroffer}, {Hodgkin}, {Hutton}, {Jasniewicz}, {Jean-Antoine-Piccolo},
  {Jordan}, {Korn}, {Krone-Martins}, {Lanzafame}, {Lebzelter}, {L{\"o}ffler},
  {Manteiga}, {Marrese}, {Mart{\'\i}n-Fleitas}, {Moitinho}, {Mora}, {Muinonen},
  {Osinde}, {Pancino}, {Pauwels}, {Petit}, {Recio-Blanco}, {Richards},
  {Rimoldini}, {Robin}, {Sarro}, {Siopis}, {Smith}, {Sozzetti}, {S{\"u}veges},
  {Torra}, {van Reeven}, {Abbas}, {Abreu Aramburu}, {Accart}, {Aerts},
  {Altavilla}, {{\'A}lvarez}, {Alvarez}, {Alves}, {Anderson}, {Andrei},
  {Anglada Varela}, {Antiche}, {Antoja}, {Arcay}, {Astraatmadja}, {Bach},
  {Baker}, {Balaguer-N{\'u}{\~n}ez}, {Balm}, {Barache}, {Barata}, {Barbato},
  {Barblan}, {Barklem}, {Barrado}, {Barros}, {Bartholom{\'e} Mu{\~n}oz},
  {Bassilana}, {Becciani}, {Bellazzini}, {Berihuete}, {Bertone}, {Bianchi},
  {Bienaym{\'e}}, {Blanco-Cuaresma}, {Boch}, {Boeche}, {Bombrun}, {Borrachero},
  {Bouquillon}, {Bourda}, {Bragaglia}, {Bramante}, {Breddels}, {Brouillet},
  {Br{\"u}semeister}, {Brugaletta}, {Bucciarelli}, {Burlacu}, {Busonero},
  {Butkevich}, {Buzzi}, {Caffau}, {Cancelliere}, {Cannizzaro}, {Carballo},
  {Carlucci}, {Carrasco}, {Casamiquela}, {Castellani}, {Castro-Ginard},
  {Charlot}, {Chemin}, {Chiavassa}, {Cocozza}, {Costigan}, {Cowell}, {Crifo},
  {Crosta}, {Crowley}, {Cuypers}, {Dafonte}, {Damerdji}, {Dapergolas}, {David},
  {David}, \& {de Laverny}}]{2018A&A...616A..10G}
{Gaia Collaboration}, {Babusiaux}, C., {van Leeuwen}, F., {et~al.} 2018, \aap,
  616, A10

\bibitem[{{Gaia Collaboration} {et~al.}(2021){Gaia Collaboration}, {Brown},
  {Vallenari}, {Prusti}, {de Bruijne}, {Babusiaux}, {Biermann}, {Creevey},
  {Evans}, {Eyer}, {Hutton}, {Jansen}, {Jordi}, {Klioner}, {Lammers},
  {Lindegren}, {Luri}, {Mignard}, {Panem}, {Pourbaix}, {Randich}, {Sartoretti},
  {Soubiran}, {Walton}, {Arenou}, {Bailer-Jones}, {Bastian}, {Cropper},
  {Drimmel}, {Katz}, {Lattanzi}, {van Leeuwen}, {Bakker}, {Cacciari},
  {Casta{\~n}eda}, {De Angeli}, {Ducourant}, {Fabricius}, {Fouesneau},
  {Fr{\'e}mat}, {Guerra}, {Guerrier}, {Guiraud}, {Jean-Antoine Piccolo},
  {Masana}, {Messineo}, {Mowlavi}, {Nicolas}, {Nienartowicz}, {Pailler},
  {Panuzzo}, {Riclet}, {Roux}, {Seabroke}, {Sordo}, {Tanga}, {Th{\'e}venin},
  {Gracia-Abril}, {Portell}, {Teyssier}, {Altmann}, {Andrae}, {Bellas-Velidis},
  {Benson}, {Berthier}, {Blomme}, {Brugaletta}, {Burgess}, {Busso}, {Carry},
  {Cellino}, {Cheek}, {Clementini}, {Damerdji}, {Davidson}, {Delchambre},
  {Dell'Oro}, {Fern{\'a}ndez-Hern{\'a}ndez}, {Galluccio}, {Garc{\'\i}a-Lario},
  {Garcia-Reinaldos}, {Gonz{\'a}lez-N{\'u}{\~n}ez}, {Gosset}, {Haigron},
  {Halbwachs}, {Hambly}, {Harrison}, {Hatzidimitriou}, {Heiter},
  {Hern{\'a}ndez}, {Hestroffer}, {Hodgkin}, {Holl}, {Jan{\ss}en}, {Jevardat de
  Fombelle}, {Jordan}, {Krone-Martins}, {Lanzafame}, {L{\"o}ffler}, {Lorca},
  {Manteiga}, {Marchal}, {Marrese}, {Moitinho}, {Mora}, {Muinonen}, {Osborne},
  {Pancino}, {Pauwels}, {Petit}, {Recio-Blanco}, {Richards}, {Riello},
  {Rimoldini}, {Robin}, {Roegiers}, {Rybizki}, {Sarro}, {Siopis}, {Smith},
  {Sozzetti}, {Ulla}, {Utrilla}, {van Leeuwen}, {van Reeven}, {Abbas}, {Abreu
  Aramburu}, {Accart}, {Aerts}, {Aguado}, {Ajaj}, {Altavilla}, {{\'A}lvarez},
  {{\'A}lvarez Cid-Fuentes}, {Alves}, {Anderson}, {Anglada Varela}, {Antoja},
  {Audard}, {Baines}, {Baker}, {Balaguer-N{\'u}{\~n}ez}, {Balbinot}, {Balog},
  {Barache}, {Barbato}, {Barros}, {Barstow}, {Bartolom{\'e}}, {Bassilana},
  {Bauchet}, {Baudesson-Stella}, {Becciani}, {Bellazzini}, {Bernet}, {Bertone},
  {Bianchi}, {Blanco-Cuaresma}, {Boch}, {Bombrun}, {Bossini}, {Bouquillon},
  {Bragaglia}, {Bramante}, {Breedt}, {Bressan}, {Brouillet}, {Bucciarelli},
  {Burlacu}, {Busonero}, {Butkevich}, {Buzzi}, {Caffau}, {Cancelliere},
  {C{\'a}novas}, {Cantat-Gaudin}, {Carballo}, {Carlucci}, {Carnerero},
  {Carrasco}, {Casamiquela}, {Castellani}, {Castro-Ginard}, {Castro Sampol},
  {Chaoul}, {Charlot}, {Chemin}, {Chiavassa}, {Cioni}, {Comoretto}, {Cooper},
  {Cornez}, {Cowell}, {Crifo}, {Crosta}, {Crowley}, {Dafonte}, {Dapergolas},
  {David}, \& {David}}]{2021A&A...649A...1G}
{Gaia Collaboration}, {Brown}, A.~G.~A., {Vallenari}, A., {et~al.} 2021, \aap,
  649, A1

\bibitem[{{Gaia Collaboration} {et~al.}(2023{\natexlab{b}}){Gaia
  Collaboration}, {Creevey}, {Sarro}, {Lobel}, {Pancino}, {Andrae}, {Smart},
  {Clementini}, {Heiter}, {Korn}, {Fouesneau}, {Fr{\'e}mat}, {De Angeli},
  {Vallenari}, {Harrison}, {Th{\'e}venin}, {Reyl{\'e}}, {Sordo}, {Garofalo},
  {Brown}, {Eyer}, {Prusti}, {de Bruijne}, {Arenou}, {Babusiaux}, {Biermann},
  {Ducourant}, {Evans}, {Guerra}, {Hutton}, {Jordi}, {Klioner}, {Lammers},
  {Lindegren}, {Luri}, {Mignard}, {Panem}, {Pourbaix}, {Randich}, {Sartoretti},
  {Soubiran}, {Tanga}, {Walton}, {Bailer-Jones}, {Bastian}, {Drimmel},
  {Jansen}, {Katz}, {Lattanzi}, {van Leeuwen}, {Bakker}, {Cacciari},
  {Casta{\~n}eda}, {Fabricius}, {Galluccio}, {Guerrier}, {Masana}, {Messineo},
  {Mowlavi}, {Nicolas}, {Nienartowicz}, {Pailler}, {Panuzzo}, {Riclet}, {Roux},
  {Seabroke}, {Gracia-Abril}, {Portell}, {Teyssier}, {Altmann}, {Audard},
  {Bellas-Velidis}, {Benson}, {Berthier}, {Blomme}, {Burgess}, {Busonero},
  {Busso}, {C{\'a}novas}, {Carry}, {Cellino}, {Cheek}, {Damerdji}, {Davidson},
  {de Teodoro}, {Nu{\~n}ez Campos}, {Delchambre}, {Dell'Oro}, {Esquej},
  {Fern{\'a}ndez-Hern{\'a}ndez}, {Fraile}, {Garabato}, {Garc{\'\i}a-Lario},
  {Gosset}, {Haigron}, {Halbwachs}, {Hambly}, {Hern{\'a}ndez}, {Hestroffer},
  {Hodgkin}, {Holl}, {Jan{\ss}en}, {Jevardat de Fombelle}, {Jordan},
  {Krone-Martins}, {Lanzafame}, {L{\"o}ffler}, {Marchal}, {Marrese},
  {Moitinho}, {Muinonen}, {Osborne}, {Pauwels}, {Recio-Blanco}, {Riello},
  {Rimoldini}, {Roegiers}, {Rybizki}, {Siopis}, {Smith}, {Sozzetti}, {Utrilla},
  {van Leeuwen}, {Abbas}, {{\'A}brah{\'a}m}, {Abreu Aramburu}, {Aerts},
  {Aguado}, {Ajaj}, {Aldea-Montero}, {Altavilla}, {{\'A}lvarez}, {Alves},
  {Anders}, {Anderson}, {Anglada Varela}, {Antoja}, {Baines}, {Baker},
  {Balaguer-N{\'u}{\~n}ez}, {Balbinot}, {Balog}, {Barache}, {Barbato},
  {Barros}, {Barstow}, {Bartolom{\'e}}, {Bassilana}, {Bauchet}, {Becciani},
  {Bellazzini}, {Berihuete}, {Bernet}, {Bertone}, {Bianchi}, {Binnenfeld},
  {Blanco-Cuaresma}, {Boch}, {Bombrun}, {Bossini}, {Bouquillon}, {Bragaglia},
  {Bramante}, {Breedt}, {Bressan}, {Brouillet}, {Brugaletta}, {Bucciarelli},
  {Burlacu}, {Butkevich}, {Buzzi}, {Caffau}, {Cancelliere}, {Cantat-Gaudin},
  {Carballo}, {Carlucci}, {Carnerero}, {Carrasco}, {Casamiquela}, {Castellani},
  {Castro-Ginard}, {Chaoul}, {Charlot}, {Chemin}, {Chiaramida}, {Chiavassa},
  {Chornay}, {Comoretto}, {Contursi}, {Cooper}, {Cornez}, {Cowell}, {Crifo},
  {Cropper}, {Crosta}, {Crowley}, {Dafonte}, {Dapergolas}, {David}, \& {de
  Laverny}}]{2023A&A...674A..39G}
{Gaia Collaboration}, {Creevey}, O.~L., {Sarro}, L.~M., {et~al.}
  2023{\natexlab{b}}, \aap, 674, A39

\bibitem[{{Gaia Collaboration} {et~al.}(2023{\natexlab{c}}){Gaia
  Collaboration}, {Vallenari}, {Brown}, {Prusti}, {de Bruijne}, {Arenou},
  {Babusiaux}, {Biermann}, {Creevey}, {Ducourant}, {Evans}, {Eyer}, {Guerra},
  {Hutton}, {Jordi}, {Klioner}, {Lammers}, {Lindegren}, {Luri}, {Mignard},
  {Panem}, {Pourbaix}, {Randich}, {Sartoretti}, {Soubiran}, {Tanga}, {Walton},
  {Bailer-Jones}, {Bastian}, {Drimmel}, {Jansen}, {Katz}, {Lattanzi}, {van
  Leeuwen}, {Bakker}, {Cacciari}, {Casta{\~n}eda}, {De Angeli}, {Fabricius},
  {Fouesneau}, {Fr{\'e}mat}, {Galluccio}, {Guerrier}, {Heiter}, {Masana},
  {Messineo}, {Mowlavi}, {Nicolas}, {Nienartowicz}, {Pailler}, {Panuzzo},
  {Riclet}, {Roux}, {Seabroke}, {Sordo}, {Th{\'e}venin}, {Gracia-Abril},
  {Portell}, {Teyssier}, {Altmann}, {Andrae}, {Audard}, {Bellas-Velidis},
  {Benson}, {Berthier}, {Blomme}, {Burgess}, {Busonero}, {Busso},
  {C{\'a}novas}, {Carry}, {Cellino}, {Cheek}, {Clementini}, {Damerdji},
  {Davidson}, {de Teodoro}, {Nu{\~n}ez Campos}, {Delchambre}, {Dell'Oro},
  {Esquej}, {Fern{\'a}ndez-Hern{\'a}ndez}, {Fraile}, {Garabato},
  {Garc{\'\i}a-Lario}, {Gosset}, {Haigron}, {Halbwachs}, {Hambly}, {Harrison},
  {Hern{\'a}ndez}, {Hestroffer}, {Hodgkin}, {Holl}, {Jan{\ss}en}, {Jevardat de
  Fombelle}, {Jordan}, {Krone-Martins}, {Lanzafame}, {L{\"o}ffler}, {Marchal},
  {Marrese}, {Moitinho}, {Muinonen}, {Osborne}, {Pancino}, {Pauwels},
  {Recio-Blanco}, {Reyl{\'e}}, {Riello}, {Rimoldini}, {Roegiers}, {Rybizki},
  {Sarro}, {Siopis}, {Smith}, {Sozzetti}, {Utrilla}, {van Leeuwen}, {Abbas},
  {{\'A}brah{\'a}m}, {Abreu Aramburu}, {Aerts}, {Aguado}, {Ajaj},
  {Aldea-Montero}, {Altavilla}, {{\'A}lvarez}, {Alves}, {Anders}, {Anderson},
  {Anglada Varela}, {Antoja}, {Baines}, {Baker}, {Balaguer-N{\'u}{\~n}ez},
  {Balbinot}, {Balog}, {Barache}, {Barbato}, {Barros}, {Barstow},
  {Bartolom{\'e}}, {Bassilana}, {Bauchet}, {Becciani}, {Bellazzini},
  {Berihuete}, {Bernet}, {Bertone}, {Bianchi}, {Binnenfeld}, {Blanco-Cuaresma},
  {Blazere}, {Boch}, {Bombrun}, {Bossini}, {Bouquillon}, {Bragaglia},
  {Bramante}, {Breedt}, {Bressan}, {Brouillet}, {Brugaletta}, {Bucciarelli},
  {Burlacu}, {Butkevich}, {Buzzi}, {Caffau}, {Cancelliere}, {Cantat-Gaudin},
  {Carballo}, {Carlucci}, {Carnerero}, {Carrasco}, {Casamiquela}, {Castellani},
  {Castro-Ginard}, {Chaoul}, {Charlot}, {Chemin}, {Chiaramida}, {Chiavassa},
  {Chornay}, {Comoretto}, {Contursi}, {Cooper}, {Cornez}, {Cowell}, {Crifo},
  {Cropper}, {Crosta}, {Crowley}, {Dafonte}, {Dapergolas}, {David}, {David},
  {de Laverny}, {De Luise}, \& {De March}}]{2023A&A...674A...1G}
{Gaia Collaboration}, {Vallenari}, A., {Brown}, A.~G.~A., {et~al.}
  2023{\natexlab{c}}, \aap, 674, A1

\bibitem[{{Gallenne} {et~al.}(2018){Gallenne}, {Pietrzy{\'n}ski}, {Graczyk},
  {Nardetto}, {M{\'e}rand}, {Kervella}, {Gieren}, {Villanova}, {Mennickent}, \&
  {Pilecki}}]{2018A&A...616A..68G}
{Gallenne}, A., {Pietrzy{\'n}ski}, G., {Graczyk}, D., {et~al.} 2018, \aap, 616,
  A68

\bibitem[{{Ghezzi} {et~al.}(2018){Ghezzi}, {Montet}, \&
  {Johnson}}]{2018ApJ...860..109G}
{Ghezzi}, L., {Montet}, B.~T., \& {Johnson}, J.~A. 2018, \apj, 860, 109

\bibitem[{{Girardi}(1999)}]{1999MNRAS.308..818G}
{Girardi}, L. 1999, \mnras, 308, 818

\bibitem[{{Girardi}(2016)}]{2016ARA&A..54...95G}
{Girardi}, L. 2016, \araa, 54, 95

\bibitem[{{Glebocki} \& {Gnacinski}(2005)}]{2005yCat.3244....0G}
{Glebocki}, R. \& {Gnacinski}, P. 2005, {VizieR Online Data Catalog: Catalog of
  Stellar Rotational Velocities (Glebocki+ 2005)}, VizieR On-line Data Catalog:
  III/244. Originally published in: 2005csss...13..571G

\bibitem[{{Gosset} {et~al.}(2025){Gosset}, {Damerdji}, {Morel}, {Delchambre},
  {Halbwachs}, {Sadowski}, {Pourbaix}, {Sozzetti}, {Panuzzo}, \&
  {Arenou}}]{2025A&A...693A.124G}
{Gosset}, E., {Damerdji}, Y., {Morel}, T., {et~al.} 2025, \aap, 693, A124

\bibitem[{{Green} {et~al.}(2019){Green}, {Schlafly}, {Zucker}, {Speagle}, \&
  {Finkbeiner}}]{2019ApJ...887...93G}
{Green}, G.~M., {Schlafly}, E., {Zucker}, C., {Speagle}, J.~S., \&
  {Finkbeiner}, D. 2019, \apj, 887, 93

\bibitem[{{Griffin}(1967)}]{1967ApJ...148..465G}
{Griffin}, R.~F. 1967, \apj, 148, 465

\bibitem[{{Han} {et~al.}(2003){Han}, {Podsiadlowski}, {Maxted}, \&
  {Marsh}}]{2003MNRAS.341..669H}
{Han}, Z., {Podsiadlowski}, P., {Maxted}, P.~F.~L., \& {Marsh}, T.~R. 2003,
  \mnras, 341, 669

\bibitem[{{Han} {et~al.}(2002){Han}, {Podsiadlowski}, {Maxted}, {Marsh}, \&
  {Ivanova}}]{2002MNRAS.336..449H}
{Han}, Z., {Podsiadlowski}, P., {Maxted}, P.~F.~L., {Marsh}, T.~R., \&
  {Ivanova}, N. 2002, \mnras, 336, 449

\bibitem[{{Heber}(2016)}]{2016PASP..128h2001H}
{Heber}, U. 2016, \pasp, 128, 082001

\bibitem[{{Heged{\H{u}}s} {et~al.}(2023){Heged{\H{u}}s}, {M{\'e}sz{\'a}ros},
  {Jofr{\'e}}, {Stringfellow}, {Feuillet}, {Garc{\'\i}a-Hern{\'a}ndez},
  {Nitschelm}, \& {Zamora}}]{2023A&A...670A.107H}
{Heged{\H{u}}s}, V., {M{\'e}sz{\'a}ros}, S., {Jofr{\'e}}, P., {et~al.} 2023,
  \aap, 670, A107

\bibitem[{{Herrero} {et~al.}(1992){Herrero}, {Kudritzki}, {Vilchez}, {Kunze},
  {Butler}, \& {Haser}}]{1992LNP...401...21H}
{Herrero}, A., {Kudritzki}, R.~P., {Vilchez}, J.~M., {et~al.} 1992, in The
  Atmospheres of Early-Type Stars, ed. U.~{Heber} \& C.~S. {Jeffery}, Vol. 401,
  21

\bibitem[{{Hojjatpanah} {et~al.}(2019){Hojjatpanah}, {Figueira}, {Santos},
  {Adibekyan}, {Sousa}, {Delgado-Mena}, {Alibert}, {Cristiani}, {Gonz{\'a}lez
  Hern{\'a}ndez}, {Lanza}, {Di Marcantonio}, {Martins}, {Micela}, {Molaro},
  {Neves}, {Oshagh}, {Pepe}, {Poretti}, {Rojas-Ayala}, {Rebolo}, {Su{\'a}rez
  Mascare{\~n}o}, \& {Zapatero Osorio}}]{2019A&A...629A..80H}
{Hojjatpanah}, S., {Figueira}, P., {Santos}, N.~C., {et~al.} 2019, \aap, 629,
  A80

\bibitem[{{Hon} {et~al.}(2021){Hon}, {Huber}, {Kuszlewicz}, {Stello}, {Sharma},
  {Tayar}, {Zinn}, {Vrard}, \& {Pinsonneault}}]{2021ApJ...919..131H}
{Hon}, M., {Huber}, D., {Kuszlewicz}, J.~S., {et~al.} 2021, \apj, 919, 131

\bibitem[{{Howell} {et~al.}(2014){Howell}, {Sobeck}, {Haas}, {Still},
  {Barclay}, {Mullally}, {Troeltzsch}, {Aigrain}, {Bryson}, {Caldwell},
  {Chaplin}, {Cochran}, {Huber}, {Marcy}, {Miglio}, {Najita}, {Smith},
  {Twicken}, \& {Fortney}}]{2014PASP..126..398H}
{Howell}, S.~B., {Sobeck}, C., {Haas}, M., {et~al.} 2014, \pasp, 126, 398

\bibitem[{{Jones} {et~al.}(2011){Jones}, {Jenkins}, {Rojo}, \&
  {Melo}}]{2011A&A...536A..71J}
{Jones}, M.~I., {Jenkins}, J.~S., {Rojo}, P., \& {Melo}, C.~H.~F. 2011, \aap,
  536, A71

\bibitem[{{Khan} {et~al.}(2023){Khan}, {Anderson}, {Miglio}, {Mosser}, \&
  {Elsworth}}]{2023A&A...680A.105K}
{Khan}, S., {Anderson}, R.~I., {Miglio}, A., {Mosser}, B., \& {Elsworth}, Y.~P.
  2023, \aap, 680, A105

\bibitem[{{Kjeldsen} \& {Bedding}(1995)}]{1995A&A...293...87K}
{Kjeldsen}, H. \& {Bedding}, T.~R. 1995, \aap, 293, 87

\bibitem[{{Lindegren} {et~al.}(2018){Lindegren}, {Hern{\'a}ndez}, {Bombrun},
  {Klioner}, {Bastian}, {Ramos-Lerate}, {de Torres}, {Steidelm{\"u}ller},
  {Stephenson}, {Hobbs}, {Lammers}, {Biermann}, {Geyer}, {Hilger}, {Michalik},
  {Stampa}, {McMillan}, {Casta{\~n}eda}, {Clotet}, {Comoretto}, {Davidson},
  {Fabricius}, {Gracia}, {Hambly}, {Hutton}, {Mora}, {Portell}, {van Leeuwen},
  {Abbas}, {Abreu}, {Altmann}, {Andrei}, {Anglada}, {Balaguer-N{\'u}{\~n}ez},
  {Barache}, {Becciani}, {Bertone}, {Bianchi}, {Bouquillon}, {Bourda},
  {Br{\"u}semeister}, {Bucciarelli}, {Busonero}, {Buzzi}, {Cancelliere},
  {Carlucci}, {Charlot}, {Cheek}, {Crosta}, {Crowley}, {de Bruijne}, {de
  Felice}, {Drimmel}, {Esquej}, {Fienga}, {Fraile}, {Gai}, {Garralda},
  {Gonz{\'a}lez-Vidal}, {Guerra}, {Hauser}, {Hofmann}, {Holl}, {Jordan},
  {Lattanzi}, {Lenhardt}, {Liao}, {Licata}, {Lister}, {L{\"o}ffler},
  {Marchant}, {Martin-Fleitas}, {Messineo}, {Mignard}, {Morbidelli}, {Poggio},
  {Riva}, {Rowell}, {Salguero}, {Sarasso}, {Sciacca}, {Siddiqui}, {Smart},
  {Spagna}, {Steele}, {Taris}, {Torra}, {van Elteren}, {van Reeven}, \&
  {Vecchiato}}]{2018A&A...616A...2L}
{Lindegren}, L., {Hern{\'a}ndez}, J., {Bombrun}, A., {et~al.} 2018, \aap, 616,
  A2

\bibitem[{{Majewski} {et~al.}(2017){Majewski}, {Schiavon}, {Frinchaboy},
  {Allende Prieto}, {Barkhouser}, {Bizyaev}, {Blank}, {Brunner}, {Burton},
  {Carrera}, {Chojnowski}, {Cunha}, {Epstein}, {Fitzgerald}, {Garc{\'\i}a
  P{\'e}rez}, {Hearty}, {Henderson}, {Holtzman}, {Johnson}, {Lam}, {Lawler},
  {Maseman}, {M{\'e}sz{\'a}ros}, {Nelson}, {Nguyen}, {Nidever}, {Pinsonneault},
  {Shetrone}, {Smee}, {Smith}, {Stolberg}, {Skrutskie}, {Walker}, {Wilson},
  {Zasowski}, {Anders}, {Basu}, {Beland}, {Blanton}, {Bovy}, {Brownstein},
  {Carlberg}, {Chaplin}, {Chiappini}, {Eisenstein}, {Elsworth}, {Feuillet},
  {Fleming}, {Galbraith-Frew}, {Garc{\'\i}a}, {Garc{\'\i}a-Hern{\'a}ndez},
  {Gillespie}, {Girardi}, {Gunn}, {Hasselquist}, {Hayden}, {Hekker}, {Ivans},
  {Kinemuchi}, {Klaene}, {Mahadevan}, {Mathur}, {Mosser}, {Muna}, {Munn},
  {Nichol}, {O'Connell}, {Parejko}, {Robin}, {Rocha-Pinto}, {Schultheis},
  {Serenelli}, {Shane}, {Silva Aguirre}, {Sobeck}, {Thompson}, {Troup},
  {Weinberg}, \& {Zamora}}]{2017AJ....154...94M}
{Majewski}, S.~R., {Schiavon}, R.~P., {Frinchaboy}, P.~M., {et~al.} 2017, \aj,
  154, 94

\bibitem[{{Marshall} {et~al.}(2006){Marshall}, {Robin}, {Reyl{\'e}},
  {Schultheis}, \& {Picaud}}]{2006A&A...453..635M}
{Marshall}, D.~J., {Robin}, A.~C., {Reyl{\'e}}, C., {Schultheis}, M., \&
  {Picaud}, S. 2006, \aap, 453, 635

\bibitem[{{Martig} {et~al.}(2015){Martig}, {Rix}, {Silva Aguirre}, {Hekker},
  {Mosser}, {Elsworth}, {Bovy}, {Stello}, {Anders}, {Garc{\'\i}a}, {Tayar},
  {Rodrigues}, {Basu}, {Carrera}, {Ceillier}, {Chaplin}, {Chiappini},
  {Frinchaboy}, {Garc{\'\i}a-Hern{\'a}ndez}, {Hearty}, {Holtzman}, {Johnson},
  {Majewski}, {Mathur}, {M{\'e}sz{\'a}ros}, {Miglio}, {Nidever}, {Pan},
  {Pinsonneault}, {Schiavon}, {Schneider}, {Serenelli}, {Shetrone}, \&
  {Zamora}}]{2015MNRAS.451.2230M}
{Martig}, M., {Rix}, H.-W., {Silva Aguirre}, V., {et~al.} 2015, \mnras, 451,
  2230

\bibitem[{{Massarotti} {et~al.}(2008){Massarotti}, {Latham}, {Stefanik}, \&
  {Fogel}}]{2008AJ....135..209M}
{Massarotti}, A., {Latham}, D.~W., {Stefanik}, R.~P., \& {Fogel}, J. 2008, \aj,
  135, 209

\bibitem[{{Maxted} {et~al.}(2001){Maxted}, {Heber}, {Marsh}, \&
  {North}}]{2001MNRAS.326.1391M}
{Maxted}, P.~F.~L., {Heber}, U., {Marsh}, T.~R., \& {North}, R.~C. 2001,
  \mnras, 326, 1391

\bibitem[{{Morton}(2015)}]{2015ascl.soft03010M}
{Morton}, T.~D. 2015, {isochrones: Stellar model grid package}, Astrophysics
  Source Code Library, record ascl:1503.010

\bibitem[{{Mucciarelli} {et~al.}(2021){Mucciarelli}, {Bellazzini}, \&
  {Massari}}]{2021A&A...653A..90M}
{Mucciarelli}, A., {Bellazzini}, M., \& {Massari}, D. 2021, \aap, 653, A90

\bibitem[{{Napiwotzki} {et~al.}(2004){Napiwotzki}, {Karl}, {Lisker}, {Heber},
  {Christlieb}, {Reimers}, {Nelemans}, \& {Homeier}}]{2004Ap&SS.291..321N}
{Napiwotzki}, R., {Karl}, C.~A., {Lisker}, T., {et~al.} 2004, \apss, 291, 321

\bibitem[{{{\O}stensen} \& {Van Winckel}(2012)}]{2012ASPC..452..163O}
{{\O}stensen}, R.~H. \& {Van Winckel}, H. 2012, in Astronomical Society of the
  Pacific Conference Series, Vol. 452, Fifth Meeting on Hot Subdwarf Stars and
  Related Objects, ed. D.~{Kilkenny}, C.~S. {Jeffery}, \& C.~{Koen}, 163

\bibitem[{{Ottoni} {et~al.}(2022){Ottoni}, {Udry}, {S{\'e}gransan}, {Buldgen},
  {Lovis}, {Eggenberger}, {Pezzotti}, {Adibekyan}, {Marmier}, {Mayor},
  {Santos}, {Sousa}, {Lagarde}, \& {Charbonnel}}]{2022A&A...657A..87O}
{Ottoni}, G., {Udry}, S., {S{\'e}gransan}, D., {et~al.} 2022, \aap, 657, A87

\bibitem[{{Paczynski}(1976)}]{1976IAUS...73...75P}
{Paczynski}, B. 1976, in IAU Symposium, Vol.~73, Structure and Evolution of
  Close Binary Systems, ed. P.~{Eggleton}, S.~{Mitton}, \& J.~{Whelan}, 75

\bibitem[{{Paxton} {et~al.}(2011){Paxton}, {Bildsten}, {Dotter}, {Herwig},
  {Lesaffre}, \& {Timmes}}]{2011ApJS..192....3P}
{Paxton}, B., {Bildsten}, L., {Dotter}, A., {et~al.} 2011, \apjs, 192, 3

\bibitem[{{Pelisoli} {et~al.}(2020){Pelisoli}, {Vos}, {Geier}, {Schaffenroth},
  \& {Baran}}]{2020A&A...642A.180P}
{Pelisoli}, I., {Vos}, J., {Geier}, S., {Schaffenroth}, V., \& {Baran}, A.~S.
  2020, \aap, 642, A180

\bibitem[{{Pietrinferni} {et~al.}(2004){Pietrinferni}, {Cassisi}, {Salaris}, \&
  {Castelli}}]{2004ApJ...612..168P}
{Pietrinferni}, A., {Cassisi}, S., {Salaris}, M., \& {Castelli}, F. 2004, \apj,
  612, 168

\bibitem[{{Queloz} {et~al.}(2001){Queloz}, {Mayor}, {Udry}, {Burnet},
  {Carrier}, {Eggenberger}, {Naef}, {Santos}, {Pepe}, {Rupprecht}, {Avila},
  {Baeza}, {Benz}, {Bertaux}, {Bouchy}, {Cavadore}, {Delabre}, {Eckert},
  {Fischer}, {Fleury}, {Gilliotte}, {Goyak}, {Guzman}, {Kohler}, {Lacroix},
  {Lizon}, {Megevand}, {Sivan}, {Sosnowska}, \&
  {Weilenmann}}]{2001Msngr.105....1Q}
{Queloz}, D., {Mayor}, M., {Udry}, S., {et~al.} 2001, The Messenger, 105, 1

\bibitem[{{Ricker} {et~al.}(2015){Ricker}, {Winn}, {Vanderspek}, {Latham},
  {Bakos}, {Bean}, {Berta-Thompson}, {Brown}, {Buchhave}, {Butler}, {Butler},
  {Chaplin}, {Charbonneau}, {Christensen-Dalsgaard}, {Clampin}, {Deming},
  {Doty}, {De Lee}, {Dressing}, {Dunham}, {Endl}, {Fressin}, {Ge}, {Henning},
  {Holman}, {Howard}, {Ida}, {Jenkins}, {Jernigan}, {Johnson}, {Kaltenegger},
  {Kawai}, {Kjeldsen}, {Laughlin}, {Levine}, {Lin}, {Lissauer}, {MacQueen},
  {Marcy}, {McCullough}, {Morton}, {Narita}, {Paegert}, {Palle}, {Pepe},
  {Pepper}, {Quirrenbach}, {Rinehart}, {Sasselov}, {Sato}, {Seager},
  {Sozzetti}, {Stassun}, {Sullivan}, {Szentgyorgyi}, {Torres}, {Udry}, \&
  {Villasenor}}]{2015JATIS...1a4003R}
{Ricker}, G.~R., {Winn}, J.~N., {Vanderspek}, R., {et~al.} 2015, Journal of
  Astronomical Telescopes, Instruments, and Systems, 1, 014003

\bibitem[{{Rowan} {et~al.}(2025){Rowan}, {Stanek}, {Kochanek}, {Thompson},
  {Jayasinghe}, {Blaum}, {Fulton}, {Ilyin}, {Isaacson}, {LeBaron}, {Lu}, \&
  {Martin}}]{2025OJAp....8E..18R}
{Rowan}, D.~M., {Stanek}, K.~Z., {Kochanek}, C.~S., {et~al.} 2025, The Open
  Journal of Astrophysics, 8, 18

\bibitem[{{Salaris} {et~al.}(2018){Salaris}, {Cassisi}, {Schiavon}, \&
  {Pietrinferni}}]{2018A&A...612A..68S}
{Salaris}, M., {Cassisi}, S., {Schiavon}, R.~P., \& {Pietrinferni}, A. 2018,
  \aap, 612, A68

\bibitem[{{S{\'e}gransan} {et~al.}(2010){S{\'e}gransan}, {Udry}, {Mayor},
  {Naef}, {Pepe}, {Queloz}, {Santos}, {Demory}, {Figueira}, {Gillon},
  {Marmier}, {M{\'e}gevand}, {Sosnowska}, {Tamuz}, \&
  {Triaud}}]{2010A&A...511A..45S}
{S{\'e}gransan}, D., {Udry}, S., {Mayor}, M., {et~al.} 2010, \aap, 511, A45

\bibitem[{{Setiawan} {et~al.}(2004){Setiawan}, {Pasquini}, {da Silva},
  {Hatzes}, {von der L{\"u}he}, {Girardi}, {de Medeiros}, \&
  {Guenther}}]{2004A&A...421..241S}
{Setiawan}, J., {Pasquini}, L., {da Silva}, L., {et~al.} 2004, \aap, 421, 241

\bibitem[{{Sneden} {et~al.}(2012){Sneden}, {Bean}, {Ivans}, {Lucatello}, \&
  {Sobeck}}]{2012ascl.soft02009S}
{Sneden}, C., {Bean}, J., {Ivans}, I., {Lucatello}, S., \& {Sobeck}, J. 2012,
  {MOOG: LTE line analysis and spectrum synthesis}, Astrophysics Source Code
  Library, record ascl:1202.009

\bibitem[{{Soto} \& {Jenkins}(2018)}]{2018A&A...615A..76S}
{Soto}, M.~G. \& {Jenkins}, J.~S. 2018, \aap, 615, A76

\bibitem[{{Soto} {et~al.}(2021){Soto}, {Jones}, \&
  {Jenkins}}]{2021A&A...647A.157S}
{Soto}, M.~G., {Jones}, M.~I., \& {Jenkins}, J.~S. 2021, \aap, 647, A157

\bibitem[{{Sousa} {et~al.}(2015){Sousa}, {Santos}, {Adibekyan}, {Delgado-Mena},
  \& {Israelian}}]{2015A&A...577A..67S}
{Sousa}, S.~G., {Santos}, N.~C., {Adibekyan}, V., {Delgado-Mena}, E., \&
  {Israelian}, G. 2015, \aap, 577, A67

\bibitem[{{Sousa} {et~al.}(2007){Sousa}, {Santos}, {Israelian}, {Mayor}, \&
  {Monteiro}}]{2007A&A...469..783S}
{Sousa}, S.~G., {Santos}, N.~C., {Israelian}, G., {Mayor}, M., \& {Monteiro},
  M.~J.~P.~F.~G. 2007, \aap, 469, 783

\bibitem[{{Stark} \& {Wade}(2003)}]{2003AJ....126.1455S}
{Stark}, M.~A. \& {Wade}, R.~A. 2003, \aj, 126, 1455

\bibitem[{{Steinmetz} {et~al.}(2006){Steinmetz}, {Zwitter}, {Siebert},
  {Watson}, {Freeman}, {Munari}, {Campbell}, {Williams}, {Seabroke}, {Wyse},
  {Parker}, {Bienaym{\'e}}, {Roeser}, {Gibson}, {Gilmore}, {Grebel}, {Helmi},
  {Navarro}, {Burton}, {Cass}, {Dawe}, {Fiegert}, {Hartley}, {Russell},
  {Saunders}, {Enke}, {Bailin}, {Binney}, {Bland-Hawthorn}, {Boeche}, {Dehnen},
  {Eisenstein}, {Evans}, {Fiorucci}, {Fulbright}, {Gerhard}, {Jauregi}, {Kelz},
  {Mijovi{\'c}}, {Minchev}, {Parmentier}, {Pe{\~n}arrubia}, {Quillen}, {Read},
  {Ruchti}, {Scholz}, {Siviero}, {Smith}, {Sordo}, {Veltz}, {Vidrih}, {von
  Berlepsch}, {Boyle}, \& {Schilbach}}]{2006AJ....132.1645S}
{Steinmetz}, M., {Zwitter}, T., {Siebert}, A., {et~al.} 2006, \aj, 132, 1645

\bibitem[{{Taylor}(2005)}]{2005ASPC..347...29T}
{Taylor}, M.~B. 2005, in Astronomical Society of the Pacific Conference Series,
  Vol. 347, Astronomical Data Analysis Software and Systems XIV, ed.
  P.~{Shopbell}, M.~{Britton}, \& R.~{Ebert}, 29

\bibitem[{{Tkachenko} {et~al.}(2020){Tkachenko}, {Pavlovski}, {Johnston},
  {Pedersen}, {Michielsen}, {Bowman}, {Southworth}, {Tsymbal}, \&
  {Aerts}}]{2020A&A...637A..60T}
{Tkachenko}, A., {Pavlovski}, K., {Johnston}, C., {et~al.} 2020, \aap, 637, A60

\bibitem[{{Tsantaki} {et~al.}(2022){Tsantaki}, {Pancino}, {Marrese},
  {Marinoni}, {Rainer}, {Sanna}, {Turchi}, {Randich}, {Gallart}, {Battaglia},
  \& {Masseron}}]{2022A&A...659A..95T}
{Tsantaki}, M., {Pancino}, E., {Marrese}, P., {et~al.} 2022, \aap, 659, A95

\bibitem[{{Ulrich}(1986)}]{1986ApJ...306L..37U}
{Ulrich}, R.~K. 1986, \apjl, 306, L37

\bibitem[{{Uzundag} {et~al.}(2022){Uzundag}, {Jones}, {Vu{\v{c}}kovi{\'c}},
  {Vos}, {Bobrick}, \& {Paladini}}]{2022A&A...668A..89U}
{Uzundag}, M., {Jones}, M.~I., {Vu{\v{c}}kovi{\'c}}, M., {et~al.} 2022, \aap,
  668, A89

\bibitem[{{Valle} {et~al.}(2024){Valle}, {Dell'Omodarme}, {Prada Moroni}, \&
  {Degl'Innocenti}}]{2024A&A...690A.323V}
{Valle}, G., {Dell'Omodarme}, M., {Prada Moroni}, P.~G., \& {Degl'Innocenti},
  S. 2024, \aap, 690, A323

\bibitem[{{Vos} {et~al.}(2020){Vos}, {Bobrick}, \&
  {Vu{\v{c}}kovi{\'c}}}]{2020A&A...641A.163V}
{Vos}, J., {Bobrick}, A., \& {Vu{\v{c}}kovi{\'c}}, M. 2020, \aap, 641, A163

\bibitem[{{Vos} {et~al.}(2018){Vos}, {N{\'e}meth}, {Vu{\v{c}}kovi{\'c}},
  {{\O}stensen}, \& {Parsons}}]{2018MNRAS.473..693V}
{Vos}, J., {N{\'e}meth}, P., {Vu{\v{c}}kovi{\'c}}, M., {{\O}stensen}, R., \&
  {Parsons}, S. 2018, \mnras, 473, 693

\bibitem[{{Vos} {et~al.}(2014){Vos}, {{\"O}stensen}, \& {Van
  Winckel}}]{2014ASPC..481..265V}
{Vos}, J., {{\"O}stensen}, R., \& {Van Winckel}, H. 2014, in Astronomical
  Society of the Pacific Conference Series, Vol. 481, 6th Meeting on Hot
  Subdwarf Stars and Related Objects, ed. V.~{van Grootel}, E.~{Green},
  G.~{Fontaine}, \& S.~{Charpinet}, 265

\bibitem[{{Vos} {et~al.}(2012){Vos}, {{\O}stensen}, {Degroote}, {De Smedt},
  {Green}, {Heber}, {Van Winckel}, {Acke}, {Bloemen}, {De Cat}, {Exter},
  {Lampens}, {Lombaert}, {Masseron}, {Menu}, {Neyskens}, {Raskin}, {Ringat},
  {Rauch}, {Smolders}, \& {Tkachenko}}]{2012A&A...548A...6V}
{Vos}, J., {{\O}stensen}, R.~H., {Degroote}, P., {et~al.} 2012, \aap, 548, A6

\bibitem[{{Vos} {et~al.}(2015){Vos}, {{\O}stensen}, {Marchant}, \& {Van
  Winckel}}]{2015A&A...579A..49V}
{Vos}, J., {{\O}stensen}, R.~H., {Marchant}, P., \& {Van Winckel}, H. 2015,
  \aap, 579, A49

\bibitem[{{Vos} {et~al.}(2013){Vos}, {{\O}stensen}, {N{\'e}meth}, {Green},
  {Heber}, \& {Van Winckel}}]{2013A&A...559A..54V}
{Vos}, J., {{\O}stensen}, R.~H., {N{\'e}meth}, P., {et~al.} 2013, \aap, 559,
  A54

\bibitem[{{Vos} {et~al.}(2017){Vos}, {{\O}stensen}, {Vu{\v{c}}kovi{\'c}}, \&
  {Van Winckel}}]{2017A&A...605A.109V}
{Vos}, J., {{\O}stensen}, R.~H., {Vu{\v{c}}kovi{\'c}}, M., \& {Van Winckel}, H.
  2017, \aap, 605, A109

\bibitem[{{Vos} {et~al.}(2019){Vos}, {Vu{\v{c}}kovi{\'c}}, {Chen}, {Han},
  {Boudreaux}, {Barlow}, {{\O}stensen}, \& {N{\'e}meth}}]{2019MNRAS.482.4592V}
{Vos}, J., {Vu{\v{c}}kovi{\'c}}, M., {Chen}, X., {et~al.} 2019, \mnras, 482,
  4592

\bibitem[{{Warfield} {et~al.}(2021){Warfield}, {Zinn}, {Pinsonneault},
  {Johnson}, {Stello}, {Elsworth}, {Garc{\'\i}a}, {Kallinger}, {Mathur},
  {Mosser}, {Beaton}, \& {Garc{\'\i}a-Hern{\'a}ndez}}]{2021AJ....161..100W}
{Warfield}, J.~T., {Zinn}, J.~C., {Pinsonneault}, M.~H., {et~al.} 2021, \aj,
  161, 100

\bibitem[{{Webbink}(1984)}]{1984ApJ...277..355W}
{Webbink}, R.~F. 1984, \apj, 277, 355

\bibitem[{{Weidner} \& {Vink}(2010)}]{2010A&A...524A..98W}
{Weidner}, C. \& {Vink}, J.~S. 2010, \aap, 524, A98

\bibitem[{{Wittenmyer} {et~al.}(2016){Wittenmyer}, {Liu}, {Wang}, {Casagrande},
  {Johnson}, \& {Tinney}}]{2016AJ....152...19W}
{Wittenmyer}, R.~A., {Liu}, F., {Wang}, L., {et~al.} 2016, \aj, 152, 19

\bibitem[{{Yu} {et~al.}(2023){Yu}, {Khanna}, {Themessl}, {Hekker}, {Dr{\'e}au},
  {Gizon}, \& {Bi}}]{2023ApJS..264...41Y}
{Yu}, J., {Khanna}, S., {Themessl}, N., {et~al.} 2023, \apjs, 264, 41

\bibitem[{{Zhao} {et~al.}(2012){Zhao}, {Zhao}, {Chu}, {Jing}, \&
  {Deng}}]{2012arXiv1206.3569Z}
{Zhao}, G., {Zhao}, Y., {Chu}, Y., {Jing}, Y., \& {Deng}, L. 2012, arXiv
  e-prints, arXiv:1206.3569

\bibitem[{{Zhou} {et~al.}(2024){Zhou}, {Bi}, {Yu}, {Li}, {Zhang}, {Li}, {Long},
  {Li}, {Sun}, \& {Ye}}]{2024ApJS..271...17Z}
{Zhou}, J., {Bi}, S., {Yu}, J., {et~al.} 2024, \apjs, 271, 17

\end{thebibliography}

\begin{appendix}

\section{Including exctintion and DR3 parallaxes}
\label{Appendix:update_sample}

\citetaliasads{2022A&A...668A..89U} used Gaia DR2 photometric and astrometric data combined with synthetic colours from MIST models to define a region in the Gaia colour-magnitude diagram containing RGB+MS candidate progenitors of long-period sdB+MS systems. This region included 2777 stars, 1932 of which lie in the southern hemisphere (declination $\leq 20^\circ$). Following the release of Gaia DR3, which provides improved parallax and interstellar extinction coefficients, we refined the sample using updated measurements. As extinction coefficients are unavailable for all stars, we employed the \texttt{mwdust} Python package \citepads{2016ApJ...818..130B} to derive $E(B-V)$ using the \texttt{Combined19} 3D dust maps, which combines the maps from \citetads{2003A&A...409..205D}, \citetads{2006A&A...453..635M}, and \citetads{2019ApJ...887...93G} to cover the full sky. We compute $A_G = 2.35E(B-V)$ \citepads{2021A&A...647A.157S} and $E(BP-RP) = (1/2)A_G$ following Gaia's documentation\footnote{\url{https://gea.esac.esa.int/archive/documentation/GDR2/Data_analysis/chap_cu8par/sec_cu8par_process/ssec_cu8par_process_priamextinction.html}}.

Originally, all the 2777 stars in the sample fulfilled the quality criteria outlined in \citetads{2018A&A...616A...2L}, namely, \texttt{parallax\_over\_error > 10}, \texttt{phot\_g\_mean\_flux\_over\_error} > 10, \texttt{phot\_bp\_mean\_flux\_over\_error} > 10 and \texttt{phot\_rp\_mean\_flux\_over\_error} > 10, ensuring astrometric and photometric errors less than 10\%. Using the updated DR3 measurements, we realised that three stars no longer fulfil these criteria. Hence, we excluded them, reducing the sample size to 2774. The three stars are located in the northern hemisphere and are not part of our observing sample.

\begin{figure}[ht!]
\centering
\includegraphics[height=0.75\linewidth, width=0.75\linewidth]{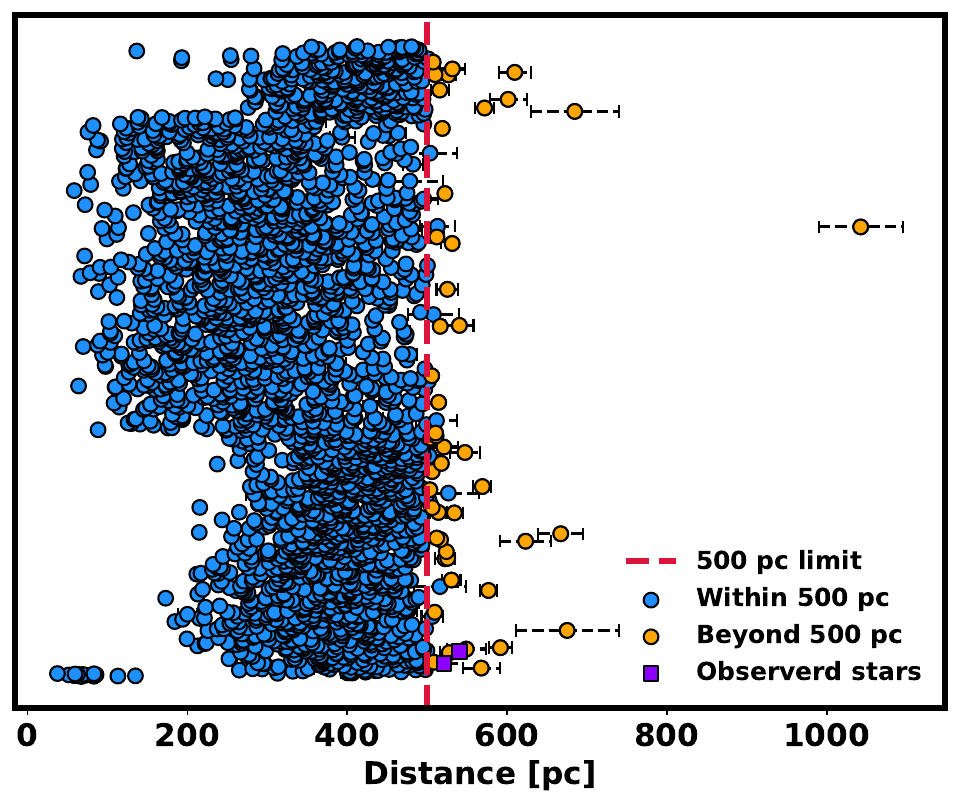}
\caption{Updated distances using DR3 parallaxes. The red dashed vertical line represents the 500 pc limit. Blue dots indicate targets with distances consistent with 500 pc within their errors. Orange dots indicate those beyond 500 pc, and purple squares mark observed targets beyond the limit. Error bars are not visible for some dots as they are smaller than the dot size.}
\label{fig:updated_distances}
\end{figure}

A second issue related to the updated parallaxes is the distance. The low errors allow us to calculate the distance as the inverse of parallax and propagate the error accordingly. However, with the updated parallaxes, some stars are beyond the 500 pc limit. This is shown in Fig. \ref{fig:updated_distances}, where most targets are within 500 pc, considering their errors (blue dots). However, 47 stars ($\sim 2\%$) exceed the 500 pc limit (orange dots). Given that only two of these (CD-662575 and HD181405) have been observed already, we included them in this analysis. However, they will not be included when the final complete observed sample is compared with the BPS study.

\begin{figure}[!ht]
\centering
\includegraphics[height=0.9\linewidth, width=0.8\linewidth]{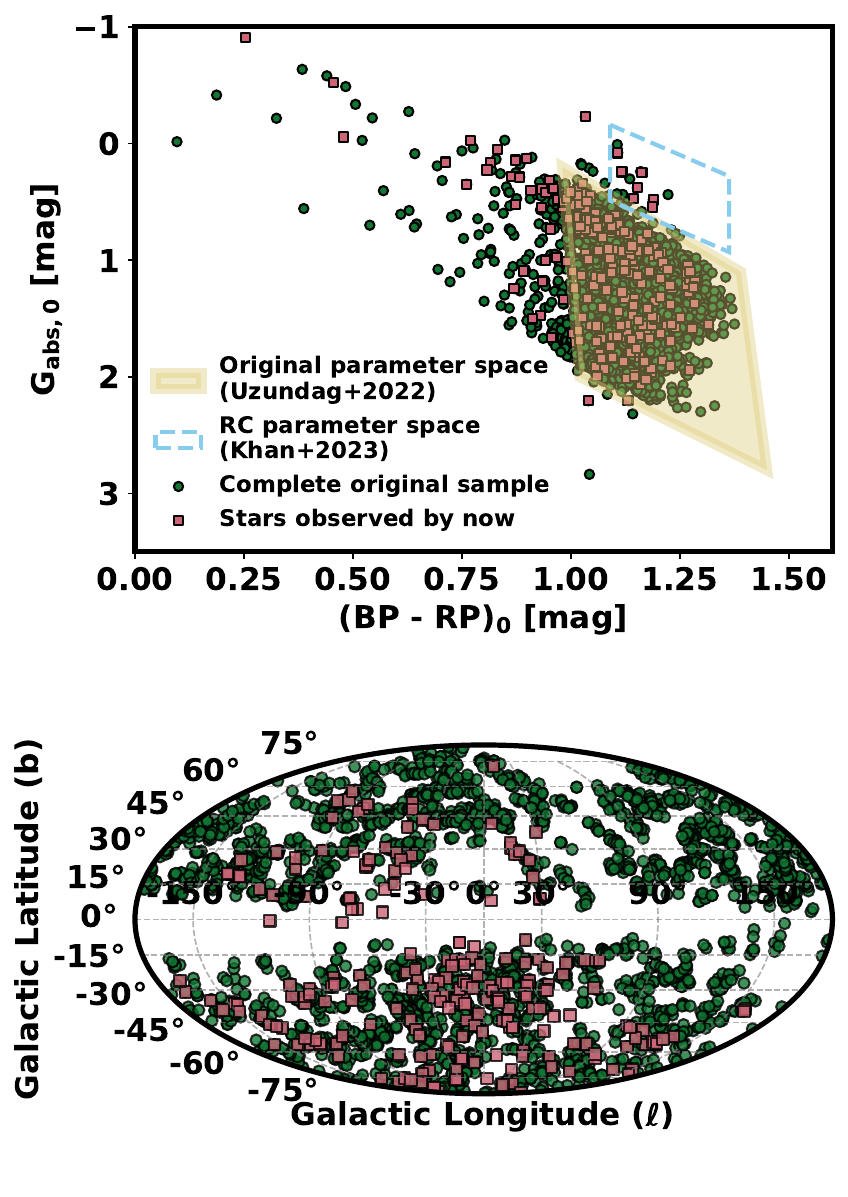}
\caption{\textbf{Upper panel}: Updated Gaia colour-magnitude distribution for the 2777 stars selected in \citetaliasads{2022A&A...668A..89U} using DR3 parallaxes and extinction coefficients. Green dots represent the original complete sample, while pink squares represent the stars observed by now. The yellow shaded box is defined in \citetaliasads{2022A&A...668A..89U} (eq. (1)–(4)), and the sky-blue dashed rectangle is the RC parameter space from \citetads{2023A&A...680A.105K}. \textbf{Lower panel}: Sky distribution (Mollweide projection) in Galactic coordinates. The colours are the same as in the upper plot.}
\label{fig:cmd_plus_sky_dist}
\end{figure}

Fig. \ref{fig:cmd_plus_sky_dist} is an update of Fig. 1 from \citetaliasads{2022A&A...668A..89U} using DR3 parallaxes and applying the interstellar reddening. The original parameter space from \citetaliasads{2022A&A...668A..89U} (eq. (1)–(4)) is defined by the yellow shaded area: the lower cut excludes MS stars, the side cuts exclude UV and IR excess objects, and the upper cut avoids RC stars. To validate the upper cut, we over-plotted the RC parameter space defined in the (G$_{\text{abs}}$,T${\text{eff}}$) plane in \citetads{2023A&A...680A.105K}, converting T${\text{eff}}$ to BP-RP using the colour-T${\text{eff}}$ relations from \citetads{2021A&A...653A..90M}. For the sky distribution (lower panel in Fig. \ref{fig:cmd_plus_sky_dist}), we used the Mollweide projection in Galactic coordinates instead of the Aitoff projection in equatorial coordinates from \citetaliasads{2022A&A...668A..89U}, better illustrating that the Galactic plane is mostly uncovered.

With updated measurements, 264 of the 2727 stars now fall outside the original parameter space defined in \citetaliasads{2022A&A...668A..89U} (Fig. \ref{fig:cmd_plus_sky_dist}, upper panel), representing $<$ 10\% of the sample. Out of the stars that have been observed by now, 54 of 228 ($\sim$24\%) lie outside the original parameter space. Since observations began before the release of DR3 (see Table \ref{tablespec1}), this was unavoidable. Some of the stars lie within the RC parameter space from \citetads{2023A&A...680A.105K} and are most likely the expected RC contaminants.

In this way, the original 2777 sample (from \citetaliasads{2022A&A...668A..89U}) has been further cleaned to 2463 stars.

\section{Comparison with CASCADES}
\label{Appendix_CASCADES}

\begin{figure*}[ht!]
\centering
\subfigure{\includegraphics[height=0.3\linewidth, width=0.3\linewidth]{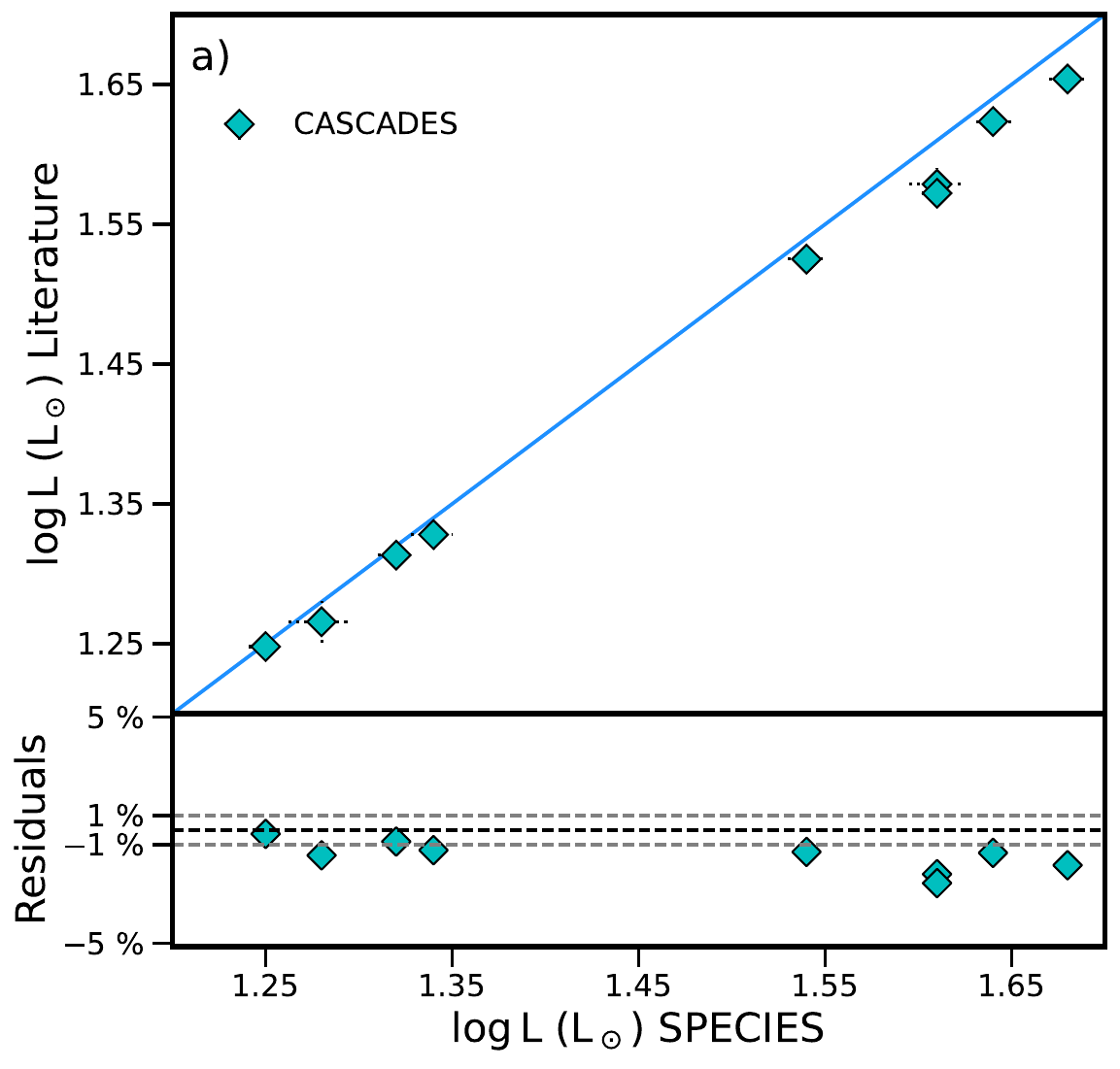}}
\subfigure{\includegraphics[height=0.3\linewidth, width=0.3\linewidth]{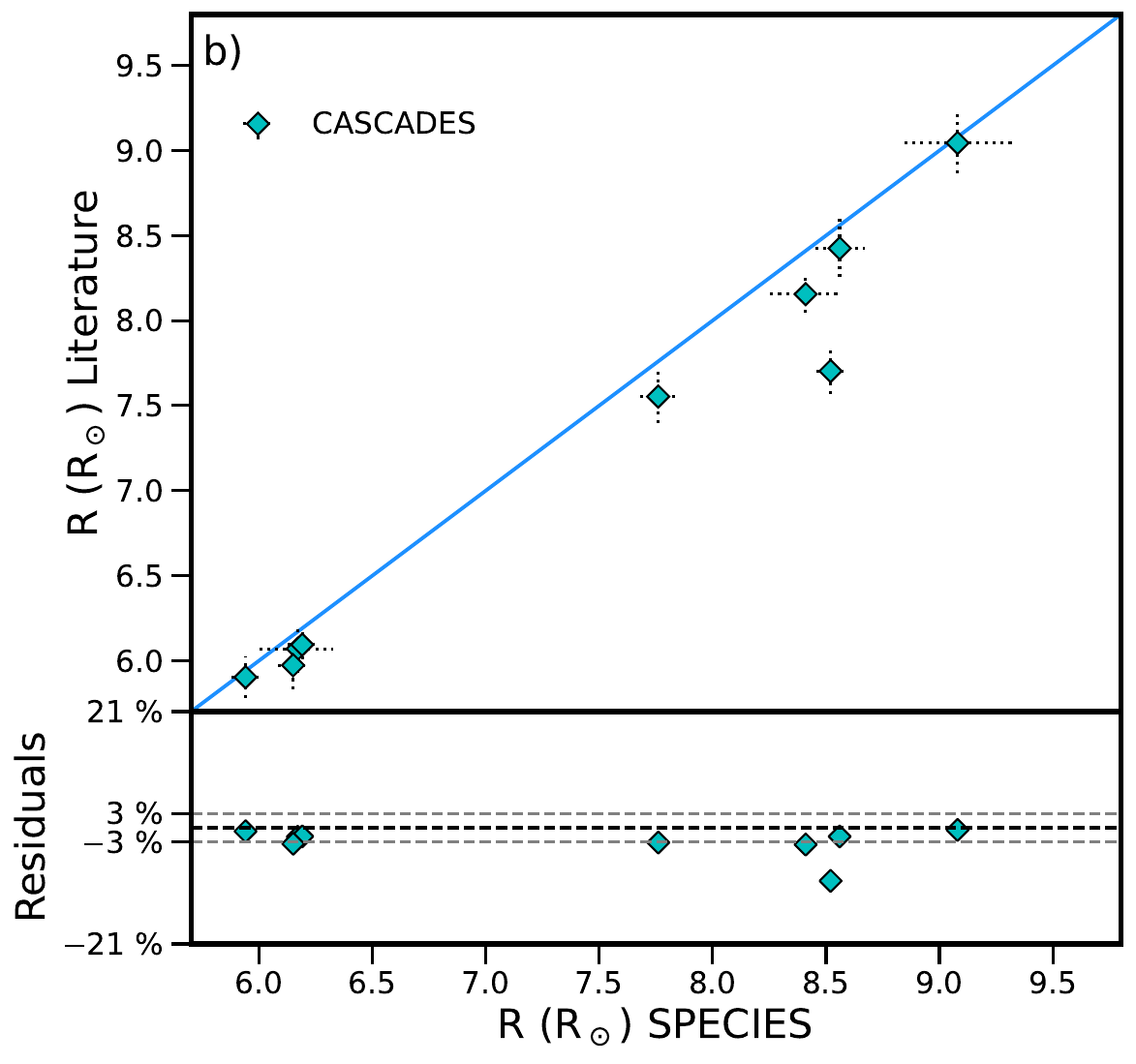}}
\subfigure{\includegraphics[height=0.3\linewidth, width=0.3\linewidth]{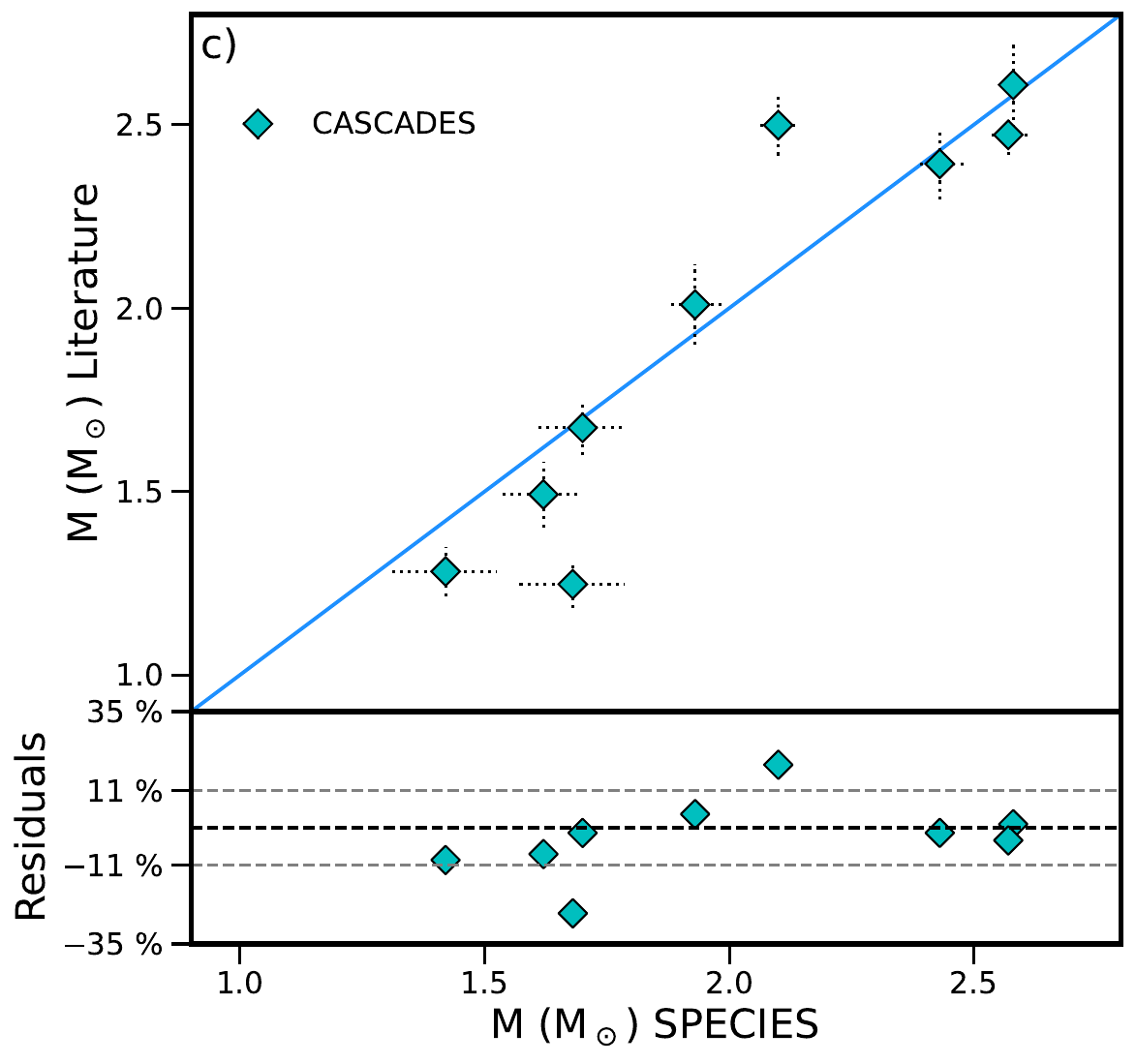}}
\caption{Comparison of stellar parameters obtained with \texttt{SPECIES} and CASCADES. The dashed grey lines in the lower panel show the 1$-\sigma$ dispersion of the residuals.}
\label{fig:comparison_cascades}
\end{figure*}

Fig. \ref{fig:comparison_cascades} shows the comparison of stellar parameters derived by SPECIES and CASCADES that are not part of Fig. \ref{fig:comparison_large_surveys}.

We can see a remarkable agreement in the parameters shown in Fig. \ref{fig:comparison_cascades}. Although the comparison suffers due to the poor number of matches, one can recognize that all the parameters agree well within the 1-$\sigma$ dispersion. As we already mentioned in sec. \ref{sec:Discussion}, this agreement is expected due to the similarities in the methodologies to derive the parameters. However, it is still interesting to remark on the relatively low dispersion in the derived $\log$L, mass, and radius, with some scatter that can be due to the different evolutionary tracks employed in both cases.

\vspace{-2.5cm}

\section{Radial velocity measurements}
\label{Appendix}
\vspace{-0.6cm}
\setlength{\tabcolsep}{1pt}
\renewcommand{\arraystretch}{1}
\begin{table}[h!]
\caption{Radial velocity measurements for all stars that are presented in this work.} \label{tab:RV_Diego}
\centering
\begin{tabular}{cccccc}
\toprule\toprule Star & Category & JD & S/N & RV & RV error \\ 
& & $-2450000$ & & (km\,s$^{-1}$) & (km\,s$^{-1}$) \\
\midrule
*BCAP         & 1        & 8650.82  & 64  & -20.962 & 0.003   \\
              &          & 9538.57  & 62  & -20.994 & 0.005   \\
              &          & 9673.89  & 62  & -20.974 & 0.003   \\
              &          & 9737.90  & 66  & -20.976 & 0.003   \\
              &          & 9739.91  & 29  & -20.973 & 0.003   \\
HD131900 & 2 & 10140.49 & 11 & -5.87   & 0.01  \\
         &                       & 10140.5  & 35 & -6      & 0.03  \\
         &                       & 10140.5  & 44 & -5.83   & 0.02  \\
HD196800 & 3 & 10224.5  & 31 & -63.653 & 0.008 \\
         &                       & 10225.49 & 44 & -63.65  & 0.007 \\
         &                       & 10227.49 & 28 & -63.644 & 0.006 \\
HD1000   & 3 & 10140.93 & 38 & -13.83  & 0.02  \\
         &                       & 10140.93 & 10 & -13.84  & 0.03  \\
HD116338 & 1 & 8651.59  & 37 & -30.528 & 0.004 \\
         &                       & 9673.73  & 43 & -29.169 & 0.003 \\
         &                       & 9737.7   & 62 & -34.554 & 0.003 \\
         &                       & 10140.55 & 62 & -28.362 & 0.004 \\
& & $\vdots$ & & & \\ 
TYC6951-496-1 & 3        & 8650.77  & 62  & 5.741   & 0.003  \\
\midrule
\end{tabular}
\tablefoot{This is a short version of the table for the online version of this manuscript. All the tables mentioned in these appendixes are available at the CDS.\\
All the stars listed here were observed using the CORALIE instrument. The category is based on the classification method from \citetaliasads{2022A&A...668A..89U}.}
\end{table}

\FloatBarrier 

\vspace{5cm}
\section{New potential binaries}
\label{Appendix:NSS}

\begin{table}[h!]
\setlength{\tabcolsep}{0pt}
\renewcommand{\arraystretch}{1}
\centering
\caption{Potential new low-mass RGB+MS binary systems found in the Gaia DR3 NSS catalogue with measured orbital parameters.} \label{tab:cross_NSS_twobody}
\begin{tabular}{ccccc}
\toprule\toprule 
Star & Solution type$^*$  & Period & $T_0$ & $e$ \\ 
     &               & (days) & (days) &    \\ 
\midrule 
HD155046       & AS & 899$\pm$8       & -401$\pm$8   & 0.19$\pm$0.01   \\
 &  & $\vdots$  &   &   \\
HD137608       & AS & 146.08$\pm$0.04 & 35$\pm$1     & 0.129$\pm$0.004 \\
\midrule
\end{tabular}
\tablefoot{$T_0$ refers to the time of periastron and $e$ to the eccentricity.\\$^*$ AS: AstroSpectroSB1}
\end{table}

\vspace{-0.4cm}

\begin{table}[h!]
\setlength{\tabcolsep}{24pt}
\renewcommand{\arraystretch}{1}
\centering
\caption{Potential new low-mass RGB+MS binary systems found in the Gaia DR3 NSS catalogue with no measured orbital parameters.} \label{tab:cross_NSS_acc}
    \begin{tabular}{cc}
\toprule\toprule 
Star & Solution type \\ 
    &  \\
\midrule
BD+11208       & FirstDegreeTrendSB1  \\
& $\vdots$ \\
HD23151        & Acceleration7       \\
\midrule
\end{tabular}

\end{table}

\FloatBarrier 
\vspace{-0.65cm}
\section{Fundamental parameters}
\label{Appendix:SPECIES}
\setlength{\tabcolsep}{1pt}
\renewcommand{\arraystretch}{1.7}
\begin{table}[h!]
\caption{$\tt SPECIES$ results for stars observed with CORALIE.}\label{tab: species_all}
\centering
\begin{tabular}{cccccccc}
\toprule\toprule
Star & T$_{\text{eff}}$& $\cdots$ & EEP & P$_\text{MS}$ & P$_\text{RGB}$ & P$_\text{HB}$ \\
     & (K) &  &  &  & \\  
\midrule
HD116338      & 5126$\pm$50 & $\cdots$ & 527.2$^{+1.2}_{-1.2}$    & 0 & 0.98 & 0.02 \\
  &  &  $\vdots$ &  &  &  &\\
TYC8519-263-1 & 4642$\pm$74 & $\cdots$ & 505.4$^{+0.8}_{-0.7}$    & 0 & 1   & 0\\
\midrule
\end{tabular}
\tablefoot{$P_{\text{MS}}$, $P_{\text{RGB}}$ and $P_{\text{HB}}$ correspond to the probability of the star being on the MS, RGB, or HB, respectively.}
\end{table}

\end{appendix}

\end{document}